# Pursuit of Truth and Beauty in Lean 4

Formally Verified Theory of Grammars, Optimization, Matroids

by

**Martin Dvořák**

March, 2026

*A thesis submitted to the*

*Graduate School*

*of the*

*Institute of Science and Technology Austria*

*in partial fulfillment of the requirements*

*for the degree of*

*Doctor of Philosophy*

Committee in charge:

Michael Sammler, Chair

Vladimir Kolmogorov

Jasmin Blanchette

Adam Topaz

Matthew Ballard

Jireh Loreaux

Alex Kontorovich

Terence Tao

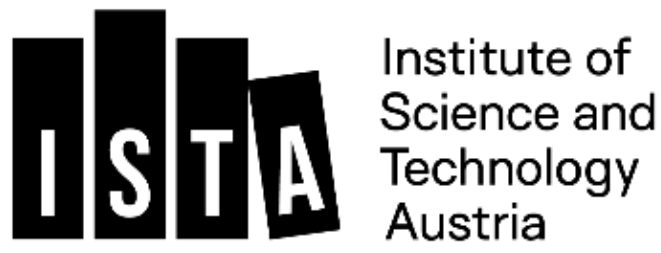

Revised versions of this thesis will be uploaded to arXiv. Please report errata and clarity issues:

martin.dvorak@matfyz.cz

The thesis of Martin Dvořák, titled *Pursuit of Truth and Beauty in Lean 4*, is approved by:

**Supervisor**: Vladimir Kolmogorov, ISTA, Klosterneuburg, Austria

Signature: ________________________________________

**Co-supervisor**: Jasmin Blanchette, Ludwig-Maximilians-Universität, München, Germany

Signature: ________________________________________

**Committee Member**: Adam Topaz, University of Alberta, Edmonton, Canada

Signature: ________________________________________

**Committee Member**: Matthew Ballard, University of South Carolina, Columbia, USA

Signature: ________________________________________

**Committee Member**: Jireh Loreaux, Southern Illinois University Edwardsville, St. Louis, USA

Signature: ________________________________________

**Committee Member**: Alex Kontorovich, Rutgers University, New Jersey, USA

Signature: ________________________________________

**Committee Member**: Terence Tao, University of California, Los Angeles, USA

Signature: ________________________________________

**Defense Chair:** Michael Sammler, ISTA, Klosterneuburg, Austria

Signature: ________________________________________

Signed page is on file

I hereby declare that this thesis is my own work and that it does not contain other people's work without it being so stated; this thesis does not contain my previous work without this being stated, and the bibliography contains all the literature that I used in writing the dissertation.

I accept full responsibility for the content and factual accuracy of this work, including the data and their analysis and presentation, and the text and citation of other work.

~~I declare that this is a true copy of my thesis, including any final revisions, as approved by my thesis committee~~, and that this thesis has not been submitted for a higher degree to any other university or institution.

I certify that any republication of materials presented in this thesis has been approved by the relevant publishers and co-authors.

Signature: ______________________

Martin Dvořák

March, 2026

Signed page is on file

## Abstract

This thesis documents a voyage towards truth and beauty via formal verification of theorems. To this end, we develop libraries in Lean 4 that present definitions and results from diverse areas of MathematiCS (i.e., Mathematics and Computer Science). The aim is to create code that is understandable, believable, useful, and elegant. The code should stand for itself as much as possible without a need for documentation; however, this text redundantly documents our code artifacts and provides additional context that isn't present in the code. This thesis is written for readers who know Lean 4 but are not familiar with any of the topics presented. We manifest truth and beauty in three formalized areas of MathematiCS.

We formalize general grammars in Lean 4 and use grammars to show closure of the class of type-0 languages under four operations; union, reversal, concatenation, and the Kleene star.

Our second stop is the theory of optimization. Farkas established that a system of linear inequalities has a solution if and only if we cannot obtain a contradiction by taking a linear combination of the inequalities. We state and formally prove several Farkas-like theorems over linearly ordered fields in Lean 4. Furthermore, we extend duality theory to the case when some coefficients are allowed to take "infinite values". Additionally, we develop the basics of the theory of optimization in terms of the framework called General-Valued Constraint Satisfaction Problems, and we prove that, if a Rational-Valued Constraint Satisfaction Problem template has symmetric fractional polymorphisms of all arities, then its basic LP relaxation is tight.

Our third stop is matroid theory. Seymour's decomposition theorem is a hallmark result in matroid theory, presenting a structural characterization of the class of regular matroids. We aim to formally verify Seymour's theorem in Lean 4. First, we build a library for working with totally unimodular matrices. We define binary matroids and their standard representations, and we prove that they form a matroid in the sense how Mathlib defines matroids. We define regular matroids to be matroids for which there exists a full representation rational matrix that is totally unimodular, and we prove that all regular matroids are binary. We define 1-sum, 2-sum, and 3-sum of binary matroids as specific ways to compose their standard representation matrices. We prove that the 1-sum, the 2-sum, and the 3-sum of regular matroids are a regular matroid, which concludes the composition direction of the Seymour's theorem. The (more difficult) decomposition direction remains unproved.

In the pursuit of truth, we focus on identifying the trusted code in each project and presenting it faithfully. We emphasize the readability and believability of definitions rather than choosing definitions that are easier to work with. In search for beauty, we focus on the philosophical framework of Roger Scruton, who emphasizes that beauty is not a mere decoration but, most importantly, beauty is the means for shaping our place in the world and a source of redemption, where it can be viewed as a substitute for religion.

## Acknowledgments

I'd like to express my gratitude to Vladimir Kolmogorov and Jasmin Blanchette for patient supervision and to David Bartl for long discussions.

When it comes to specific projects, I owe thanks to the following people:

- Kate Kočická for discussing ideas about the Kleene star construction
- Patrick Johnson, Floris van Doorn, and Damiano Testa for their small yet very valuable contributions to grammars in Lean 3
- Damiano Testa for help with linters in Lean 4
- Henrik Böving for help with generalization from extended rationals to extended linearly ordered fields and implementing the notation
- Kevin Buzzard for help with API for extended linearly ordered fields
- Richard Copley for pointing out that our conversion from linearly ordered fields to extended linearly ordered fields should be bundled as a homomorphism
- Antoine Chambert-Loir for consultations about linear programming
- Andrew Yang for a proof of `Finset.univ_sum_of_zero_when_not`
- Emilie Shad for help with proving `ValuedCSP.Instance.solutionVCSPtoBLP_cost`
- Yaël Dillies for a proof of `Multiset.toList_map_sum`
- Damiano Testa for a proof of `Finset.univ_sum_multisetToType`
- Pietro Monticone for help with `leanproject`
- Riccardo Brasca for a proof of `Matrix.one_linearIndependent`
- Johan Commelin and Edward van de Meent for their advice about proving `Matrix.fromBlocks_isTotallyUnimodular`
- Aaron Liu for advice about handling `HEq`
- Yaël Dillies for advice about inverting functions
- Christian Merten for advice about subtypes and submodules
- Jireh Loreaux for advice about singular matrices

## About the Author

Martin Dvořák completed a BSc in Computer Science with focus on Mathematical Linguistics and an MSc in Artificial Intelligence with focus on Intelligent Agents, both at the Faculty of Mathematics and Physics of the Charles University in Prague, Czechia.

## List of Collaborators and Publications

### Papers written during my Ph.D. that are used in this thesis

Martin Dvorak and Jasmin Blanchette (2023). *Closure Properties of General Grammars — Formally Verified* (paper in ITP 2023 — the 14th International Conference on Interactive Theorem Proving)

- I did all implementation and wrote most of the text (but not most of the Introduction).

Martin Dvorak and Vladimir Kolmogorov (2024). *Duality Theory in Linear Optimization and its Extensions — Formally Verified* (will be published in Annals of Formalized Mathematics (acceptance obtained on 2025-09-23))

- I did all implementation and wrote most of the text.

Martin Dvorak, Tristan Figueroa-Reid, Rida Hamadani, Byung-Hak Hwang, Evgenia Karunus, Vladimir Kolmogorov, Alexander Meiburg, Alexander Nelson, Peter Nelson, Mark Sandey, Ivan Sergeev (2025). *Composition Direction of Seymour's Theorem for Regular Matroids — Formally Verified* (technical report)

- Overview of contributions is on the next page.

### Papers written during my Ph.D. but not used in this thesis

Martin Dvorak and Sara Nicholson (2021). *Massively Winning Configurations in the Convex Grabbing Game on the Plane* (paper in CCCG 2021 — the 33rd Canadian Conference on Computational Geometry)

Martin Dvorak and Vladimir Kolmogorov (2024). *Generalized Minimum 0-Extension Problem and Discrete Convexity* (paper in Mathematical Programming — A Publication of the Mathematical Optimization Society)

Matthew Bolan, Joachim Breitner, Jose Brox, Mario Carneiro, Martin Dvorak, Andres Goens, Aaron Hill, Harald Husum, Zoltan Kocsis, Bruno Le Floch, Lorenzo Luccioli, Alex Meiburg, Pietro Monticone, Giovanni Paolini, Marco Petracci, Bernhard Reinke, David Renshaw, Marcus Rossel, Cody Roux, Jeremy Scanvic, Shreyas Srinivas, Anand Rao Tadipatri, Terence Tao, Vlad Tsyrklevich, Daniel Weber, Fan Zheng (2025). *The Equational Theories Project: Advancing Collaborative Mathematical Research at Scale* (technical report)

**Seymour project**
(authors ordered from the biggest contribution to smaller for every item)

Conceptualization: Ivan Sergeev, Vladimir Kolmogorov

Blueprint: Ivan Sergeev, Martin Dvorak, Mark Sandey, Pietro Monticone

Implementation:

- Basic
  - Basic Martin Dvorak
  - Conversions Martin Dvorak
  - Fin Martin Dvorak
  - FunctionDecompose Martin Dvorak
  - FunctionToHalfSum Martin Dvorak
  - Sets Martin Dvorak
  - SignTypeCast Martin Dvorak, Tristan Figueroa-Reid
  - SubmoduleSpans Peter Nelson, Martin Dvorak
- Matrix
  - Basic Martin Dvorak
  - Conversions Martin Dvorak
  - Determinants Martin Dvorak
  - LinearIndependence Peter Nelson, Rida Hamadani, MD, Riccardo Brasca
  - LinearIndependenceBlock Martin Dvorak
  - PartialUnimodularity Martin Dvorak
  - Pivoting Martin Dvorak, Ivan Sergeev, Tristan Figueroa-Reid
  - Signing Martin Dvorak
  - SubmoduleBasis Martin Dvorak
  - Support Martin Dvorak
  - TotalUnimodularity Martin Dvorak
  - TotalUnimodularityTest Tristan Figueroa-Reid, Martin Dvorak
- Matroid
  - Basic Martin Dvorak
  - Duality Ivan Sergeev, Martin Dvorak, Evgenia Karunus
  - FromMatrix Martin Dvorak, Byung-Hak Hwang, Ivan Sergeev
  - Graphicness Ivan Sergeev, Tristan Figueroa-Reid, Martin Dvorak
  - R10 Tristan Figueroa-Reid, Martin Dvorak
  - Regularity Martin Dvorak, Ivan Sergeev, Tristan Figueroa-Reid
  - StandardRepresentation MD, IS, Tristan Figueroa-Reid, Christian Merten
  - Sum1 Martin Dvorak, Byung-Hak Hwang
  - Sum2 Ivan Sergeev, Martin Dvorak, Tristan Figueroa-Reid
  - Sum3
    - Matrix level Ivan Sergeev, Evgenia Karunus, Alexander Meiburg, MD
    - StdRepr level Martin Dvorak
    - Matroid level Martin Dvorak
- Results Martin Dvorak
- Seymour Martin Dvorak

Writing: Ivan Sergeev, Martin Dvorak, Alexander Nelson, Byung-Hak Hwang, Mark Sandey, Rida Hamadani, Tristan Figueroa-Reid, Alexander Meiburg

# List of Abbreviations

| | |
|---|---|
| **iff** | IF and only iF[1] |
| **ite** | If Then Else |
| **LHS** | Left-Hand Side |
| **RHS** | Right-Hand Side |
| **ITP** | Interactive Theorem Proving |
| **PR** | Pull Request |
| **API** | Application Programming Interface |
| **IDE** | Integrated Development Environment |
| **OOP** | Object-Oriented Programming |
| **LP** | Linear Program |
| **SMT** | Satisfiability Modulo Theories |
| **VCSP** | Valued Constraint Satisfaction Problem |
| **AI** | Artificial Intelligence |
| **CFG** | Context-Free Grammar |

[1] In meta-theoretic statements, we say the full phrase "if and only if".

# Pursuit of Truth and Beauty in Lean 4

Night had fallen over Klosterneuburg, and the world outside had gone quiet. In the distance, the Danube moved unseen, its slow current whispering beneath the hum of the streetlamps. Inside, the faint light of the monitor trembled across the desk. A cursor blinked — steady, expectant — as if holding its breath. He sat there, fingers hovering over the keyboard, not yet typing. Somewhere deep in the processor, Lean was sleeping, waiting to be awakened by another tactic call.

It had been hours since he had last spoken aloud. The office smelled faintly of cold tea and half-eaten cookies forgotten beside the keyboard. Lines of code stretched down the screen like verses from an undeciphered scripture. He knew, in a small way, that he was not alone. Across centuries, others had stared into similar darkness — not at pixels, but at the stars.

He imagined one of them. A man in the dust of Athens stood at the edge of a wavering circle formed by firelight. A few students leaned in as he spoke, tracing logic in the air with his hand. The flame flickered across his face, turning each idea into a dance of shadows. Every now and then, he bent down and scratched something in the sand. At one point he paused, then said: “There is something that does not move.” They watched him, unsure of what he meant, unaware that the thought carried in that trembling light would one day become a line of reasoning, inscribed not in dust but in code.

The cursor blinked again — once, twice — as if it too were waiting for him to remember what he was doing. The hum of the computer filled the silence, like a modern firelight whispering through circuits instead of flames. He thought of those tracings in the sand, of how fragile they must have been — a breath of wind, a stray step, and they would vanish. And yet, the idea they held managed to survive every storm.

He rested his hands on the keyboard. Each symbol he wrote was another attempt at finding the ultimate truth. Only now, the medium was made of logic — crystalline and incorruptible. Lean would not let him pretend; it would not mistake gesture for proof. It demanded the same thing the philosopher once had: "Show me why it must really be the way you say it is!"

He paused before pressing Enter, watching his reflection shimmer faintly on the screen. Was he teaching the machine to reason, or was the machine teaching him to see the truth? It had no interest in his intuition, only in what endured its exacting demands.

He pressed the key. Lean lingered for a moment — a silent patience — and then answered. No red, no protest. The state of the proof in the Infoview on the right side of the screen changed. Exactly as he wanted.

He knew there would be errors ahead, tangles of logic yet to unwind. But for now, the silence between him and the machine felt full — enough to call it peace. Tomorrow, the proof would continue, line by line, in the patient conversation between the mind and the machine. For where clarity deepens, truth takes form — and in the form, he hopes to find beauty.

## 1 Introduction

My Ph.D. thesis presents four repositories of formalized MathematiCS written in Lean 4.18.0, building on top of the Lean mathematical library Mathlib [1] revision `aa936c36e8484abd300` (dated 2025-04-01):

- Duality theory in linear optimization[2], paper about which was submitted to the Annals of Formalized Mathematics, where the paper has been accepted but not published yet.
- Valued Constraint Satisfaction Problems[3], which was never published academically.
- Seymour project[4], which has a technical report on arXiv [2].
- Grammars[5], paper about which (in particular, about its Lean 3 version[6]) was published at the ITP 2023 conference [3].

Throughout the thesis, we will see the word MathematiCS, whose spelling (with capital C and capital S) is motivated by my desire to not draw a line between Mathematics and Computer Science, so I write MathematiCS to refer to the entire area that would be conventionally split into two separate departments at most universities.

This thesis puts Lean first, which presents a huge advantage over reading the above-mentioned papers, which frequently mix formalisms and informalisms together, neither of which is developed in sufficient detail in those papers. This thesis avoids all forms of informalism and properly explains everything using only Lean code and plain English. As a result, reading this thesis is precisely tailored for students and professionals who are fluent in Lean. At the same time, this thesis is completely useless for people who don't know Lean, as this thesis neither explains Lean for beginners nor presents any body of knowledge that would be digestible without understanding Lean. The target audience is a curious reader who has already mastered Lean (the language) and wants to learn about new-for-the-reader areas of MathematiCS. I chose this audience because there are currently no works tailored specifically for them.

While this thesis doesn't assume any prior exposure to the areas of MathematiCS studied in it, this thesis still takes into account readers who already know these areas from informal MathematiCS — the formal definitions are written so that they can usually be easily identified with their informal counterparts. Since the readers' ability to trust our results is our top priority, we usually made the formal definitions faithful to textbook definitions even in situations where it made the formal development more difficult. A good formalization practice, adherent to the principles of intellectual honesty, is to write all definitions and state the main results before starting the work on proofs — this way, the objectives are fixed in advance, so that our definitions and statements reflect the intended mathematical content rather than the path of least resistance in proving the theorems. Not every time were we able to stick to this philosophy; in the process of developing our projects, we committed certain misformalizations that would make it literally impossible to prove the results, hence the definitions and/or the statements of the theorems had to be modified.

This thesis is not optimized for casual readers who just want to skim the thesis to know what it is about — they should rather read the Abstract and stop there. Instead, I focus on delivering the best possible experience for voracious readers who will read the thesis cover-to-cover and hopefully remember something from it for the rest of their lives. Some readers will be surprised that, in addition to saying what was done, my thesis focuses on capturing the vibe, what it felt like doing it.

---

[2] https://github.com/madvorak/duality/tree/v3.5.0

[3] https://github.com/madvorak/vcsp/tree/v8.2.0

[4] https://github.com/Ivan-Sergeyev/seymour/tree/v1.2.0

[5] https://github.com/madvorak/chomsky/tree/v1.2.0

[6] https://github.com/madvorak/grammars

Because formatting of this thesis is less restricted than papers submitted to established journals and conferences, which always have rigid formatting templates, writing this thesis gave me more freedom to customize the graphical appearance and allowed me to achieve better typesetting than in the above-mentioned papers.

This thesis needs to be read on white background. On some computers, when people choose a dark theme, they get documents displayed as white text on black background. However, this thesis really needs to be read as black text on white background, with code snippets using three different colors (dark gray, dark blue, dark red) but still on white background; otherwise, the lexical highlighting will destroy the experience rather than enhance it.

## *1.1 Contents of this thesis*

We start with a bit of history (Section 1.2), then I present the goals of my Ph.D. (Section 1.3). Chapter 2 describes everything that isn't my work but that laid the foundation for my work. Chapter 3 (on optimization theory) presents my two projects that I worked on in 2023/2024. Chapter 4 (on matroid theory) presents a project that I worked on in 2024/2025 with a growing group of collaborators, reaching the point of 12 authors at the end. Chapter 5 (on the theory of grammars) presents a project that I developed in 2022 and translated from Lean 3 to Lean 4 at the end of 2025.

Unless stated otherwise, everything is a formalization of known MathematiCS.

Code snippets (whether repeating our code or mentioning code of upstream projects) are intended to be accurate if possible and feasible, but many small amendments have been done to allow understanding them on paper or generally when read outside of IDE.

- `Type` is written in place of `Type u`, `Sort u`, `Type*`, `Sort*`, and so on (the only exception being the discussion of axioms).
- Decidability is omitted unless a part of a structure.
- When a definition or a lemma/theorem statement contains a proof, the proof is usually replaced by `sorry` in the thesis, while the repository contains a full implementation.
- Names of proofs are omitted unless referenced in the text.
- Arguments inserted via the `variable` command are visibly displayed when quoting out of context.
- Various small simplifications for better readability have been made.

In the text, we sometimes use unusual hyphenation (incorrect from the viewpoint of the English language). For example, we write "nonassociative-semiring" (with a hyphen) but "noncommutative semiring" (as two words) because every noncommutative semiring is a semiring whereas nonassociative-semiring isn't a semiring (since the definition of a semiring requires associativity of multiplication but not commutativity of multiplication). The example above (the nonassociative-semiring) is an example of so-called semantic extension [4]. Hyphenating such phrases is highly unusual; compare it, for example, with "almond milk" or "sea horse"—they are never hyphenated.

## *1.2 What is formalization and where it came from*

*Formalization* is expressing mathematical truths *formally*, as a consequence of certain *axioms*. The motivation is that, if we *assume* axioms to be true, everything that follows from them using valid logical rules is also true. Let's push the question [what valid logical rules are] aside for a while and look at a bit of history.

The first documented effort at formalization is Euclid's formalization of geometry in his book Stoikheia [5] at around 300 BCE [6]. Attempts to formalize the rest of mathematics are much more recent. In 1870s, Georg Cantor [7] introduced *set theory* as an attempt to lay foundations for all mathematics. This so-called naive set theory led to paradoxes, out of which Russell's paradox [8] is probably the most famous. These paradoxes were hopefully fixed by Ernst Zermelo [9] [10] and Abraham Fraenkel [11] in what is today known as ZF set theory. Another attempt to fix the problems found in naive set theory led to the development of *type theory* by Alfred North Whitehead and Bertrand Russell in their three-volume book series Principia Mathematica [12]. Working with their type theory is infamously hard; for example, it took them 360 pages of theory building until they were able to prove `1+1=2`. Principia Mathematica was also criticized by Kurt Gödel [13] for not having a precise statement of the syntax of their formalism. Nevertheless, Principia Mathematica was a huge leap forward in how precisely its ideas were expressed compared to pre–first-order logic set theory [14].

Let's focus on the logical rules now. Propositional logic was known since 350 BCE by Aristotle and later by Stoic philosophers Chrysippus and Sextus Empiricus [15], but propositional logic was too weak to express anything of interest to mathematicians. First-order logic, which enriched propositional logic with quantifiers, laid the foundations for axiomatic set theory and, thereby, for a very large portion of mathematics. Unfortunately, the first-order logic came very late in comparison [16]; it was first suggested by Charles Peirce [17][7] in 1885 but widely established by David Hilbert and Wilhelm Ackermann [18] in 1928. Other names [16] strongly connected to beginnings of the first-order logic are Leopold Löwenheim, Paul Bernays, and Thoralf Skolem. In my personal opinion, the first time mathematics really became formal was when the axiomatic set theory got encoded in the first-order logic. In this spirit, we can say that set theory became formal before type theory became formal.

Kurt Gödel, who was a prominent member of the Vienna Circle [19], established both the strengths and limitations of the first-order logic [20]. His completeness theorem [21] from 1929 says that a first-order sentence is provable if and only if it is true. His first incompleteness theorem [22] from 1930 says that no consistent enumerable extension of Robinson arithmetic can prove all theorems about the standard model of natural numbers. His second incompleteness theorem [23] from 1931 says that no consistent enumerable extension of Peano arithmetic can prove its own consistency (in fact, the theory can prove its own consistency if and only if it is inconsistent).[8] Since the incompleteness theorems are often misused [24], I would like to clarify a few things. First, many people wonder how come that the completeness theorem and the first incompleteness theorem don't contradict each other. It is because the completeness theorem talks about the logical system, i.e., about how the inference rules relate to the semantics of the first-order logic. Here, a sentence is provable if and only if it is true, where "true" means that it holds in all models. In contrast, the incompleteness theorems talk about properties of theories. If said consistent enumerable extension of Robinson arithmetic cannot prove a certain theorem about the standard model of natural numbers then, by the completeness theorem, we know that this theory has a model where the putative theorem doesn't hold. As a consequence of both theorems (the completeness and the first incompleteness), we learn that there is no enumerable extension of Robinson arithmetic with exactly one model; if we allow the standard model of natural numbers, we inevitably allow nonstandard models of natural numbers, too. There is no contradiction in it. Second, many people think that the second incompleteness theorem implies the existence of god(s). However, to the best of my knowledge, nobody has yet presented a coherent argument to justify such a wild deduction.

---

[7] It may be not immediately clear that what Charles Peirce outlines in his paper *On the Algebra of Logic* is the first-order logic. He speaks in the language of objects, signs, icons, indices, tokens, and minds.

[8] In fact, Gödel analyzed a broader class of systems than discussed here. However, let's not digress too much.

In the 1940s, first digital computers appeared as an attempt to automate tedious calculations. Early machines such as ENIAC[9] [25], EDSAC[10] [26], and UNIVAC[11] [27] were room-sized devices operating with vacuum tubes, and programming them required low-level manipulation of hardware instructions. At roughly the same time, Alan Turing's theoretical work on computability [28] and John von Neumann's architectural model [29] provided a conceptual framework for understanding computation itself. Initially, computers were viewed primarily as fast calculators (numbers on the input, numbers on the output).

The rise of computers in the mid-20th century seemed to promise a new era for mathematics. After computers mastered numerical calculations, people began to think that computers would be able to prove theorems automatically. Soon people realized that the research of automated theorem provers was mostly unsuccessful. The space of all possible things mathematicians can do when proving theorems is enormous. What computers are much better at is to check a mathematical proof written by humans in a language understandable by computers. However, writing the entire proof in a formal language in a single shot is too difficult. The solution is interactive theorem proving (ITP) where the computer keeps track of what is already known and the user specifies the next step of the proof. Lately, many features of automated theorem proving have been soaked into ITP; nowadays, the computer not only keeps track of what is already known and what remains to be done but also suggests the next step and proves simple subgoals automatically. The perception of the role the computer plays is gradually shifting from a strict judge who has no mercy with user's errors to a digital assistant who helps the user to eventually reach the goal that was set by the user [30].

Automath [31], the first notable ITP language, was based on type theory and gave rise to Nuprl [32], Rocq [33], Agda [34], and Lean [35] [36] [37] [38] [39] [40] [41] [42] [43]. These systems don't use the old type theory of Whitehead and Russell [12]. They use so-called dependent type theories [44], where types can be parametrized both by types and by values. For example, a list can be parametrized by the type `Char` and by the value `5`, resulting in a type of five-letter words. Automath uses the dependently typed λ-calculus by De Bruijn [45]. Nuprl and Agda are based on the Martin-Löf type theory [46]. Rocq and Lean are based on the calculus of inductive constructions [47] [48]. However, historically the most important system is probably Mizar [49], which is based on set theory (in particular, the Tarski-Grothendieck set theory [50] [51]). The minimalistic system Metamath [52] is also based on set theory (but it is the Zermelo-Fraenkel set theory with the axiom of choice [53], which is simpler). Next to the ITP languages based on set theory and type theory, there are also the ITP languages HOL Light [54] and Isabelle/HOL [55] [56], which are based on the higher-order logic [57]. While their type system (simply typed λ-calculus) is less expressive than dependent type theories and has issues with formalizing some parts of set-theoretic MathematiCS, the system Isabelle/HOL features a very powerful proof-automation tactic Sledgehammer [58], which might be the main reason why Isabelle/HOL is still widely used today.

While almost all ITP languages have mathematical libraries, Lean has a singular mathematical library called Mathlib [1]. Mathlib is highly interconnected. Mathlib strives for high generality and high quality of code (high quality means that not only everything is formally verified but also carefully reviewed by human experts). Mathlib is regularly updated to new Lean versions, which requires continuous hard work from its maintainers. Additional effort is made to ensure that its compilation isn't too slow. Mathlib aspires to lay the groundwork for interaction with current mathematical research.

[9] Electronic Numerical Integrator And Computer

[10] Electronic Delay Storage Automatic Calculator

[11] UNIVersal Automatic Computer

## *1.3 Objectives*

> This above all; to thine own self be true.
> And it must follow, as the night the day.
> Thou canst not then be false to any man.
>
> — Polonius in Hamlet [59]

The main objectives of my Ph.D. are truth and beauty. Second comes my desire to obtain the Ph.D. title. The second objective, however, adds a lot of additional requirements on my work. One of them is nontriviality — Ph.D. is awarded only to people who have completed a big chunk of work in the area of their specialization. I partially empathize with this sentiment — there are many truths that could be written, even written in a beautiful way, but nobody would be interested in reading them, because it is a triviality that everybody qualified to read it already knows. As a result, the choice of problems to work on during my Ph.D. is a balance between what is achievable within a reasonable time (fully achievable, i.e., no `sorry` left) and what is noteworthy once achieved. A popular way to plan a Ph.D. is to set goals towards publications and later turn each publication into one chapter of the Ph.D. thesis. For most of the time, it was my approach, too, but I didn't want to end up writing a cumulative thesis. Instead, I intended to present all areas [that I studied] within a unified framework (both on the conceptual level and accompanied by code artifacts in the same Lean version). And even though the topics will greatly vary between the chapters, the longing for truth and beauty will stay the central theme of my work.

### 1.3.1 Truth

People have always wondered what is true.

The first humans looked up at the night sky and asked whether the lights above were gods, fires, or distant worlds. They asked why the sun returns after darkness, why the sea obeys the moon, and why rivers never cease their flow. They wondered whether the rhythm of the seasons was the will of unseen powers or the order of nature itself. They told stories to make sense of birth and death, of fortune and suffering, weaving meaning into the silence that reason had not yet learned to speak.

Over time, as stories multiplied, doubt began to stir within them. People started to question whether the myths themselves could bear the weight of truth, or whether truth belonged not to the tales nor to the priests' rituals that accompanied them but to the principles they dimly revealed — the unspoken regularities by which the heavens moved and seasons turned, waiting to be discovered beneath the veil of fable. Wonder gradually turned inward, from the gods to the mind that conceived them. The focus shifted from the deeds of gods to the hidden order behind them — the reality that endures when all stories end. Out of this questioning, philosophy was born as an attempt to preserve the awe of myth while submitting it to the discipline of reason. The inquiry turned from the changing things of the world to the enduring conditions that make knowledge attainable and truth possible.

Plato [60] sought truth in forms more perfect than the things that reflect them, while Aristotle [61] grounded it in the correspondence between thought and the world. The search for truth moved from mythos to reason, yet the longing remained the same — to find in the world something that speaks back to the mind, revealing that truth is not fabricated but discovered.

Over time, this longing learned discipline. The passion for clarity gave rise to method, and the love of wisdom gently hardened into structure. Philosophy gave birth to logic, mathematics, and the sciences. Here, truth ceased to be a matter of persuasion and became a matter of proof. Still, even in its most abstract form, the desire remained the same — to find statements that could stand on their own, independent of the speaker, the century, or the culture.

The methods have changed, but the spirit of inquiry has not. Each generation inherits the same question — how can one know that what seems true is actually true? How can one be sure one isn't deceived by imperfect senses or argumentation fallacies? What began as wonder before the stars now takes shape within systems of symbols and rules. In this long lineage, formalization stands as the latest expression of humanity's desire to make truth unassailable and to anchor our understanding in the enduring bedrock of logic.

At first glance, justifying that I have succeeded in the pursuit of truth seems pretty easy. Mathematical proof is widely considered to be the highest form of undeniable truth, and formal proofs are the pinnacle of this effort.

However, having my result formally verified is not the only requirement for achieving the truth. There is also room for potential deception in the description of my results and naming of the definitions. I believe that a requirement for ultimate truth is not only the absence of lies but also the absence of deception.

I should be ready for a potential reader who might assume that I don't present my results in good faith. What if I proved something trivial but then masked it with misleading notation or misnamed definitions so that it looks like a different theorem — one that would look like a difficult and valuable result?

There are voices in the Lean community saying that definitions don't matter much, only API is important. I disagree. First of all, API needs to be built around something. Technically, one could build API only, not backed by definitions, but create axioms instead. Since axioms can lead to inconsistencies (as a result of which everything is provable), I cannot recommend this approach. The only viable approach is to start with definitions. Once definitions are written, unlimited layers of abstractions can be built around them. Now that we have established that we need some definitions, I want to argue that we need good definitions. From the pragmatic point of view, present IDEs allow the user to click "Go to definition", and it is good if the user can read and understand what she finds there. No matter how well the API is developed, the definition is still the first thing the user will see if she follows the most convenient way to examine the notion in question. From the philosophical point of view, we need a basis of trust in the definitions.

The trust in definitions brings us to the topic of *trusted code*. While Lean ensures correctness in the sense of logical consistency, there are certain parts of the code that the reader must check herself in order to make sure that not only the proofs are correct but also the very things we proved are correct. We refer to the statements of the final results and the definitions they (transitively) depend on as trusted code [62]. There are many other definitions in the code, which are used only to prove main results and not to state them (they are often used in the statements of auxiliary lemmas, but if they are ill-defined or mis-stated, it creates issues only for what the proofs of the main results depend on and not for what the statements of the main results depend on), which means that they don't have to be checked before we can believe the main results — we say that these definitions are not part of the trusted code (some people, preferring to rather say what something is than to say what something isn't, would say that these definitions are implementation details; however, the phrase "implementation detail" is used with many different flavours, subjectively depending on what each person considers to be "implementation" and "detail", hence we refrain from the phrase "implementation detail" altogether; we will only distinguish what is and what isn't part of the trusted code, where the subjectivity goes only as far as deciding what the main results are).

In the Seymour project (Chapter 4), trusted code is presented in a file `Seymour.lean` separate from the implementation, because the implementation is too large and too complicated for a casual reader to browse. In the other presented projects, finding the trusted code should be easy enough for a reader who knows what to search for, directions for which are present in this thesis in abundance.

The text of this thesis primarily emphasizes definitions, followed by theorems, while giving minimal attention to proofs. The reader is expected to run the Lean compiler to check that the proofs are correct. Occasionally, we will comment on proofs, too, but only when we want to highlight a particular proof technique or explain how the proof is decomposed into lemmas.

### 1.3.2 Beauty

I originally didn't think of myself as of an artistic person. Drawing was my least favourite activity already in kindergarten, turning into a full-blown hatred towards art classes in primary school. If you asked the young me why I was drawn to MathematiCS, I would certainly tell you that I like it because it is very useful. Most of all, it was very useful for me as a young aspiring Pokémon trainer, since mathematical knowledge allowed me to optimize Pokémon strategies like nothing else could. I didn't think of MathematiCS in terms of beauty.

The second thing that sparked my interest in MathematiCS was its precision. In the first year of my undergraduate studies at the Charles University in Prague, I learnt how every word has to be defined, how every claim needs to be justified, and how every solution must be presented in a way that allows the reader to trace all steps leading to it. Doing homework was my daily practice that brought deep satisfaction and the sense of meaning in my life, especially when I later combined being a student with being a teacher at the same time. However, in later years of my studies, I was spending more and more time with programming, and less time with equations and mathematical proofs. In the world of code, almost everything was explicit, so the sense of precision was preserved, but the ideas represented by the code were usually not as interesting as the ideas studied in more theoretical courses. The precision was still there, but its purpose had changed — it served practical goals rather than the search for truth for its own sake.

In my Ph.D., I jumped back into a more theoretical direction. However, once textbooks were replaced by research papers, MathematiCS no longer looked as I remembered it. The meaning of notation was often left to context; many proof steps were omitted. I would spend hours, days, sometimes weeks reconstructing the missing links. Occasionally I discovered outright errors. When I wrote to the authors, they often didn't bother to correct their mistakes and reupload a revised version. Reading papers became an exercise in endurance and tolerance for frustration rather than understanding.

It was only when I found Lean — a world that combines the precision of programming languages with the conceptual depth of abstract mathematical ideas — when MathematiCS regained its appeal and my life regained its meaning. And, as my love for Lean was growing stronger, my loathing for traditional mathematical notations was becoming more visible. The pen-and-paper MathematiCS began to feel like meeting an ex-lover who had often hurt my feelings yet with whom I had stayed for too long. At times, the repulsion I felt at implicit multiplication (denoted by mere juxtaposition of factors) became a visceral reaction.

And this story brings me to the second main philosophical dimension of my work — beauty.

Roger Scruton [63] said on the subject of beauty:

> "We live in a world which has been, in many ways, uglified — and it is the world we want to redeem, so that we are part of it once again. And our fulfillment is as if it were reflected back to us in the things we encounter, and that is really part of what I mean by 'redemption'. And that is the function of the aesthetic."

Scruton's words point to a deep human need — to experience the world as meaningful and harmonious rather than fractured and hostile. It isn't merely a decorative claim. Scruton reminds us that ugliness is not merely about appearance but about alienation. Ugliness, in this sense, arises when form ceases to be transparent to meaning, when the surface no longer carries the depth it ought to express. Beauty redeems by reuniting form and content, by allowing us once again to recognize ourselves in what we encounter.

Mathematical writing always navigates the balance between clarity and economy, between the desire for perfection and the pressures of time and publication. For many of us, the written culture of contemporary MathematiCS has grown inhospitable. Papers often trade precision for brevity, hide essential steps behind references or tradition, and present arguments in a style that assumes an audience already initiated. For the reader, it often results in estrangement; the ideas may be profound, but the form obscures rather than reveals them. What should be a path towards clarity becomes an experience of disorientation.

The practice of theorem proving in Lean offers a form of redemption. In this medium, no detail is lost — every assumption is stated, every inference justified, every algebra laid bare. In this setting, beauty is not an ornament but a structure. It is the alignment of thought and expression. It is the absence of gaps where understanding could slip away. What once felt elusive becomes visible; what was hidden in the shadows of "it is clear that…" now stands plainly in the light. To engage with Lean is to step into a world where MathematiCS has regained its rigor and transparency. What emerges is a landscape in which the reader can truly see the grand outline of the proof and the fine texture of its details, coexisting in harmony.

This harmony is deeply satisfying because it answers the very longing Scruton describes. To work within Lean is to encounter MathematiCS that reflects back to us the fulfillment of understanding, the joy of seeing each part in its rightful place. The uglification of opacity and omission is replaced by the beauty of transparency and precision. In Lean, the proofs don't only convince; they allow us to dwell within MathematiCS as something whole, intelligible, and beautiful.

That said, clarity of thought is not all there is on the subject of beauty in MathematiCS. There is a full stack of form, from the tactile to the transcendental, starting from a good font and a nice color scheme, through helpful notation, up to the most abstract mathematical beauty.

Beauty begins with what first meets the eye. The font is the opening gesture, the frame in which everything else evinces. A good font makes no demands; it distinguishes `l` from `1` and `I` and `|`. Its grace lies in its invisibility. It doesn't call attention to itself, but allows the reader to attend to the structure of an argument without the friction of deciphering marks.

I chose JuliaMono [64] because it has exceptional Unicode coverage, the symbols are easily distinguishable from each other, and the majority of its symbols look similar to corresponding symbols in other fonts, which makes the transition from reading other fonts to reading JuliaMono relatively easy. The letter `r` is probably the only character that looks a bit weird in it. The way I perceive it, JuliaMono is a font whose qualities whisper rather than shout. I use JuliaMono both in IDE and for code snippets in this thesis.

Upon this quiet stage, lexical highlighting introduces color. Here the page takes on depth; variables, constants, operators, and keywords separate into distinct voices. What was once monochrome becomes luminous; what was once bleak becomes alive. Meaning begins to shimmer towards the surface. The syntax itself starts to breathe, and the machinery of logic turns to melody where each symbol finds its own rhythm in the polyphony of reason. The eye learns the grammar before the mind does; perception leads understanding. Color, then, is not embellishment but orientation. It teaches the gaze where to rest and where to move, letting the structure of reasoning appear not as a wall of text but as a landscape that can be traversed, not just parsed. Color softens the entry into precision, giving warmth to the rigor. In Lean, lexical highlighting becomes a kind of pedagogy of the senses. It trains us to see patterns as music rather than machinery.

One common mistake in the design of a color scheme is setting the default color of the text to black on white background or to white on dark background. Black on white exhausts the full range of contrast, leaving no headroom for emphasis — highlighted symbols then appear weaker and, as a result, their intended prominence is diminished or even reversed. In IDE, I use light pastel colors on dark background[12]. In this text, I use dark colors on white background to optimize this thesis for printing. Subtlety respects hierarchy — the colors sing rather than scream, guiding the eye without distraction.

If color is the music of syntax, then notation is a choreography of thought. It is where the aesthetic of MathematiCS meets the architecture of language. A good notation does not merely abbreviate; it liberates. It gives form to intuition, allowing complex ideas to move with the lightness of a single symbol. When designed with care, it carries meaning like a poem carries emotion — precise, structured, yet full of resonance.

Custom notation in Lean extends this beauty into the formal realm. It allows the mathematician not only to express an idea but to sculpt the very language in which the idea lives. Each symbol, each binder, each operator becomes a decision about how thought should flow. Poor notation interrupts; good notation moves thoughts forward. When the notation fits its purpose, one doesn't just read computer code; one reads MathematiCS — pure and whole.

In Lean, the discipline of formal precision meets the artistry of expressive design. Here we see that beauty and rigor are not adversaries but companions. The formalist's demand that every symbol have meaning and the aesthete's desire that meaning take elegant shape, turn out to be two faces of one pursuit.

Another subjective element of writing code in Lean is that I think it is better to not name variables that are used only once. Fortunately, Lean allows one to forgo unnecessary names. Temporary constructions or one-off functions need not be forced into permanence; they can exist just long enough to carry the proof forward. This freedom reduces clutter, letting the mind follow the current of the argument with less distraction. In doing so, it honors both clarity and elegance — the proof breathes naturally, and the eye lingers where the progress truly resides.

---

[12] https://github.com/madvorak/vscode-lean4-colors

Another principle I try to follow in Lean is that what belongs logically together should also appear visually together. For example, the Mathlib definition `Matroid.disjointSum` doesn't comply with this principle because, when called in practical settings, it will look like `M.disjointSum N hMN` for example, making `M` and `N` stand far from each other, while `N` and `hMN` are close to each other visually, without a good motivation for such visual presentation. In contrast, the Mathlib definition `Matrix.submatrix` perfectly follows this principle because calling it like `A.submatrix f g` makes the matrix stand on one side and both indexing functions stand on the other side (together). The same principle extends to larger settings beyond a single line of code, such as grouping related definitions together and grouping similar lemmas together. In this alignment, the eye perceives the harmony of the argument even before the mind has traced each step. Logical and visual proximity converge, and the code itself becomes a landscape in which understanding flows naturally.

When this visual rhythm stretches through longer passages, it becomes apparent that spacing plays a structural role. For greater visual separation, two consecutive empty lines work well — they signal a genuine shift in thought, a new layer of abstraction, or a pause for the reader to reorient. However, they should be used sparingly. Just as excessive ornament dilutes beauty, excessive spacing erodes form. The code must breathe, but not lose cohesion. When spacing reflects conceptual hierarchy rather than mere whim, the reader senses the architecture of the argument before even reading the contents. In this balance between air and density, visual design becomes a silent helper in reasoning.

Beyond notation and visual grouping, beauty in Lean also emerges from the principles of good software engineering. Clear folder and file structures, reusable definitions, and well-designed abstractions are not merely pragmatic conveniences; they are expressions of elegance. When the code is organized thoughtfully, proofs become easier to read, maintain, and extend, and the relationships between ideas are revealed rather than obscured. The discipline of engineering — once seen as purely functional — becomes another source of aesthetic pleasure. Simplicity, coherence, and composability combine with each other to form a system in which logic and intuition move in harmony, and the mind can dwell within a landscape that is both rigorous and graceful.

Ultimately, beauty in Lean is felt in the smooth passage from thought to expression. When the language, notation, and design decisions allow an idea to be encoded almost as quickly as it is conceived, the mind encounters minimal resistance. Friction between intuition and formalization lightens, the current of reasoning moves more freely. The act of formalization becomes a medium rather than a barrier; one can move seamlessly from insight to proof, from concept to code, experiencing the ideas themselves with minimal distraction or interruption.

Lean fully embodies the sense of wholeness. It enforces coherence not only as a constraint, but as a promise, that what is written will stand, that what is proved will endure. The formal language becomes a vessel for the eternal language of mathematics, uniting the mechanical and the creative. From the curve of a glyph to the architecture of a theory, from syntax highlighting to theorem hierarchies, beauty runs continuously through the stack — one harmony, perceived at different scales.

# 2 Preliminaries

This chapter reviews preliminaries shared by multiple projects. Almost everything mentioned in this chapter is a part of the trusted code, so read carefully.

Sections of this chapter are organized according to conceptual grouping rather than compilation order. In some unfortunate cases, dependencies between the definitions happen to go against the chronological order of the text. I apologize for this imperfection. I acknowledge that it is a compromise which will make some readers unhappy — it makes me unhappy, too; believe me.

The majority of content in Preliminaries comes from Mathlib [1], with some basic declarations being distributed with Lean itself (especially in the first few sections of this chapter; and also the axioms explained in the last section are part of the core). In rare cases, we will also elaborate on our custom extensions of the standard API, but most of it will be postponed to respective chapters (divided by topics). Occasionally, the line between Preliminaries and our own projects becomes slightly blurry, since some parts of our projects have been (and more parts possibly will be) upstreamed to Mathlib.

## *2.1 Types and subtypes*

Every *term* in Lean has a *type*. This section reviews several ways of forming new types from existing types.

### 2.1.1 Prod

A *product type* is the type-theoretic analogue of a Cartesian product. Lean defines it as follows:

```
structure Prod (α β : Type) where
  fst : α
  snd : β
```

We typically use notation `α × β` as a shortcut for `Prod α β` which is right-associative; notation `α × β × γ` means `Prod α (Prod β γ)` and so on. In the context of

```
α β : Type
a : α
b : β
```

we can write `(a, b)` as a syntactic sugar for `Prod.mk a b` and it is right-associative; in the context of

```
α β γ : Type
a : α
b : β
c : γ
```

we can write `(a, b, c)` and it gets translated to:

```
Prod.mk a (Prod.mk b c) : α × β × γ
```

### 2.1.2 Sum

A *sum type* is the type-theoretic analogue of a disjoint union. Lean defines it as follows:

```
inductive Sum (α β : Type) where
  | inl (_ : α) : Sum α β
  | inr (_ : β) : Sum α β
```

We typically use notation `α ⊕ β` as a shortcut for `Sum α β` which is also right-associative. Furthermore, we define custom notation for its two constructors:

```
prefix:max "◩" => Sum.inl
prefix:max "◪" => Sum.inr
```

The semantics is that `◩` denotes the left or top variant whereas `◪` denotes the right or bottom variant. This notation can be chained without the need for parentheses; for example, `◪◩◩◩◪0` is a shorthand for:

```
Sum.inr (Sum.inl (Sum.inl (Sum.inl (Sum.inr 0))))
```

A function from a sum type can be implemented by providing functions from respective types to the same type:

```
def Sum.elim {α β γ : Type} (f : α → γ) (g : β → γ) : α ⊕ β → γ :=
  (·.casesOn f g)
```

`Sum.casesOn` is an autogenerated recursor without any recursive arguments. Its second and third explicit arguments denote how the `Sum.inl` terms and the `Sum.inr` terms are mapped, respectively.

If the codomain is also a sum type and the two functions "never mix", the implementation can be simplified using:

```
def Sum.map {α α' β β' : Type} (f : α → α') (g : β → β') :
    α ⊕ β → α' ⊕ β' :=
  Sum.elim (Sum.inl ∘ f) (Sum.inr ∘ g)
```

### 2.1.3 Option

An *option type* is essentially a sum of a given type with a singleton:

```
inductive Option (α : Type) where
  | none : Option α
  | some (_ : α) : Option α
```

The special value `none` is typically used to denote an invalid result.

If we want to map the contained value but only when it is valid (i.e., keeping `none` intact), we apply the following function:

```
def Option.map {α β : Type} (f : α → β) : Option α → Option β
  | some x => some (f x)
  | none   => none
```

In specific cases, `Option` is named `WithBot` or `WithTop` to denote distinct semantics:

```
def WithBot (α : Type) := Option α
def WithTop (α : Type) := Option α
```

The special values are of `WithBot` and `WithTop` are denoted by `⊥` and `⊤` respectively, while the `Option.some` constructor is often denoted as type coercion.

When we know that their values are not the special value, we can cast them back to the original type:

```
def WithBot.unbot {α : Type} : ∀ x : WithBot α, x ≠ ⊥ → α | (x : α), _ => x
def WithTop.untop {α : Type} : ∀ x : WithTop α, x ≠ ⊤ → α | (x : α), _ => x
```

### 2.1.4 Subtype

A *subtype* is the type-theoretic analogue of a subset. Lean defines it as follows:

```
structure Subtype {α : Type} (p : α → Prop) where
  val : α
  property : p val
```

We use notation `{ a : α // p a }` to denote the subtype of `α` where the condition `p` holds. In many expressions, `.val` is an automatic conversion.

## *2.2 Numbers*

While I believe that numbers are not the most fundamental notion in MathematiCS, it is the notion I decided to start with. We will review basic types of numbers, from natural to real, how they are defined in Lean or Mathlib.

### 2.2.1 Nat

Most people know *natural numbers* already since kindergarten. Lean defines them as follows:

```
inductive Nat where
  | zero : Nat
  | succ (n : Nat) : Nat
```

The inductive definition corresponds to the unary encoding of natural numbers. In practice, instead of writing nested constructors, we use the following symbols to denote natural numbers (so-called Arabic numerals [65] [66]):

```
Nat.zero   = 0
Nat.succ 0 = 1
Nat.succ 1 = 2
Nat.succ 2 = 3
Nat.succ 3 = 4
Nat.succ 4 = 5
Nat.succ 5 = 6
Nat.succ 6 = 7
Nat.succ 7 = 8
Nat.succ 8 = 9
```

After `9` the numbers start using multiple digits, using the well-known rules, so-called decimal (positional) notation [65] [66]. In this thesis, single-digit numbers will be sufficient for most of the time.

Mathlib equips `Nat` with the usual symbol `ℕ` which we will use every time we work with natural numbers.

The *addition* is defined on natural numbers inductively, the same way as in Robinson arithmetic [67]:

```
def Nat.add : ℕ → ℕ → ℕ
  | a, 0          => a
  | a, Nat.succ b => Nat.succ (Nat.add a b)
```

For example, `3 + 2 = 5` is calculated as follows:

```
Nat.add 3 2 = Nat.add 3 (Nat.succ 1) = Nat.succ (Nat.add 3 1) =
Nat.succ (Nat.add 3 (Nat.succ 0)) = Nat.succ (Nat.succ (Nat.add 3 0)) =
Nat.succ (Nat.succ 3) = Nat.succ 4 = 5
```

The *multiplication* is defined on natural numbers inductively, the same way as in Robinson arithmetic [67]:

```
def Nat.mul : ℕ → ℕ → ℕ
  | _, 0          => 0
  | a, Nat.succ b => Nat.add (Nat.mul a b) a
```

For example, `3 * 2 = 6` is calculated as follows:

```
Nat.mul 3 2 = Nat.mul 3 (Nat.succ 1) = Nat.add (Nat.mul 3 1) 3 =
Nat.add (Nat.mul 3 (Nat.succ 0)) 3 = Nat.add (Nat.add (Nat.mul 3 0) 3) 3 =
Nat.add (Nat.add 0 3) 3 = Nat.add 3 3 = 6
```

The *exponentiation* ("to the power of") of natural numbers is defined in a similar way:

```
def Nat.pow (m : ℕ) : ℕ → ℕ
  | 0          => 1
  | Nat.succ n => Nat.mul (Nat.pow m n) m
```

The *predecessor* of natural numbers is defined as follows:

```
def Nat.pred : ℕ → ℕ
  | 0          => 0
  | Nat.succ a => a
```

Note that the predecessor of `1` is `0` and the predecessor of `0` is `0` as well.

The *subtraction* of natural numbers is defined as follows:

```
def Nat.sub : ℕ → ℕ → ℕ
  | a, 0          => a
  | a, Nat.succ b => Nat.pred (Nat.sub a b)
```

For example, `7 - 2 = 5` is calculated as follows:

```
Nat.sub 7 2 = Nat.sub 7 (Nat.succ 1) = Nat.pred (Nat.sub 7 1) =
Nat.pred (Nat.sub 7 (Nat.succ 0)) = Nat.pred (Nat.pred (Nat.sub 7 0)) =
Nat.pred (Nat.pred 7) = Nat.pred (Nat.pred (Nat.succ 6)) = Nat.pred 6 =
Nat.pred (Nat.succ 5) = 5
```

Note that subtracting a number from a smaller number always gives zero. This operation is sometimes called "truncated subtraction" [68]. For example, `1 - 3 = 0` is calculated as follows:

```
Nat.sub 1 3 = Nat.sub 1 (Nat.succ 2) = Nat.pred (Nat.sub 1 2) =
Nat.pred (Nat.sub 1 (Nat.succ 1)) = Nat.pred (Nat.pred (Nat.sub 1 1)) =
Nat.pred (Nat.pred (Nat.sub 1 (Nat.succ 0))) =
Nat.pred (Nat.pred (Nat.pred (Nat.sub 1 0))) =
Nat.pred (Nat.pred (Nat.pred 1)) =
Nat.pred (Nat.pred (Nat.pred (Nat.succ 0))) =
Nat.pred (Nat.pred 0) = Nat.pred 0 = 0
```

The *comparison* ≤ of natural numbers is defined as the least relation ≤ satisfying `n ≤ n` and `n ≤ m → n ≤ m + 1` for all natural numbers:

```
inductive Nat.le (n : ℕ) : ℕ → Prop
  | refl          : Nat.le n n
  | step {m : ℕ} : Nat.le n m → Nat.le n (Nat.succ m)
```

The *strict comparison* < of natural numbers is defined so that `n < m ↔ n + 1 ≤ m` holds by definition for all natural numbers:

```
def Nat.lt (n m : ℕ) : Prop :=
  Nat.le (Nat.succ n) m
```

The *division* of natural numbers has a very different definition (essentially, asking how many times we can subtract the divisor from the dividend):

```
def Nat.div (x y : ℕ) : ℕ :=
  if 0 < y ∧ y ≤ x then
    Nat.div (x - y) y + 1
  else
    0
```

Note that division by zero always gives zero and division by other numbers rounds down. For example, 8 / 3 = 2 is calculated as follows:

```
Nat.div 8 3 = Nat.div (8 - 3) 3 + 1 = Nat.div 5 3 + 1 =
(Nat.div (5 - 3) 3 + 1) + 1 = (Nat.div 2 3 + 1) + 1 = (0 + 1) + 1 = 2
```

The following infix operators are provided for binary operations on natural numbers:

| **Lean name** | **Operator** | **English name** |
|---|---|---|
| `Nat.add` | `+` | addition |
| `Nat.sub` | `-` | (truncated) subtraction |
| `Nat.mul` | `*` | multiplication |
| `Nat.div` | `/` | (floored) division |
| `Nat.pow` | `^` | exponentiation |

Note that computation in unary encoding is very inefficient. Fortunately, natural numbers have special support in both the kernel and the compiler, so that computation with natural numbers is performed with binary encoding in practice. Note the presence of the holy trinity here—the mathematical definition uses unary encoding, the executable code uses binary encoding, and displaying the numbers for users uses decimal encoding.

If you want to restrict natural numbers to be less than `n` then the following structure is for you:

```
structure Fin (n : ℕ) where
  val  : ℕ
  isLt : val < n
```

We will employ `Fin` in all chapters, usually for indexing. We will also occasionally use `ZMod` which, for all positive natural numbers `n`, the type `ZMod n` is just `Fin n` with more instances on it (for example, it forms a ring with addition; moreover, if `n` is a prime, `Fin n` forms a field). In particular, we will write `Z2` for `ZMod 2` and `Z3` for `ZMod 3` to shorten expressions and eliminate one level of parentheses.

In some unfortunate situations, we have a term of the type `Fin n` when we need a term of the type `Fin m`, where `m` and `n` are equal but not definitionally equal. Mathlib provides a conversion function:

```
def Fin.cast {n m : ℕ} (hnm : n = m) (i : Fin n) : Fin m :=
  ⟨i.val, hnm ▸ i.isLt⟩
```

### 2.2.2 Int

*Integers* are defined as two cases (nonnegative, negative):

```
inductive Int where
  | ofNat   : Nat → Int
  | negSucc : Nat → Int
```

The constructor `Int.ofNat` gives nonnegative integers whereas the constructor `Int.negSucc` gives negative integers (shifted by one, e.g., `Int.negSucc 3` represents the integer `-4`). Notation `-[n +1]` will be used to denote `Int.negSucc n` in the rest of this subsection.

Mathlib equips `Int` with the usual symbol `ℤ` which we will use every time we work with integers.

Before we define any operation on integers, we define the integer difference of two natural numbers (which is perhaps the more familiar definition of subtracting natural numbers):

```
def Int.subNatNat (m n : ℕ) : ℤ :=
  match (n - m : ℕ) with
  | 0          => Int.ofNat (m - n)
  | (Nat.succ k) => Int.negSucc k
```

The addition of integers is defined by four cases:

```
def Int.add (m n : ℤ) : ℤ :=
  match m, n with
  | Int.ofNat m, Int.ofNat n => Int.ofNat (m + n)
  | Int.ofNat m, -[n +1]     => Int.subNatNat m n.succ
  | -[m +1]    , Int.ofNat n => Int.subNatNat n m.succ
  | -[m +1]    , -[n +1]     => Int.negSucc (m + n).succ
```

The multiplication of integers is also defined by four cases:

```
def Int.negOfNat : ℕ → ℤ
  | 0          => 0
  | Nat.succ m => Int.negSucc m
```

```
def Int.mul (m n : ℤ) : ℤ :=
  match m, n with
  | Int.ofNat m, Int.ofNat n => Int.ofNat (m * n)
  | Int.ofNat m, -[n +1]     => Int.negOfNat (m * n.succ)
  | -[m +1]    , Int.ofNat n => Int.negOfNat (m.succ * n)
  | -[m +1]    , -[n +1]     => Int.ofNat (m.succ * n.succ)
```

The exponentiation of integers is defined in almost the same way as exponentiation of natural numbers:

```
def Int.pow (m : ℤ) : ℕ → ℤ
  | 0          => 1
  | Nat.succ n => Int.pow m n * m
```

The subtraction of integers is defined as adding the opposite number (where "-" denotes the unary minus a.k.a. `Int.neg` defined immediately below):

```
def Int.neg (n : ℤ) : ℤ :=
  match n with
  | Int.ofNat n   => Int.negOfNat n
  | Int.negSucc n => Int.ofNat n.succ
```

```
def Int.sub (m n : ℤ) : ℤ := m + (- n)
```

The division of integers is defined according to the E-rounding convention [69]:

```
def Int.ediv : ℤ → ℤ → ℤ
  | Int.ofNat m, Int.ofNat n      => Int.ofNat (m / n)
  | Int.ofNat m, -[n +1]          => - Int.ofNat (m / n.succ)
  | -[_ +1]    , 0                => 0
  | -[m +1]    , Int.ofNat n.succ => -[m / n.succ +1]
  | -[m +1]    , -[n +1]          => Int.ofNat (m / n.succ).succ
```

Note that all occurrences of / above denote division of natural numbers, no recursive definition.

The following infix operators are provided for binary operations on integers:

| **Lean name** | **Operator** | **English name** |
|---|---|---|
| `Int.add` | `+` | addition |
| `Int.sub` | `-` | subtraction |
| `Int.mul` | `*` | multiplication |
| `Int.ediv` | `/` | (Euclidean) division |
| `Int.pow` | `^` | to the power of natural number |

For cultural reasons, the operators for `Int.neg` and `Int.sub` have the same symbol.

Again, Lean has special support for integers, so that they can be used efficiently in computation, apart from their usage in theorem proving.

On a related note, you might encounter `ℤˣ` in the code. It refers to the type containing only two integers `1` and `-1` similarly to `Fin 2` containing only two natural numbers `0` and `1` but with multiplication as the main operation. The full definition is more general:

```
structure Units (α : Type) [Monoid α] where
  val : α
  inv : α
  val_inv : val * inv = 1
  inv_val : inv * val = 1

attribute [coe] Units.val

postfix:1024 "ˣ" => Units
```

### 2.2.3 Rat

*Rational numbers* are probably the easiest type of numbers to work with (due to well-behaved both subtraction and division). Lean defines them as follows (the numerator is an integer, and the denominator is a nonzero natural number coprime with the numerator):

```
structure Rat where mk' ::
  num : Int
  den : Nat
  den_nz : den ≠ 0
  reduced : num.natAbs.Coprime den
```

Mathlib equips `Rat` with the usual symbol `ℚ` which we will use every time we work with rational numbers.

`Rat.normalize` takes a numerator and a denominator, which may share a common divisor, and produces a valid rational number, where the numerator and the denominator are coprime. Then `mkRat` is just a wrapper of the type `(ℤ → ℕ → ℚ)`. With type coercions on the RHS we have:

```
lemma mkRat_eq_div (n : ℤ) (d : ℕ) : mkRat n d = n / d
```

We have the following binary operations on rationals and their infix operators:

| **Lean name** | **Operator** | **English name** |
|---|---|---|
| `Rat.add` | `+` | addition |
| `Rat.sub` | `-` | subtraction |
| `Rat.mul` | `*` | multiplication |
| `Rat.div` | `/` | division |
| `Rat.instPowNat` | `^` | to the power of natural number |

The addition of rational numbers is defined in a way that satisfies the following identity:

```
theorem Rat.add_def' (a b : ℚ) :
    a + b = mkRat (a.num * b.den + b.num * a.den) (a.den * b.den)
```

The subtraction of rational numbers is defined in a way that satisfies the following identity:

```
theorem Rat.sub_def' (a b : ℚ) :
    a - b = mkRat (a.num * b.den - b.num * a.den) (a.den * b.den)
```

The multiplication of rational numbers is defined in a way that satisfies the following identity:

```
theorem mkRat_mul_mkRat (n₁ n₂ : ℤ) (d₁ d₂ : ℕ) :
    mkRat n₁ d₁ * mkRat n₂ d₂ = mkRat (n₁ * n₂) (d₁ * d₂)
```

The inverse (denoted by $^{-1}$ after the term) of a rational number is defined in a way that satisfies:

```
theorem Rat.mul_inv_cancel (a : ℚ) : a ≠ 0 → a * a⁻¹ = 1
```

The division of rational numbers is defined so that, by definition:

```
theorem Rat.div_def (a b : ℚ) : a / b = a * b⁻¹
```

The exponentiation of rational numbers is defined so that, by definition:

```
lemma Rat.pow_def (q : ℚ) (n : ℕ) :
    q ^ n = ⟨q.num ^ n, q.den ^ n, sorry, sorry⟩
```

### 2.2.4 Real

*Real numbers* are defined using Cauchy sequences of rational numbers:

```
structure Real where ofCauchy ::
  cauchy : CauSeq.Completion.Cauchy (_ : ℚ → ℚ)
```

Mathlib equips `Real` with the usual symbol ℝ which we will use every time we work with real numbers.

Because we will not work with real numbers apart from a few illustrative examples, which are not part of the trusted code, we will not review how operations on real numbers are defined.

## *2.3 Collections*

Collection types are essential to most programmers. They are also important for formalization of theorems. We will review them from bottom up, starting with lists. All of these collections are generic types (i.e., parametrized by another type, which determines what can be stored in the collection).

### 2.3.1 List

The typical way to represent a general finite amount of items is to have a *list*. Lean defines lists inductively:

```
inductive List (α : Type) where
  | nil : List α
  | cons (head : α) (tail : List α) : List α
```

For `List.nil` (the empty list), Lean provides the usual notation `[]` for convenience. For

`List.cons` (a list that has a *head* (the first element) and a *tail* (a list of all remaining elements)), Lean provides a notation `::` as a right-associative infix operator. In the context of

```
α : Type
a : α
l : List α
```

we write `a :: l` to denote `List.cons a l` (where `l` may be empty). Furthermore, Lean allows to occupy the square brackets with elements delimited by commas. For example, the following five lines all represent the same list of three elements:

```
[1, 2, 3]
1 :: [2, 3]
1 :: (2 :: [3])
1 :: 2 :: [3]
1 :: 2 :: 3 :: []
```

The *concatenation* of lists is defined as follows:

```
def List.append {α : Type} : List α → List α → List α
  | []    , s => s
  | a :: l, s => a :: List.append l s
```

The infix operator `++` denotes `List.append` from now on. For example, the following five lines all represent the same list of five elements:

```
1 :: 2 :: 3 :: 4 :: 5 :: []
[1, 2, 3, 4, 5]
[1, 2, 3] ++ [4, 5]
[1, 2] ++ 3 :: [4, 5]
1 :: [2, 3, 4] ++ [5]
```

The *length* of a list is an easy recursive definition:

```
def List.length {α : Type} : List α → ℕ
  | []     => 0
  | _ :: s => s.length + 1
```

The *map* is one of the most useful functions on lists—it applies given function to every element of a list:

```
def List.map {α β : Type} (f : α → β) : List α → List β
  | []     => []
  | a :: s => f a :: s.map f
```

The *filtering map* is given a “partial” function and combines mapping with omitting a part of the list (in particular, the elements that are mapped to `none` are omitted):

```
def List.filterMap {α β : Type} (f : α → Option β) : List α → List β
  | []   => []
  | a :: s =>
    match f a with
    | none   => s.filterMap f
    | some b => b :: s.filterMap f
```

The *reverse* list (i.e., a list starting with given list's last element and ending with given list's first element) could be defined as follows:

```
private def List.rev {α : Type} : List α → List α
  | []   => []
  | a::l => l.rev ++ [a]
```

However, it would lead to quadratic time complexity. Instead, a list is reversed using the following (in the auxiliary definition, `r` works as an "accumulator"—a part that has already been reversed) linear-time definition:

```
def List.reverseAux {α : Type} : List α → List α → List α
  | []  , r => r
  | a::l, r => l.reverseAux (a::r)

def List.reverse {α : Type} (s : List α) : List α :=
  s.reverseAux []
```

We can also have a list of lists. One thing we can do with them is to *join* them into one normal list:

```
def List.flatten {α : Type} : List (List α) → List α
  | []     => []
  | l :: L => l ++ L.flatten
```

An important binary relation on lists is ~ (to be a *permutation* of):

```
inductive Perm : List α → List α → Prop
  | nil : Perm [] []
  | cons (x : α) {l₁ l₂ : List α} : Perm l₁ l₂ → Perm (x :: l₁) (x :: l₂)
  | swap (x y : α) (l : List α) : Perm (y :: x :: l) (x :: y :: l)
  | trans {l₁ l₂ l₃ : List α} : Perm l₁ l₂ → Perm l₂ l₃ → Perm l₁ l₃
```

The four rules defining list permutations can be summarized as follows:

- empty list is a permutation of empty list: `[] ~ []`
- if `a` is an element and `x` and `y` are lists such that `x ~ y` then we have: `a :: x ~ a :: y`
- if `a` and `b` are elements and `x` is a list then we have: `b :: a :: x ~ a :: b :: x`
- if `x`, `y`, `z` are lists such that `x ~ y` and `y ~ z` then we have: `x ~ z`

It is easy to convince oneself that these four rules are necessary. It is harder to believe that they are sufficient.

Lean proves that ~ is an *equivalence relation*, in the sense of being reflexive, symmetric, and transitive:

```
structure Equivalence {α : Type} (r : α → α → Prop) : Prop where
  refl  : ∀ x : α, r x x
  symm  : ∀ {x y : α}, r x y → r y x
  trans : ∀ {x y z : α}, r x y → r y z → r x z
```

Furthermore, Lean proves that lists together with ~ form a setoid (i.e., a bundled equivalence):

```
class Setoid (α : Type) where
  r : α → α → Prop
  iseqv : Equivalence r
```

```
instance List.isSetoid (α : Type) : Setoid (List α) := Setoid.mk Perm sorry
```

### 2.3.2 Multiset

If we don't care about the order in which items come but care about multiplicity, we use a *multiset*. Mathlib defines multisets as the quotient of lists over list permutations:

```
def Multiset (α : Type) : Type :=
  Quotient (List.isSetoid α)
```

By definition, it is a finite multiset (as every list is finite). Its *cardinality* ("size") is defined as follows:

```
def Multiset.card {α : Type} : Multiset α → ℕ :=
  Quot.lift List.length (fun _ _ => Perm.length_eq)
```

We will again need a map function. Its implementation in Mathlib is a bit cryptic:

```
def Multiset.map {α β : Type} (f : α → β) (s : Multiset α) : Multiset β :=
  Quot.liftOn s
    (fun l : List α => (l.map f : Multiset β))
    (fun _ _ p => Quot.sound (p.map f))
```

Reading the definition probably didn't illuminate what it actually means. Hence, before we proceed to trust the definition, we will examine its API to reassure ourselves that it really does what we think it does:

```
theorem Multiset.map_singleton {α β : Type} (f : α → β) (a : α) :
    ({a} : Multiset α).map f = {f a}
```

```
theorem Multiset.map_cons {α β : Type} (f : α → β) (a : α)
    (s : Multiset α) :
    Multiset.map f (a ::ₘ s) = f a ::ₘ Multiset.map f s
```

The operator `::ₘ` on multisets is similar to the operator `::` on lists (see `Multiset.cons_coe` for the exact correspondence between them). Indeed, `Multiset.map` does what we expect from it.

We will eventually need summing up multisets of numbers. Mathlib defines it as follows:

```
def Multiset.sum {α : Type} [AddCommMonoid α] : Multiset α → α :=
  Multiset.foldr (· + ·) 0
```

Similarly, the product of a multiset is defined as follows:

```
def Multiset.prod {α : Type} [CommMonoid α] : Multiset α → α :=
  Multiset.foldr (· * ·) 1
```

For readers unfamiliar with the `fold` functions[13], which can be tricky to understand (especially when we don't work with lists per se), we recommend basing our trust in the correctness of the two definitions above on the four theorems below:

```
theorem Multiset.sum_singleton {α : Type} [AddCommMonoid α] (a : α) :
    Multiset.sum {a} = a
```

```
theorem Multiset.sum_cons {α : Type} [AddCommMonoid α]
    (a : α) (s : Multiset α) :
    Multiset.sum (a ::ₘ s) = a + Multiset.sum s
```

```
theorem Multiset.prod_singleton {α : Type} [CommMonoid α] (a : α) :
    Multiset.prod {a} = a
```

```
theorem Multiset.prod_cons {α : Type} [CommMonoid α]
    (a : α) (s : Multiset α) :
    Multiset.prod (a ::ₘ s) = a * Multiset.prod s
```

At the same time, it shouldn't come as a surprise that the sum of the empty multiset is `0` and its product is `1`.

### 2.3.3 Finset

If we care neither about order nor about multiplicity of items, we use a *finset* (not to be mistaken for a finite set, which may represent the same collection but is a different type). Mathlib defines finsets as multisets without duplicity:

```
inductive Pairwise {α : Type} (R : α → α → Prop) : List α → Prop
  | nil : Pairwise []
  | cons : ∀ {a : α} {l : List α},
      (∀ a' ∈ l, R a a') → Pairwise l → Pairwise (a :: l)
```

```
def List.Nodup {α : Type} : List α → Prop := Pairwise (· ≠ ·)
```

```
def Multiset.Nodup {α : Type} (s : Multiset α) : Prop :=
  Quot.liftOn s List.Nodup sorry
```

```
structure Finset (α : Type) where
  val : Multiset α
  nodup : Nodup val
```

A finset is *nonempty* iff it contains an element:

```
def Finset.Nonempty {α : Type} (s : Finset α) : Prop := ∃ x : α, x ∈ s
```

The *sum* of a finset is defined as follows:

```
def Finset.sum {α β : Type} [AddCommMonoid β] (s : Finset α) (f : α → β) :
    β :=
  (s.val.map f).sum
```

[13] https://en.wikipedia.org/wiki/Fold_(higher-order_function)

The *product* of a finset is defined as follows:

```
def Finset.prod {α β : Type} [CommMonoid β] (s : Finset α) (f : α → β) :
    β :=
  (s.val.map f).prod
```

Note that, unlike the two functions on multisets, `Finset.sum` and `Finset.prod` incorporate mapping in themselves.

Moreover, Mathlib defines so-called big operators ∑ for `Finset.sum` and ∏ for `Finset.prod` (popular from pen-and-paper notation) with the following syntax:

```
example {α β : Type} [AddCommMonoid β] (s : Finset α) (f : α → β) :
    ∑ i ∈ s, f i = s.sum f

example {α β : Type} [CommMonoid β] (s : Finset α) (f : α → β) :
    ∏ i ∈ s, f i = s.prod f
```

For example, we have:

```
theorem Real.log_prod {α : Type} (s : Finset α) (f : α → ℝ)
    (_ : ∀ x ∈ s, f x ≠ 0) :
    log (∏ i ∈ s, f i) = ∑ i ∈ s, log (f i)
```

The big operators ∑ and ∏ have the same precedence, and it is strictly between the precedence of `*` and the precedence of `+`. For example, two identities [70] are written as follows:

```
∑ k ∈ K, (a k + b k) =  ∑ k ∈ K, a k  +  ∑ k ∈ K, b k
∏ k ∈ K,  a k * b k  = (∏ k ∈ K, a k) * (∏ k ∈ K, b k)
```

The big operators ∑ and ∏ shouldn't be confused for the type-theoretical primitives Σ and Π (note that the big operators are taller, pointier, and have overall different feeling to them).

The *infimum* of a function on a finset (or, better, on a nonempty finset, respectively) is defined, assuming the operator ⊓ denotes the minimum of two elements (or, more generally, the greatest elements below the two elements) as follows:

```
variable {α β : Type} [SemilatticeInf α]

def Finset.inf [OrderTop α] (s : Finset β) (f : β → α) : α :=
  s.fold (· ⊓ ·) ⊤ f

def Finset.inf' (s : Finset β) (_ : s.Nonempty) (f : β → α) : α :=
  WithTop.untop (s.inf ((↑) ∘ f)) sorry
```

Again, for readers unfamiliar with `fold`, we examine the descriptive theorems:

```
theorem Finset.inf'_singleton (f : β → α) {b : β} :
    Finset.inf' {b} sorry f = f b

theorem Finset.inf'_cons {s : Finset β} (hs : s.Nonempty) (f : β → α)
    {b : β} {hb : b ∉ s} :
    (Finset.cons b s hb).inf' sorry f = f b ⊓ s.inf' hs f
```

The supremum is defined similarly, but we will not need it.

The conversions `LinearOrder.toLattice` and `Lattice.toSemilatticeInf` from Mathlib make the definitions useful for us; we will utilize infima only for linear orders.

### 2.3.4 Fintype

A type is *fintype* iff there is a finset that contains all possible values of given type:

```
class Fintype (α : Type) where
  elems : Finset α
  complete : ∀ x : α, x ∈ elems
```

This class is defined constructively, i.e., its `elems` can be retrieved. The idiomatic way is to call `Finset.univ` where `α` is implicit.

The big operators ∑ and ∏ have also exist for `Fintype` with the following syntax:

```
example {α β : Type} [AddCommMonoid β] [Fintype α] (f : α → β) :
    ∑ i : α, f i = Finset.univ.sum f

example {α β : Type} [CommMonoid β] [Fintype α] (f : α → β) :
    ∏ i : α, f i = Finset.univ.prod f
```

## *2.4 Not collections*

There are other (generic) types that conceptually represent collections of things but aren't collections per se. For example, a matrix can be thought of as a two-dimensional array, but the actual implementation is that a matrix is a binary function (without memoization of data).

### 2.4.1 Finite

First, we need to define *equiv*, i.e., a bundled bijection (don't mistake `Equiv` for a logical equivalence; don't mistake `Equiv` for an equivalence relation):

```
def LeftInverse {α β : Type} (g : β → α) (f : α → β) : Prop :=
  ∀ x : α, g (f x) = x

def RightInverse {α β : Type} (g : β → α) (f : α → β) : Prop :=
  LeftInverse f g

structure Equiv (α β : Type) where
  toFun : α → β
  invFun : β → α
  left_inv : LeftInverse invFun toFun
  right_inv : RightInverse invFun toFun

infixl:25 " ≃ " => Equiv
```

Mathlib also defines a *permutation* on a type as an equiv with itself:

```
abbrev Equiv.Perm (α : Type) := Equiv α α
```

Section 4.12 says more about working with equivs.

A type is *finite* iff it has one-to-one correspondence with the canonical type on `n` elements:

```
class inductive Finite (α : Type) : Prop
  | intro {n : ℕ} : α ≃ Fin n → Finite α
```

`Finite` is a nonconstructive analogue of `Fintype`.

### 2.4.2 Set

A *set* in Lean is just a unary predicate:

```
def Set (α : Type) := α → Prop
```

In the context of

```
α : Type
a : α
S : Set α
```

we use the syntactic sugar `a ∈ S` to denote `S a`. Another symbol that comes with sets is a *subset*. In the context of

```
α : Type
S T : Set α
```

we have (definitionally):

```
(S ⊆ T) = ∀ x : α, x ∈ S → x ∈ T
```

There is also a *strict subset*. In the same context, we have (definitionally):

```
(S ⊂ T) = (S ⊆ T ∧ ¬ T ⊆ S)
```

An equivalent description of a strict subset (in the same context) is as follows:

```
S ⊂ T ↔ S ⊆ T ∧ ∃ x ∈ T, x ∉ S
```

In the context of

```
α : Type
p : α → Prop
```

there is a syntactic sugar `{ x : α | p x }` that just means `p` typed as a set. For example, the set `{ x : ℝ | x ^ x < 8 }` is the set defined by the predicate `(fun x : ℝ => x ^ x < 8)`.

The *union* of sets is the set of terms that belong to at least one of the sets:

```
theorem Set.mem_union {α : Type} (x : α) (S T : Set α) :
    x ∈ S ∪ T ↔ x ∈ S ∨ x ∈ T
```

The *intersection* of sets is the set of terms that belong to both sets:

```
theorem Set.mem_inter_iff {α : Type} (x : α) (S T : Set α) :
    x ∈ S ∩ T ↔ x ∈ S ∧ x ∈ T
```

Sometimes we need to convert a set to a type. Mathlib defines it as a subtype:

```
def Set.Elem {α : Type} (S : Set α) : Type := { x : α // x ∈ S }
```

`Set.Elem` is an implicit coercion, but it works automatically only in the most basic situations, hence we often end up writing `Set.Elem` explicitly.

Similarly to finite types, Mathlib also defines finite sets (again, nonconstructively):

```
def Set.Finite {α : Type} (S : Set α) : Prop := Finite S.Elem
```

If we want a constructive version, we can write `Fintype S.Elem` or `Fintype S` for short. We will use it in indexing matrices (defined later) by sets.

We declare a notation `¬∪` for `Insert.insert` which, in turn, in case of `Set.insert` means the following (and this special case will be sufficient for understanding `¬∪` in the entire thesis):

```
example {α : Type} (a : α) (S : Set α) :
    a ¬∪ S = { x : α | x = a ∨ x ∈ S }
```

### 2.4.3 Function

A *function* is just a `Pi` type (a "dependent function") in which the type of the output doesn't depend on the value of the input (hence the adjective "dependent" removed). We denote functions by the right-associative infix operator `→` (and, by Curry-Howard correspondence, the same symbol denotes implications; though, for the comfort of the reader, we will color the arrow in red when we mean the logical connective). We assume that the reader already understands functions (including the operator `∘` for function composition, `Function.swap` for flipping its arguments, and the anonymous function notation `·`), as they are a standard part of Lean, and we will explore other declarations defined on top of functions.

A function is *injective* iff it doesn't have any collision:

```
def Function.Injective {α β : Type} (f : α → β) : Prop :=
  ∀ a₁ a₂ : α, f a₁ = f a₂ → a₁ = a₂
```

For example, multiplying natural numbers by two is injective but dividing natural numbers by two isn't.

Let's talk about vectors…

We distinguish two types of vectors; implicit vectors and explicit vectors. Implicit vectors (called just "vectors") are members of a vector space; they don't have any internal structure. Explicit vectors are functions from coordinates to values (or just "vectors" when it is clear from context that our vector is a map) and, if the values are from a field, they are also vectors in the former sense. We will discuss implicit vectors later (see Section 2.5.4 on modules). Let's now focus on explicit vectors.

The type of coordinates doesn't have to be ordered and doesn't have to be finite. However, finite vectors have some advantages. For example, they allow us to define the *dot product*:

```
def dotProduct {m α : Type} [Fintype m] [Mul α] [AddCommMonoid α]
    (v w : m → α) : α :=
  ∑ i : m, v i * w i
```

It comes with infix notation:

```
infixl:72 " ·ᵥ " => dotProduct
```

Note that the `+` used inside the `∑` must form an abelian monoid; otherwise, we wouldn't obtain a well-defined result without ordering the coordinates, which we don't want to require.

Sometimes we talk about vector families. A vector *family* is a function from an indexing type to a type of vectors (they can be explicit vectors or implicit vectors). Explicit vectors and vector families are just informal notions to discuss the math we do; they aren't specific Lean types.

The following instance defines how explicit vectors are compared, i.e., element-wise "less or equal to" (it is, in fact, more general than we need because the following instance talks about `Pi` types, i.e., the type of `x i` can depend on the value of `i`):

```
instance Pi.hasLe {ι : Type} {π : ι → Type} [∀ i, LE (π i)] :
  LE (∀ i, π i) where le x y := ∀ i : ι, x i ≤ y i
```

When the indexing type is `Fin n`, we can conveniently write an exclamation mark followed by square brackets with elements (in the canonical order) delimited by commas. The following example illustrates how this syntactic sugar works:

```
example : ![’a’, ’b’, ’c’, ’d’] = fun i : Fin 4 => match i with
  | 0 => ’a’
  | 1 => ’b’
  | 2 => ’c’
  | 3 => ’d’
```

Furthermore, `List.ofFn` can convert the vector indexed by `Fin n` to a list of length `n` (starting with the image of zero). Combining the last two things gives us:

```
example : List.ofFn ![1, 2, 3] = [1, 2, 3]
```

### 2.4.4 Matrix

A *matrix* is a curried [71] binary function:

```
def Matrix (m n α : Type) := m → n → α
```

When a matrix happens to be finite (i.e., both `m` and `n` are finite) and its entries are numeric, we like to represent it by a table of numbers.

The *transposition* is `Function.swap` but on matrices:

```
def Matrix.transpose (M : Matrix m n α) : Matrix n m α :=
  fun x y => M y x
```

We write $^\top$ to denote a transposed matrix. Therefore, by definition:

```
example {m n α : Type} (M : Matrix m n α) (i : m) (j : n) :
    M i j = Mᵀ j i
```

A *submatrix* is defined as a function:

```
def Matrix.submatrix {m m' n n' α : Type}
    (A : Matrix m n α) (f : m' → m) (g : n' → n) :
    Matrix m' n' α :=
  fun i j => A (f i) (g j)
```

Therefore, by definition:

```
example {m m' n n' α : Type}
    (A : Matrix m n α) (f : m' → m) (g : n' → n) (i : m') (j : n') :
    (A.submatrix f g) i j = A (f i) (g j)
```

Note that submatrix can repeat and/or reorder rows and/or columns. For example,

$$\begin{pmatrix} 1 & 2 & 3 \\ 4 & 5 & 6 \\ 7 & 8 & 9 \end{pmatrix}$$

has

$$\begin{pmatrix} 3 & 3 & 1 & 1 \end{pmatrix}$$

as a submatrix (the reindexing functions being `![0]` for rows and `![2, 2, 0, 0]` for columns).

A product *matrix times vector* is defined as follows:

```
def Matrix.mulVec {m n α : Type} [Fintype n] [NonUnitalNonAssocSemiring α]
    (M : Matrix m n α) (v : n → α) : m → α
  | i => (fun j : m => M i j) ⬝ᵥ v
```

```
infixr:73 " *ᵥ " => Matrix.mulVec
```

It isn't clear to me why `[NonUnitalNonAssocSemiring α]` is required for multiplying matrix by vector, when `[Mul α] [AddCommMonoid α]` was sufficient for the dot product. I suppose it is a historical accident.

A product *matrix times matrix*, denoted by `*` as normal multiplication, is also defined via the dot product:

```
instance {l m n α : Type} [Fintype m] [Mul α] [AddCommMonoid α] :
    HMul (Matrix l m α) (Matrix m n α) (Matrix l n α) where
  hMul M N :=
    fun i : l => fun k : n => (fun j : m => M i j) ⬝ᵥ (fun j : m => N j k)
```

Applying the definitions gives us:

```
theorem Matrix.mul_apply {l m n α : Type}
    [Fintype m] [Mul α] [AddCommMonoid α]
    (M : Matrix l m α) (N : Matrix m n α) (i : l) (k : n) :
    (M * N) i k = ∑ j : m, M i j * N j k
```

The *determinant* is defined as follows:

```
def Matrix.detRowAlternating {n R : Type} [Fintype n] [CommRing R] :
    (n → R) [⋀^n]→ₗ[R] R :=
  MultilinearMap.alternatization
    ((MultilinearMap.mkPiAlgebra R n R).compLinearMap LinearMap.proj)
```

```
abbrev Matrix.det {n R : Type} [Fintype n] [CommRing R]
    (M : Matrix n n R) : R :=
  Matrix.detRowAlternating M
```

Since the definition above is very far from being self-contained, instead of unfolding it further, I suggest we base our trust in correctness of the definition on the following identity:

```
theorem Matrix.det_apply {n R : Type} [Fintype n] [CommRing R]
    (M : Matrix n n R) :
    M.det = ∑ σ : Equiv.Perm n, σ.sign • ∏ i : n, M (σ i) i
```

In this definition, `Equiv.Perm.sign {α : Type} [Fintype α] : Perm α →* ℤˣ` refers to the (unique) surjective group homomorphism from `Equiv.Perm α` to the group with two elements; the function returns `1` for even permutations; the function returns `-1` for odd permutations.

For example, the determinant of a 2×2 matrix is calculated as follows:

```
theorem Matrix.det_fin_two {R : Type} [CommRing R]
    (A : Matrix (Fin 2) (Fin 2) R) :
    A.det = A 0 0 * A 1 1 - A 0 1 * A 1 0
```

Since there are six permutations on three elements, the determinant of a $3 \times 3$ matrix is harder to calculate:

```
theorem Matrix.det_fin_three {R : Type} [CommRing R]
    (A : Matrix (Fin 3) (Fin 3) R) :
    A.det =
      A 0 0 * A 1 1 * A 2 2
    - A 0 0 * A 1 2 * A 2 1
    - A 0 1 * A 1 0 * A 2 2
    + A 0 1 * A 1 2 * A 2 0
    + A 0 2 * A 1 0 * A 2 1
    - A 0 2 * A 1 1 * A 2 0
```

Note that `Matrix.det` requires a finite square matrix, but its indices don't have to be ordered.

For matrices whose both dimensions are `Fin` size, we use similar notation to the notation for explicit vectors; however, this time there are two exclamation marks before the opening bracket and there are two levels of delimitation (a semicolon separates rows, a comma separates values in a row). For example `!![1, 2; 3, 4]` represents:

$$\begin{pmatrix} 1 & 2 \\ 3 & 4 \end{pmatrix}$$

Let's now review some important special cases of matrices.

The *zero matrix* has zeros everywhere:

```
theorem Matrix.zero_apply {m n α : Type} [Zero α] (i : m) (j : n) :
    (0 : Matrix m n α) i j = 0
```

For example:

$$\begin{pmatrix} 0 & 0 & 0 \\ 0 & 0 & 0 \end{pmatrix}$$

The *unit matrix* has ones on the main diagonal and zeros elsewhere:

```
theorem Matrix.one_apply_eq {n α : Type} [Zero α] [One α] (i : n) :
    (1 : Matrix n n α) i i = 1
```

```
theorem Matrix.one_apply_ne {n α : Type} [Zero α] [One α] {i j : n}
    (_ : i ≠ j) :
    (1 : Matrix n n α) i j = 0
```

For example:

$$\begin{pmatrix} 1 & 0 & 0 \\ 0 & 1 & 0 \\ 0 & 0 & 1 \end{pmatrix}$$

In three of the four presented projects, we will need to work with block matrices. Let's review all definitions we will need for it. `Matrix.fromSomething` functions are frequently used. `Matrix.toSomething` functions are rarely used. Their implicit type arguments are reordered for readability in the text (hence `recall` will not work here).

Stacking matrices vertically:

```
def Matrix.fromRows {m₁ m₂ n α : Type}
    (A₁ : Matrix m₁ n α) (A₂ : Matrix m₂ n α) :
```

```
    Matrix (m₁ ⊕ m₂) n α :=
  Sum.elim A₁ A₂
```

Visually: `Matrix.fromRows A₁ A₂` = $\begin{pmatrix} A_1 \\ A_2 \end{pmatrix}$

Conversely, we can extract its parts as follows:

```
def Matrix.toRows₁ {m₁ m₂ n α : Type} (A : Matrix (m₁ ⊕ m₂) n α) :
    Matrix m₁ n α :=
  (A ◩· ·)
```

```
def Matrix.toRows₂ {m₁ m₂ n α : Type} (A : Matrix (m₁ ⊕ m₂) n α) :
    Matrix m₂ n α :=
  (A ◪· ·)
```

Stacking matrices horizontally:

```
def Matrix.fromCols {m n₁ n₂ α : Type}
    (A₁ : Matrix m n₁ α) (A₂ : Matrix m n₂ α) :
    Matrix m (n₁ ⊕ n₂) α :=
  fun i : m => Sum.elim (A₁ i) (A₂ i)
```

Visually: `Matrix.fromCols A₁ A₂` = $\begin{pmatrix} A_1 & A_2 \end{pmatrix}$

Conversely, we can extract its parts as follows:

```
def Matrix.toCols₁ {m n₁ n₂ α : Type} (A : Matrix m (n₁ ⊕ n₂) α) :
    Matrix m n₁ α :=
  (A · ◩·)
```

```
def Matrix.toCols₂ {m n₁ n₂ α : Type} (A : Matrix m (n₁ ⊕ n₂) α) :
    Matrix m n₂ α :=
  (A · ◪·)
```

Making a matrix from four (two times two) blocks:

```
def Matrix.fromBlocks {m₁ m₂ n₁ n₂ α : Type}
    (A₁₁ : Matrix m₁ n₁ α) (A₁₂ : Matrix m₁ n₂ α)
    (A₂₁ : Matrix m₂ n₁ α) (A₂₂ : Matrix m₂ n₂ α) :
    Matrix (m₁ ⊕ m₂) (n₁ ⊕ n₂) α :=
  Sum.elim
    (fun i₁ : m₁ => Sum.elim (A₁₁ i₁) (A₁₂ i₁))
    (fun i₂ : m₂ => Sum.elim (A₂₁ i₂) (A₂₂ i₂))
```

Visually: `Matrix.fromBlocks A₁₁ A₁₂ A₂₁ A₂₂` = $\begin{pmatrix} A_{11} & A_{12} \\ A_{21} & A_{22} \end{pmatrix}$

It could be (definitionally) equally defined as follows:

```
def Matrix.fromBlocks {m₁ m₂ n₁ n₂ α : Type}
    (A₁₁ : Matrix m₁ n₁ α) (A₁₂ : Matrix m₁ n₂ α)
    (A₂₁ : Matrix m₂ n₁ α) (A₂₂ : Matrix m₂ n₂ α) :
    Matrix (m₁ ⊕ m₂) (n₁ ⊕ n₂) α :=
  Matrix.fromRows (Matrix.fromCols A₁₁ A₁₂) (Matrix.fromCols A₂₁ A₂₂)
```

Conversely, we can extract individual blocks as follows:

```
def Matrix.toBlocks₁₁ {m₁ m₂ n₁ n₂ α : Type}
    (A : Matrix (m₁ ⊕ m₂) (n₁ ⊕ n₂) α) :
    Matrix m₁ n₁ α :=
  (A ◩· ◩·)
```

```
def Matrix.toBlocks₁₂ {m₁ m₂ n₁ n₂ α : Type}
    (A : Matrix (m₁ ⊕ m₂) (n₁ ⊕ n₂) α) :
    Matrix m₁ n₂ α :=
  (A ◩· ◪·)
```

```
def Matrix.toBlocks₂₁ {m₁ m₂ n₁ n₂ α : Type}
    (A : Matrix (m₁ ⊕ m₂) (n₁ ⊕ n₂) α) :
    Matrix m₂ n₁ α :=
  (A ◪· ◩·)
```

```
def Matrix.toBlocks₂₂ {m₁ m₂ n₁ n₂ α : Type}
    (A : Matrix (m₁ ⊕ m₂) (n₁ ⊕ n₂) α) :
    Matrix m₂ n₂ α :=
  (A ◪· ◪·)
```

Summary and sanity check:

```
theorem Matrix.fromBlocks_toBlocks {m₁ m₂ n₁ n₂ α : Type}
    (A : Matrix (m₁ ⊕ m₂) (n₁ ⊕ n₂) α) :
    Matrix.fromBlocks A.toBlocks₁₁ A.toBlocks₁₂ A.toBlocks₂₁ A.toBlocks₂₂ =
    A
```

Sometimes we need to assume that a matrix is *invertible* (also called nonsingular). A general definition of invertible elements in a monoid is based on `Units` (introduced in Section 2.2.2) as follows:

```
def IsUnit {M : Type} [Monoid M] (a : M) : Prop :=
  ∃ u : Mˣ, (u : M) = a
```

Unfortunately, the Mathlib terminology `Units` and `IsUnit` might evoke the notion of the unit matrix, which isn't what is being defined here (though, the symbol `1` inside the definition refers to the unit matrix, of course).

Two equivalent descriptions of invertible elements in a monoid are as follows:

```
lemma isUnit_iff_exists {M : Type} [Monoid M] {a : M} :
    IsUnit a ↔ ∃ b : M, a * b = 1 ∧ b * a = 1
```

```
theorem isUnit_iff_exists_and_exists {M : Type} [Monoid M] {a : M} :
    IsUnit a ↔ (∃ b : M, a * b = 1) ∧ (∃ c : M, c * a = 1)
```

When it comes to matrices, Mathlib provides a convenient characterization of invertibility; a square matrix is invertible iff its determinant is invertible:

```
theorem Matrix.isUnit_iff_isUnit_det {n α : Type} [Fintype n] [CommRing α]
    (A : Matrix n n α) :
    IsUnit A ↔ IsUnit A.det
```

For matrices over rationals, for example, it means that a matrix is invertible iff its determinant is nonzero. We will need invertibility for matrices over `Z2`, where invertibility boils down to the same condition (or, equivalently, that its determinant is `1`).

## *2.5 Algebraic classes*

Lean has an extremely powerful system of *typeclasses*. They are similar to interfaces known from OOP but much more powerful. One could hardly find a more powerful typeclass system than Lean has. Typeclasses in Lean can require existence of:

(1) data (constants, functions, operators)

(2) proofs (required properties of the data)

Using typeclasses, we may, for example, define a function that takes a finite set equipped with addition and computes its total sum, but only if its addition is commutative and associative. Typeclasses are an essential tool for using Lean effectively. This section reviews typeclasses defined in Mathlib that will be used later. In particular, we build the algebraic hierarchy up to linearly ordered vector spaces.

Before we move to abstract algebra, let's review one very basic yet important typeclass; type is *nonempty* if there is a term of given type:

```
class inductive Nonempty (α : Sort u) : Prop where
  | intro (_ : α) : Nonempty α
```

Contrast with type being *inhabited*, which requires us to provide a concrete witness:

```
class Inhabited (α : Sort u) where
  default : α
```

For example, for most types of numbers, the `default` element is the number `0`.

The importance of typeclasses for formal mathematics and the depth of the typeclass-built algebraic hierarchy in Mathlib stem from the tendency of modern mathematics to develop towards higher and higher levels of abstraction.

At its beginning, mathematics was dealing with concrete objects. Consider the following example:

> Mary had three apples and got two more apples. How many apples does Mary have?

Or consider the following example:

> Tom had three marbles and got two more marbles. How many marbles does Tom have?

Today we all know that both questions boil down to calculating three plus two. We would solve it as follows:

```
3 + 2 = 5
```

To the earliest (pre)mathematicians, the abstraction "three plus two" would make no sense. You need to have three of something, not "three". You can possess apples and marbles, not numbers. Yet today, we are fully comfortable with numbers without knowing what they stand for, what real-world objects we count with them.

In the second abstraction step, we replace numbers by letters that represent some unspecified numbers. For example, we would consider the following equality (for `a` and `b` being some integers, let's say) correct:

```
2 * a - (3 * b - a) = 3 * (a - b)
```

For example, when `a = 9` and `b = 8`, the LHS becomes `2 * 9 - (3 * 8 - 9) = 18 - 15 = 3` and the RHS becomes `3 * (9 - 8) = 3 * 1 = 3`, which are equal.

Or, when `a = 1` and `b = -1`, the LHS becomes `2 * 1 - (3 * (-1) - 1) = 2 - (-4) = 6` and the RHS becomes `3 * (1 - (-1)) = 3 * 2 = 6`, which are equal, again.

We have just experienced the power of abstraction; we can perform the abstract simplification of the algebraic expression once, and then every time we want to calculate the LHS for some numbers, we can calculate the RHS instead, arriving to the same answer.

So far, we wouldn't need typeclasses to formalize the truths we have discussed. However, as the reader knows, there is more to mathematics than elementary algebra. Not only numbers can be replaced by variables, also the operation can be unspecified. Yet, we are still able to say something in such general settings. For example, we can prove that in a monoid (an algebra where `*` is associative and `1` is neutral, as we will review in a moment), if an element has both a left inverse and a right inverse, they are identical:

```
x * y = 1 = y * z → x = z
```

This implication holds in every monoid, no matter if `*` denotes the multiplication of rational numbers, the multiplication of complex numbers, the multiplication of square real matrices, the composition of functions, the symmetric difference of sets, or the addition of continuous real functions.

Mathlib defines monoids as a `class` and, every time we work with a type that is an `instance` of monoid, all theorems about monoids become automatically applicable. This way, Lean users take full advantage of discoveries from abstract algebra.

To recapitulate, the four stages of abstraction in mathematics go as follows:

1) We have concrete numbers of concrete objects, and we perform concrete operations with them.
2) We have concrete numbers of general objects, and we perform concrete operations with them.
3) We have general numbers of general objects, and we perform concrete operations with them.
4) We have a general number of general objects, or perhaps not numbers at all but something more abstract, and we perform general operations with them.

We will not elaborate on further levels of abstractions, such as category theory, because we will not utilize them; though Mathlib provides a rich support for category theory [72], too.

### 2.5.1 Binary operations

An *additive semigroup* is a structure on any type with addition (denoted by the infix `+` operator) where the addition is associative:

```
class AddSemigroup (G : Type) extends Add G where
  add_assoc : ∀ a b c : G, (a + b) + c = a + (b + c)
```

A *semigroup*, similarly, is a structure on any type with multiplication (denoted by the infix ∗ operator) where the multiplication is associative:

```
class Semigroup (G : Type) extends Mul G where
  mul_assoc : ∀ a b c : G, (a ∗ b) ∗ c = a ∗ (b ∗ c)
```

An *additive monoid* is an additive semigroup with the “zero” element that is neutral with respect to addition from both left and right, equipped with a scalar multiplication by the natural numbers:

```
class AddZeroClass (M : Type) extends Zero M, Add M where
  zero_add : ∀ a : M, 0 + a = a
  add_zero : ∀ a : M, a + 0 = a
```

```
class AddMonoid (M : Type) extends AddSemigroup M, AddZeroClass M where
  nsmul : ℕ → M → M
  nsmul_zero : ∀ x : M, nsmul 0 x = 0
  nsmul_succ : ∀ (n : ℕ) (x : M), nsmul (n + 1) x = nsmul n x + x
```

Note that the requirement of `nsmul` and its properties doesn’t restrict the class of additive monoids. They can be (minimally[14]) defined as follows:

```
private class AddMonoid' (M : Type) extends AddSemigroup M, AddZeroClass M
```

A *monoid*, similarly, is a semigroup with the “one” element that is neutral with respect to multiplication from both left and right, equipped with a power to the natural numbers:

```
class MulOneClass (M : Type) extends One M, Mul M where
  one_mul : ∀ a : M, 1 ∗ a = a
  mul_one : ∀ a : M, a ∗ 1 = a
```

```
class Monoid (M : Type) extends Semigroup M, MulOneClass M where
  npow : ℕ → M → M
  npow_zero : ∀ x : M, npow 0 x = 1
  npow_succ : ∀ (n : ℕ) (x : M), npow (n + 1) x = npow n x ∗ x
```

Note that the requirement of `npow` and its properties doesn’t restrict the class of monoids. They can be (minimally) defined as follows:

```
private class Monoid' (M : Type) extends Semigroup M, MulOneClass M
```

A *subtractive monoid* (an additive monoid with subtraction) is an additive monoid that adds two more operations (unary and binary minus) that satisfy some basic properties (please note that “adding minus itself gives zero” is not required yet; it will be required, e.g., in an additive group):

[14] The file https://github.com/madvorak/preliminaries/blob/main/Preliminaries.lean demonstrates that the shorter requirements really suffice. The same applies for all subsequent occurrences of “can be (minimally) defined”— check it in the file that I am not lying to you.

```
class SubNegMonoid (G : Type) extends AddMonoid G, Neg G, Sub G where
  sub_eq_add_neg : ∀ a b : G, a - b = a + -b
  zsmul : ℤ → G → G
  zsmul_zero' : ∀ a : G, zsmul 0 a = 0
  zsmul_succ' (n : ℕ) (a : G) : zsmul n.succ a = zsmul n a + a
  zsmul_neg' (n : ℕ) (a : G) : zsmul (Int.negSucc n) a = -(zsmul n.succ a)
```

Note that the requirement of `zsmul` and its properties doesn't restrict the class of subtractive monoids. They can be (minimally) defined as follows:

```
private class SubNegMonoid' (G : Type) extends
    AddMonoid G, Neg G, Sub G where
  sub_eq_add_neg : ∀ a b : G, a - b = a + -b
```

A *division monoid*, similarly, is a monoid that adds two more operations (inverse and divide) that satisfy some basic properties (please note that "multiplication by an inverse gives one" is not required yet):

```
class DivInvMonoid (G : Type) extends Monoid G, Inv G, Div G where
  div_eq_mul_inv : ∀ a b : G, a / b = a * b⁻¹
  zpow : ℤ → G → G
  zpow_zero' : ∀ a : G, zpow 0 a = 1
  zpow_succ' (n : ℕ) (a : G) : zpow n.succ a = zpow n a * a
  zpow_neg' (n : ℕ) (a : G) : zpow (Int.negSucc n) a = (zpow n.succ a)⁻¹
```

Note that the requirement of `zpow` and its properties doesn't restrict the class of division monoids. They can be (minimally) defined as follows:

```
private class DivInvMonoid' (G : Type) extends Monoid G, Inv G, Div G where
  div_eq_mul_inv : ∀ a b : G, a / b = a * b⁻¹
```

An *additive group* is a subtractive monoid in which the unary minus acts as a left inverse with respect to addition:

```
class AddGroup (A : Type) extends SubNegMonoid A where
  neg_add_cancel : ∀ a : A, -a + a = 0
```

An *abelian magma* is a structure on any type that has commutative addition:

```
class AddCommMagma (G : Type) extends Add G where
  add_comm : ∀ a b : G, a + b = b + a
```

A *commutative magma*, similarly, is a structure on any type with commutative multiplication:

```
class CommMagma (G : Type) extends Mul G where
  mul_comm : ∀ a b : G, a * b = b * a
```

An *abelian semigroup* is an abelian magma and an additive semigroup at the same time:

```
class AddCommSemigroup (G : Type) extends AddSemigroup G, AddCommMagma G
```

A *commutative semigroup*, similarly, is a commutative magma and a semigroup at the same time:

```
class CommSemigroup (G : Type) extends Semigroup G, CommMagma G
```

An *abelian monoid* is an additive monoid and an abelian semigroup at the same time:

```
class AddCommMonoid (M : Type) extends AddMonoid M, AddCommSemigroup M
```

A *commutative monoid*, similarly is a monoid and a commutative semigroup at the same time:

```
class CommMonoid (M : Type) extends Monoid M, CommSemigroup M
```

An *abelian group* is an additive group and an abelian monoid at the same time:

```
class AddCommGroup (G : Type) extends AddGroup G, AddCommMonoid G
```

A *distrib* is a structure on any type with an addition and a multiplication where both the left distributivity and the right distributivity hold:

```
class Distrib (R : Type) extends Mul R, Add R where
  left_distrib  : ∀ a b c : R, a * (b + c) = a * b + a * c
  right_distrib : ∀ a b c : R, (a + b) * c = a * c + b * c
```

A *nonunital-nonassociative-semiring* is an abelian monoid with a distributive multiplication and a well-behaved zero:

```
class MulZeroClass (M₀ : Type) extends Mul M₀, Zero M₀ where
  zero_mul : ∀ a : M₀, 0 * a = 0
  mul_zero : ∀ a : M₀, a * 0 = 0
```

```
class NonUnitalNonAssocSemiring (α : Type) extends
    AddCommMonoid α, Distrib α, MulZeroClass α
```

A *nonunital-semiring* (also called semirung) is a nonunital-nonassociative-semiring that forms a semigroup with zero (i.e., the semigroup-with-zero requirement makes it associative):

```
class SemigroupWithZero (S₀ : Type) extends Semigroup S₀, MulZeroClass S₀
```

```
class NonUnitalSemiring (α : Type) extends
    NonUnitalNonAssocSemiring α, SemigroupWithZero α
```

An *additive monoid with one* is an additive monoid equipped with the symbol “one” and an embedding of natural numbers:

```
class AddMonoidWithOne (R : Type) extends NatCast R, AddMonoid R, One R where
  natCast_zero : natCast 0 = 0
  natCast_succ : ∀ n : ℕ, natCast (n + 1) = (natCast n) + 1
```

Note that the requirement of `NatCast` and its properties doesn’t restrict the class of additive monoids with one. They can be (minimally) defined as follows:

```
private class AddMonoidWithOne' (R : Type) extends AddMonoid R, One R
```

An additive monoid with one has *characteristic zero* if the canonical map from natural numbers is injective (and it is a mixin, hence it can be used on stronger classes, too, without defining it again):

```
class CharZero (R : Type) [AddMonoidWithOne R] : Prop where
  cast_injective : Function.Injective (Nat.cast : ℕ → R)
```

An *abelian monoid with one* is an additive monoid with one and an abelian monoid at the same time:

```
class AddCommMonoidWithOne (R : Type) extends
    AddMonoidWithOne R, AddCommMonoid R
```

An *additive group with one* is an additive monoid with one and an additive group, and embeds all integers:

```
class AddGroupWithOne (R : Type) extends
    IntCast R, AddMonoidWithOne R, AddGroup R where
  intCast_ofNat : ∀ n : ℕ, intCast (n : ℕ) = Nat.cast n
  intCast_negSucc : ∀ n : ℕ, intCast (Int.negSucc n) = - Nat.cast (n + 1)
```

Note that the requirement of `IntCast` and its properties doesn't restrict the class of additive groups with one. They can be (minimally) defined as follows:

```
private class AddGroupWithOne' (R : Type) extends
    AddMonoidWithOne R, AddGroup R
```

A *nonassociative-semiring* is a nonunital-nonassociative-semiring that has a well-behaved multiplication by both zero and one and forms an abelian monoid with one:

```
class MulZeroOneClass (M₀ : Type) extends MulOneClass M₀, MulZeroClass M₀
```

```
class NonAssocSemiring (α : Type) extends
    NonUnitalNonAssocSemiring α, MulZeroOneClass α, AddCommMonoidWithOne α
```

A *semiring* is a nonunital-semiring and a nonassociative-semiring at the same time, and forms a monoid with zero:

```
class MonoidWithZero (M₀ : Type) extends
    Monoid M₀, MulZeroOneClass M₀, SemigroupWithZero M₀
```

```
class Semiring (α : Type) extends
    NonUnitalSemiring α, NonAssocSemiring α, MonoidWithZero α
```

A *ring* is a semiring and an abelian group at the same time that has "one" that behaves well:

```
class Ring (R : Type) extends Semiring R, AddCommGroup R, AddGroupWithOne R
```

A *commutative ring* is a ring (guarantees commutative addition) and a commutative monoid (guarantees commutative multiplication) at the same time:

```
class CommRing (α : Type) extends Ring α, CommMonoid α
```

A *division ring* is a *nontrivial* (i.e., at least two elements) ring whose multiplication forms a division monoid, whose nonzero elements have multiplicative inverses, whose zero is inverse to itself (if you find the equality `0⁻¹ = 0` disturbing, read the blog post [73] that explains it), and embeds rational numbers:

```
class Nontrivial (α : Type) : Prop where
  exists_pair_ne : ∃ x y : α, x ≠ y
```

```
class DivisionRing (K : Type) extends Ring K, DivInvMonoid K, Nontrivial K,
    NNRatCast K, RatCast K where
  mul_inv_cancel : ∀ a : K, a ≠ 0 → a * a⁻¹ = 1
  inv_zero : (0 : K)⁻¹ = 0
```

Note that the requirement of `NNRatCast` and `RatCast` and their properties doesn't restrict the class of division rings. They can be (minimally) defined as follows:

```
private class DivisionRing' (R : Type) extends
    Ring R, DivInvMonoid R, Nontrivial R where
  mul_inv_cancel : ∀ a : R, a ≠ 0 → a * a⁻¹ = 1
  inv_zero : (0 : R)⁻¹ = 0
```

A *field* is a commutative ring and a division ring at the same time:

```
class Field (K : Type) extends CommRing K, DivisionRing K
```

### 2.5.2 Binary relations

A *preorder* is a reflexive & transitive relation on any structure with binary relational symbols $\leq$ and $<$ where the strict comparison $a < b$ is equivalent to $a \leq b \land \neg(b \leq a)$ given by the relation $\leq$ which is neither required to be symmetric nor required to be antisymmetric:

```
class Preorder (α : Type) extends LE α, LT α where
  le_refl : ∀ a : α, a ≤ a
  le_trans : ∀ a b c : α, a ≤ b → b ≤ c → a ≤ c
  lt_iff_le_not_le : ∀ a b : α, a < b ↔ a ≤ b ∧ ¬(b ≤ a)
```

A *partial order* is an antisymmetric preorder (hence it is a reflexive & antisymmetric & transitive relation):

```
class PartialOrder (α : Type) extends Preorder α where
  le_antisymm : ∀ a b : α, a ≤ b → b ≤ a → a = b
```

A *linear order* (sometimes called a total order) is a partial order where every two elements are comparable (technical details are omitted):

```
class LinearOrder (α : Type) extends PartialOrder α where
  le_total (a b : α) : a ≤ b ∨ b ≤ a
  min_def (a b : α) : a ⊓ b = if a ≤ b then a else b
  max_def (a b : α) : a ⊔ b = if a ≤ b then b else a
  (...)
```

### 2.5.3 Binary operations and relations

An *ordered abelian monoid* is an abelian monoid with a partial order that respects the addition:

```
class OrderedAddCommMonoid (α : Type) extends
    AddCommMonoid α, PartialOrder α where
  add_le_add_left : ∀ a b : α, a ≤ b → ∀ c : α, c + a ≤ c + b
```

An *ordered cancellative abelian monoid* is an ordered abelian monoid where the operation of adding the same number from the left to both sides of an inequality can be cancelled (and, because of commutativity of addition, adding the same number from the right to both sides of an inequality can be cancelled as well):

```
class OrderedCancelAddCommMonoid (α : Type) extends
    OrderedAddCommMonoid α where
  le_of_add_le_add_left : ∀ a b c : α, a + b ≤ a + c → b ≤ c
```

An *ordered abelian group* is an abelian group with partial order that respects addition:

```
class OrderedAddCommGroup (α : Type) extends
    AddCommGroup α, PartialOrder α where
  add_le_add_left : ∀ a b : α, a ≤ b → ∀ c : α, c + a ≤ c + b
```

An *ordered ring* is a ring and an ordered abelian group where zero is less or equal to one and the product of nonnegative elements is nonnegative:

```
class OrderedRing (α : Type) extends Ring α, OrderedAddCommGroup α where
  zero_le_one : 0 ≤ (1 : α)
  mul_nonneg : ∀ a b : α, 0 ≤ a → 0 ≤ b → 0 ≤ a * b
```

An *ordered commutative ring* is an ordered ring and a commutative ring at the same time:

```
class OrderedCommRing (α : Type) extends OrderedRing α, CommRing α
```

A *strictly ordered ring* is a nontrivial ring whose addition behaves as an ordered abelian group, where zero is less or equal to one (in fact, zero must be strictly less than one, thanks to other requirements), and the product of two strictly positive elements is strictly positive:

```
class StrictOrderedRing (α : Type) extends
    Ring α, OrderedAddCommGroup α, Nontrivial α where
  zero_le_one : 0 ≤ (1 : α)
  mul_pos : ∀ a b : α, 0 < a → 0 < b → 0 < a * b
```

A *linearly ordered abelian monoid* is an ordered abelian monoid whose order is linear:

```
class LinearOrderedAddCommMonoid (α : Type) extends
    OrderedAddCommMonoid α, LinearOrder α
```

A *linearly ordered cancellative abelian monoid* is an ordered cancellative abelian monoid and linearly ordered abelian monoid at the same time:

```
class LinearOrderedCancelAddCommMonoid (α : Type) extends
    OrderedCancelAddCommMonoid α, LinearOrderedAddCommMonoid α
```

A *linearly ordered abelian group* is an ordered abelian group whose order is linear:

```
class LinearOrderedAddCommGroup (α : Type) extends
    OrderedAddCommGroup α, LinearOrder α
```

A *linearly ordered ring* is a strictly ordered ring where every two elements are comparable:

```
class LinearOrderedRing (α : Type) extends
    StrictOrderedRing α, LinearOrder α
```

A *linearly ordered commutative ring* is a linearly ordered ring and commutative monoid at the same time:

```
class LinearOrderedCommRing (α : Type) extends
    LinearOrderedRing α, CommMonoid α
```

In the Duality project, we define a *linearly ordered division ring* as a linearly ordered ring that is a division ring at the same time:

```
class LinearOrderedDivisionRing (R : Type) extends
    LinearOrderedRing R, DivisionRing R
```

A *linearly ordered field* is defined in Mathlib as a linearly ordered commutative ring that is a field at the same time:

```
class LinearOrderedField (α : Type) extends
    LinearOrderedCommRing α, Field α
```

Note that `LinearOrderedDivisionRing` is not a part of the algebraic hierarchy provided by Mathlib, hence `LinearOrderedField` does not inherit `LinearOrderedDivisionRing`. To compensate for it, we provide a custom instance that converts `LinearOrderedField` to `LinearOrderedDivisionRing` as follows:

```
instance LinearOrderedField.toLinearOrderedDivisionRing {F : Type}
    [instF : LinearOrderedField F] :
  LinearOrderedDivisionRing F := { instF with }
```

Note that everything in this subsection completely changed when Mathlib switched from Lean 4.18.0 to 4.19.0 (now they are all mixins). We review how they are implemented in the version of Mathlib that is based on Lean 4.18.0, which is the version we use in all subsequent chapters. While I prefer the older (more bundled) version of classes that combine binary operations with binary relations, it was noted [74] that their refactoring to mixins notably sped up compilation of Mathlib.

### 2.5.4 Modules

Given types `α` and `β` such that `α` has a scalar action on `β` (denoted by the infix `•` operator) and `α` forms a monoid, Mathlib defines a *multiplicative action* where `1` of the type `α` gives the identity action on `β` and multiplication in the monoid associates with the scalar action:

```
class MulAction (α : Type) (β : Type) [Monoid α] extends SMul α β where
  one_smul : ∀ b : β, (1 : α) • b = b
  mul_smul : ∀ (x y : α) (b : β), (x * y) • b = x • y • b
```

For a *distributive multiplicative action*, we furthermore require the latter type to form an additive monoid and two more properties are required; applying any action to the zero element preserves the zero element, and the multiplicative action is distributive with respect to addition:

```
class DistribMulAction (M A : Type) [Monoid M] [AddMonoid A] extends
    MulAction M A where
  smul_zero : ∀ a : M, a • (0 : A) = 0
  smul_add : ∀ (a : M) (x y : A), a • (x + y) = a • x + a • y
```

We can finally review the definition of a *module*. Here, the former type must form a semiring and the latter type an abelian monoid. A module requires a distributive multiplicative action and two additional properties; addition in the semiring distributes with the multiplicative action, and applying the zero action to any element gives the zero element:

```
class Module (R : Type) (M : Type) [Semiring R] [AddCommMonoid M] extends
    DistribMulAction R M where
  add_smul : ∀ (r s : R) (x : M), (r + s) • x = r • x + s • x
  zero_smul : ∀ x : M, (0 : R) • x = 0
```

Note the class `Module` does not extend the class `Semiring`; instead, it requires `Semiring` as an argument. The abelian monoid is also required as an argument in the definition. We call such a class “mixin”. Thanks to this design, we don’t need to define subclasses of `Module` in order to require “more than a module”. Instead, we use subclasses in the respective arguments, i.e., we require “more than a semiring” and/or “more than an abelian monoid”. For example, if we replace

```
[Semiring R] [AddCommMonoid M] [Module R M]
```

in assumptions by

```
[Field R] [AddCommGroup M] [Module R M]
```

we require `M` to be a vector space over `R` (a field). We don’t need to extend `Module` in order to define what a vector space is.

At one point in this thesis (the theorem `fintypeFarkasBartl` in Section 3.1.2 and its proof in Section 3.1.3), we will need to work with linearly ordered vector spaces. To formalize that `V` is a linearly ordered vector space, we need `R` to be a linearly ordered division ring, we need `V` to be a linearly ordered abelian group, and we need `V` to form a module over `R`. Furthermore, we need a relationship between how `R` is ordered and how `V` is ordered. For that, we use another mixin, defined in Mathlib as follows:

```
class PosSMulMono (α β : Type)
    [SMul α β] [Preorder α] [Preorder β] [Zero α] :
    Prop where
  elim {a : α} (_ : 0 ≤ a) {b₁ b₂ : β} (_ : b₁ ≤ b₂) : a • b₁ ≤ a • b₂
```

The following list of assumption formalizes the notion of `V` being a linearly ordered vector space over `R` whose multiplication needn’t be commutative:

```
{R V : Type} [LinearOrderedDivisionRing R] [LinearOrderedAddCommGroup V]
[Module R V] [PosSMulMono R V]
```

Since I don’t know whether “vector space” is the correct terminology in situations where the underlying division ring doesn’t have commutative multiplication (i.e., it isn’t a vector space over a field), I will refrain from saying “vector space” in the next chapter.

## *2.6 Axioms*

Unlike set theory, which is usually defined inside the first-order logic, the calculus of inductive constructions replaces both logic and set theory on its own. The type system naturally gives rise to logical rules via the Curry-Howard correspondence [75] [76]. However, the resulting logical framework is not as strong as what first-order logic provides. For example, the calculus of inductive constructions alone cannot prove the law of excluded middle. Similarly, Lean’s kernel cannot cancel double negation. In order to cover the entire range of reasoning that mathematicians do, Lean needs to be equipped with extra axioms. There are three standard axioms that are accepted as legitimate tools in so-called *classical reasoning*.

First, we have the *propositional extensionality*:

```
axiom propext {a b : Prop} :
  (a ↔ b) → a = b
```

It states that, when two propositions imply each other, they are equal. This axiom shouldn’t stir any controversy at all. If we have equivalent propositions, we can substitute one for the other in any scenario [77]. Mathematicians do it all the time without thinking about it.

Second, we have the *axiom of choice*:

```
axiom Classical.choice {α : Sort u} :
  Nonempty α → α
```

This creatio-ex-nihilo axiom constructs data out of a (nonconstructive) promise that something of given type exists. Traditionally, the axiom of choice was more nuanced. In other formal systems, for example, the axiom of choice is described as "given any family of nonempty sets, it is possible to construct a new set by taking one element from each set". Its appeal lies in its application to an infinite family of nonempty sets, constructing a new infinite set by skipping over infinitely many decisions about how to choose elements from respective sets [78]. Its introduction quickly became one of the most debated moments in the foundations of mathematics. Gina Garcia Tarrach [79] summarized it as follows:

> "Zermelo's publication was immediately controversial. Its consequences were not only mathematical, but also philosophical, for it postulated the existence of certain mathematical objects that were not explicitly defined (the axiom explicits no rule by which the choices are made), and therefore questioned the very notion of what a mathematical object is and what does it mean that it exists. The discussion about whether or not the Axiom of Choice and what it implied should be accepted arose a heated debate between constructivist mathematicians and non-constructivist ones."

In the end, however, most mathematicians accepted the axiom of choice for how useful it is. For example, the axiom of choice is used to prove that every vector space has a basis or to prove that every connected graph has a spanning tree. In Lean, the axiom of choice is even more important than in set-theory-based mathematics because, in Lean, the axiom of choice is used to prove necessities of classical reasoning like the law of excluded middle and the double negation elimination [77].

Third, we have the *quotient soundness*:

```
axiom Quot.sound {α : Sort u} {r : α → α → Prop} {a b : α} (_ : r a b) :
  Quot.mk r a = Quot.mk r b
```

In the language of cosets, this axiom can be explained as "elements lying in the same equivalence class represent the same coset". The notion of congruence thereby gets elevated to the notion of equality, which is more powerful [77]. For a full explanation of quotients and their implementation in Lean, we highly recommend a blog post [80] from the Xena project.

# 3 Optimization theory

Optimization theory (or mathematical optimization) is the study of how to select the best option from a set of available alternatives according to some criteria [81]. Generally speaking, we have some variables that can be assigned some values, some constraints on the values of those variables, and an objective function that is supposed to be either minimized or maximized.

Optimization theory is a major area of applied MathematiCS. We can imagine maximizing the profit or minimizing the prediction error. We can also think of practical situations with many different constraints that must be all satisfied at the same time. For example, a delivery company might optimize its routes to minimize total travel cost while ensuring that every package is delivered within its promised window and the weight of the cargo never exceeds the vehicle's capacity. We could go on and on.

Depending on the domain of said variables, we speak of either continuous optimization or discrete optimization. In continuous optimization, we usually work with real numbers. In discrete optimization, we typically deal with bitstrings, with set systems (such as graphs), or with maps between two countable sets. These areas aren't disjoint, and indeed, the VCSP theory discussed later unifies elements of both continuous and discrete optimization (albeit usually studied only in the latter context).

Section 3.1 builds towards the theory of linear programming. Section 3.2 focuses on a more general theory of optimization, less practical and more abstract; however, later parts reveal its connection to linear programming again. Linear programming is a highly practical area, with applications ranging from optimization of power plants [82] and energy storage [83] to logistics [84], public transport [85], telecommunications [86], financial planning [87], or even the design of radiation therapy for cancer [88].

In several places in this chapter, it will be convenient to refer to nonnegative numbers that are bundled, i.e., instead of saying "here is a number" and "here is a proof that the number is nonnegative", we want to have a type that allows only nonnegative numbers. We define it as a subtype:

```
abbrev NNeg (F : Type) [OrderedAddCommMonoid F] := { a : F // 0 ≤ a }
syntax:max ident noWs "≥0" : term
```

This subtype is equipped with a notation. Whenever `F` is an ordered abelian monoid, we can write `F≥0` to denote the type of nonnegative `F` values. We automatically obtain a conversion from `F≥0` to `F` but, to convert explicit vectors with `F≥0` entries to explicit vectors with `F` entries, we need to implement the conversion manually:

```
def coeNN {I R : Type} [OrderedAddCommMonoid R] : (I → R≥0) → (I → R) :=
  (Subtype.val ∘ ·)
```

Furthermore, we register it as an implicit coercion:

```
instance {I R : Type} [OrderedAddCommMonoid R] : Coe (I → R≥0) (I → R) :=
  ⟨coeNN⟩
```

Note that Lean distinguishes between nonnegative (explicit) vectors and (explicit) vectors of nonnegative elements. A nonnegative vector is a tuple "function from coordinates to values" and a proof "for every coordinate, the output is nonnegative". A vector of nonnegative elements is a function from coordinates to tuples "value, proof of nonnegativity". In the language of type theory, the former is a `Sigma` type of `Pi` types, whereas the latter is a `Pi` type of `Sigma` types.

## 3.1 Linear duality

Gyula Farkas [89] [90] established that a system of linear equalities has a nonnegative solution iff we cannot obtain a contradiction by taking a linear combination of the equalities:

```
theorem equalityFarkas {I J F : Type} [Fintype I] [Fintype J]
    [LinearOrderedField F]
    (A : Matrix I J F) (b : I → F) :
    (∃ x : J → F, 0 ≤ x ∧ A *ᵥ x = b) ≠
    (∃ y : I → F, 0 ≤ Aᵀ *ᵥ y ∧ b ⬝ᵥ y < 0)
```

Geometric interpretation of `equalityFarkas` is as follows. The column vectors of `A` generate a cone (in the |`I`|-dimensional Euclidean space) from the origin. The point `b` either lies inside this cone (in which case, the entries of `x` give nonnegative coefficients which, when applied to the column vectors of `A`, give a vector from the origin to the point `b`), or there exists a hyperplane that contains the origin and strictly separates `b` from this cone (in which case, `y` gives a normal vector of this hyperplane).

Our project makes the following contributions in Lean 4:

- We formalize and prove several Farkas-like theorems and the strong LP duality.
- We extend the theory to settings when some coefficients are allowed to be infinitely large.

### 3.1.1 Homomorphisms

Before we state `equalityFarkas` in more general settings, we need to learn how Mathlib defines linear maps.

An *additive homomorphism* is a function that behaves well with respect to addition:

```
structure AddHom (M N : Type) [Add M] [Add N] where
  toFun : M → N
  map_add' : ∀ x y : M, toFun (x + y) = toFun x + toFun y
```

A *linear map* is defined as follows:

```
structure LinearMap {R S : Type} [Semiring R] [Semiring S] (σ : R →+* S)
    (M : Type) (M' : Type) [AddCommMonoid M] [AddCommMonoid M']
    [Module R M] [Module S M'] extends
    AddHom M M', MulActionHom σ M M'
```

Unfortunately, understanding the definition above is too difficult because both the semiring homomorphism and the scalar action homomorphism (which are required by this definition) are very complicated. Fortunately, we will need only a certain special case of linear maps. This special case is denoted by `M →ₗ[R] M'` where both `M` and `M'` are `R`-modules, in which `σ` can be ignored (it is identity), and it can be best understood via the lens of the following structure:

```
structure IsLinearMap (R : Type) {M M' : Type} [Semiring R]
    [AddCommMonoid M] [AddCommMonoid M'] [Module R M] [Module R M']
    (f : M → M') : Prop where
  map_add : ∀ x y : M, f (x + y) = f x + f y
  map_smul : ∀ (c : R) (x : M), f (c • x) = c • f x
```

We easily see that the function `f` is required to be compatible with `+` (addition) and `•` (the distributive multiplicative action) from the two modules[15]. A linear map (in the general sense) from "a function that happens to be a linear map" (in the narrow sense) is then constructed as follows:

```
notation:25 M " →ₗ[" R:25 "] " M':0 => LinearMap (RingHom.id R) M M'
```

```
def IsLinearMap.mk' {R : Type} {M M' : Type} [Semiring R]
    [AddCommMonoid M] [AddCommMonoid M'] [Module R M] [Module R M']
    (f : M → M') (_ : IsLinearMap R f) :
    M →ₗ[R] M' where
  toFun := f
  map_add' := sorry
  map_smul' := sorry
```

Conversely, every linear map in the `M →ₗ[R] M'` sense `IsLinearMap` in the expected way:

```
theorem LinearMap.isLinear {R : Type} {M M' : Type} [Semiring R]
    [AddCommMonoid M] [AddCommMonoid M'] [Module R M] [Module R M']
    (fₗ : M →ₗ[R] M') :
    IsLinearMap R fₗ
```

This way, we have learnt everything we need to know about linear maps for the purposes of Section 3.1, without studying the complicated general definitions.

### 3.1.2 Generalizations

The next theorem generalizes `equalityFarkas` to structures where multiplication doesn't have to be commutative; furthermore, it supports infinitely many equations:

```
theorem coordinateFarkasBartl {I J R : Type} [Fintype J]
    [LinearOrderedDivisionRing R]
    (A : (I → R) →ₗ[R] J → R) (b : (I → R) →ₗ[R] R) :
    (∃ x : J → R, 0 ≤ x ∧ ∀ w : I → R, ∑ j : J, A w j • x j = b w) ≠
    (∃ y : I → R, 0 ≤ A y ∧ b y < 0)
```

In the next theorem [91], the partially ordered module `I → R` is replaced by a general `R`-module `W`:

```
theorem almostFarkasBartl {J R W : Type} [Fintype J]
    [LinearOrderedDivisionRing R] [AddCommGroup W] [Module R W]
    (A : W →ₗ[R] J → R) (b : W →ₗ[R] R) :
    (∃ x : J → R, 0 ≤ x ∧ ∀ w : W, ∑ j : J, A w j • x j = b w) ≠
    (∃ y : W, 0 ≤ A y ∧ b y < 0)
```

In the most general theorem [92], stated below, certain occurrences of `R` are replaced by a linearly ordered `R`-module `V` whose order respects the order on `R`:

[15] Operations from `M` are on the LHS. Operations from `M'` are on the RHS.

```
theorem fintypeFarkasBartl {J R V W : Type} [Fintype J]
    [LinearOrderedDivisionRing R]
    [LinearOrderedAddCommGroup V] [Module R V] [PosSMulMono R V]
    [AddCommGroup W] [Module R W]
    (A : W →ₗ[R] J → R) (b : W →ₗ[R] V) :
    (∃ x : J → V, 0 ≤ x ∧ ∀ w : W, ∑ j : J, A w j • x j = b w) ≠
    (∃ y : W, 0 ≤ A y ∧ b y < 0)
```

Note that `fintypeFarkasBartl` subsumes `scalarFarkas` as well as the other versions, since `R` can be viewed as a linearly ordered module over itself.

We have hereby stated a three-fold generalization of the original Farkas' result. Let's prove it! Our proof, described below, is based on a modern algebraic proof by David Bartl [92]. We first prove a tiny-bit-less-general version `finFarkasBartl` which uses `Fin n` (i.e., indexing by natural numbers between `0` inclusive and `n` exclusive) instead of an arbitrary (unordered) finite type `J`. In the end, we obtain `fintypeFarkasBartl` from `finFarkasBartl` using some boring mechanisms regarding equivs between finite types.

### 3.1.3 Proving finFarkasBartl

In this subsection, we will prove:

```
theorem finFarkasBartl {R V W : Type} {n : ℕ}
    [LinearOrderedDivisionRing R]
    [LinearOrderedAddCommGroup V] [Module R V] [PosSMulMono R V]
    [AddCommGroup W] [Module R W]
    (A : W →ₗ[R] Fin n → R) (b : W →ₗ[R] V) :
    (∃ x : Fin n → V, 0 ≤ x ∧ ∀ w : W, ∑ j : Fin n, A w j • x j = b w) ≠
    (∃ y : W, 0 ≤ A y ∧ b y < 0)
```

We first rephrase the goal to:

```
(∃ x : Fin n → V, 0 ≤ x ∧ ∀ w : W, ∑ j : Fin n, A w j • x j = b w) ↔
(∀ y : W, 0 ≤ A y → 0 ≤ b y)
```

Implication from left to right is immediately satisfied by the following term:

```
fun ⟨x, hx, hb⟩ y hy =>
  hb y ▸ Finset.sum_nonneg (fun i _ => smul_nonneg (hy i) (hx i))
```

Implication from right to left will be proved by induction on `n` with generalized `A` and `b`.

In case `n = 0` we immediately have:

```
A_tauto : ∀ w : W, 0 ≤ A w
```

We have an assumption:

```
hAb : ∀ y : W, 0 ≤ A y → 0 ≤ b y
```

We set `x` to be the empty vector family. Now, for every `w : W`, we must prove:

```
∑ j : Fin 0, A w j • (0 : Fin 0 → V) j = b w
```

We simplify the goal to:

```
0 = b w
```

We exploit the fact that `V` is ordered and prove the equality as two inequalities.

Inequality `0 ≤ b w` is directly satisfied by:

```
hAb w (A_tauto w)
```

Inequality `b w ≤ 0` is easily reduced to:

```
hAb (-w) (A_tauto (-w))
```

The induction step is stated as a lemma:

```
lemma industepFarkasBartl {R V W : Type} {m : ℕ}
    [LinearOrderedDivisionRing R]
    [LinearOrderedAddCommGroup V] [Module R V] [PosSMulMono R V]
    [AddCommGroup W] [Module R W]
    (ih : ∀ A₀ : W →ₗ[R] Fin m → R, ∀ b₀ : W →ₗ[R] V,
      (∀ y₀ : W, 0 ≤ A₀ y₀ → 0 ≤ b₀ y₀) →
        (∃ x₀ : Fin m → V, 0 ≤ x₀ ∧ ∀ w₀ : W,
            ∑ i₀ : Fin m, A₀ w₀ i₀ • x₀ i₀ = b₀ w₀))
    {A : W →ₗ[R] Fin m.succ → R} {b : W →ₗ[R] V}
    (hAb : ∀ y : W, 0 ≤ A y → 0 ≤ b y) :
    ∃ x : Fin m.succ → V,
      0 ≤ x ∧ ∀ w : W, ∑ i : Fin m.succ, A w i • x i = b w
```

We define

```
a : W →ₗ[R] Fin m → R
```

as the first `m` rows of `A` (i.e., `A` without the last row):

```
a := (fun w : W => fun i : Fin m => A w i)
```

To prove `industepFarkasBartl` we first consider the easy case:

```
is_easy : ∀ y : W, 0 ≤ a y → 0 ≤ b y
```

From `ih a b is_easy` we obtain:

```
x : Fin m → V
hx : 0 ≤ x
hxb : ∀ w₀ : W, ∑ i₀ : Fin m, a w₀ i₀ • x i₀ = b w₀
```

The goal in the easy case is satisfied by this vector family:

```
(fun i : Fin m.succ => if i < m then x i else 0)
```

Easy case analysis shows that the vector family is nonnegative.

Now we need to prove:

```
∀ w : W,
  ∑ i : Fin m.succ,
    A w i • (fun i : Fin m.succ => if i < m then x i else 0) i =
  b w
```

We simplify the goal to:

```
∀ w : W, ∑ i : Fin m, A w i • x i = b w
```

This is exactly `hxb`.

Now for the hard case; negation of `is_easy` gives us:

```
y' : W
hay' : 0 ≤ a y'
hby' : b y' < 0
```

Let `y` be (flipped and) rescaled `y'` as follows:

```
y : W := (A y' m)⁻¹ • y'
```

From `hAb` with `hby'` and `hay'` we get:

```
hAy' : A y' m < 0
```

Therefore `hAy'.ne : A y'm ≠ 0` implies that `y` has the property that motivated the rescaling above:

```
hAy : A y m = 1
```

From `hAy` we have:

```
hAA : ∀ w : W, A (w - (A w m • y)) m = 0
```

Using `hAA` and `hAb` we prove:

```
hbA : ∀ w : W, 0 ≤ a (w - (A w m • y)) → 0 ≤ b (w - (A w m • y))
```

From `hbA` we have:

```
hbAb : ∀ w : W, 0 ≤ (a - (A · m • a y)) w → 0 ≤ (b - (A · m • b y)) w
```

We observe that these two terms (appearing in `hbAb` we just proved) are linear maps:

```
(a - (A · m • a y))
(b - (A · m • b y))
```

Therefore, we can plug them into `ih` and provide `hbAb` as the last argument. We obtain:

```
x' : Fin m → V
hx' : 0 ≤ x'
hxb' : ∀ w₀ : W, ∑ i₀ : Fin m, (a - (A · m • a y)) w₀ i₀ • x' i₀ =
                 (b - (A · m • b y)) w₀
```

We claim that our lemma is satisfied by the following vector family:

```
(fun i : Fin m.succ =>
  if i < m then x' i else b y - ∑ j : Fin m, a y i • x' j)
```

Let us show the nonnegativity first.

Nonnegativity of everything except of the last vector follows from `hx'`.

Now, to prove nonnegativity of the last vector, from `hAy'` we have:

```
hAy'' : (A y' m)⁻¹ ≤ 0
```

From `hAy''` with `hay'` we have:

```
hay : a y ≤ 0
```

From `hAy''` with `hby'` converted to nonstrict inequality we have:

```
hby : 0 ≤ b y
```

We need to prove:

```
∑ j : Fin m, a y j • x' j ≤ b y
```

It follows from `hay j` with `hx' j` and `hby` using basic properties of inequalities.

The only remaining task is to show:

```
∀ w : W,
  ∑ i : Fin m.succ,
    (A w i • (if i < m then x' i else b y - ∑ j : Fin m, a y j • x' j)) =
  b w
```

Given general `w : W` we make a key observation (using `hxb' w`):

```
haAa : ∑ i : Fin m, (a w i - A w m * a y i) • x' i = b w - A w m • b y
```

With the help of `haAa` we transform the goal to:

```
∑ i : Fin m.succ,
  (A w i • (if i < m then x' i else b y - ∑ j : Fin m, a y j • x' j)) =
∑ i : Fin m, (a w i - A w m * a y i) • x' i + A w m • b y
```

We distribute `•` over `ite` so that the goal becomes:

```
∑ i : Fin m.succ,
  (if i < m then A w i • x' i else
    A w i • (b y - ∑ j : Fin m, a y j • x' j)) =
∑ i : Fin m, (a w i - A w m * a y i) • x' i + A w m • b y
```

We split the LHS into two parts:

```
∑ i : Fin m, (a w i • x' i) + A w m • (b y - ∑ j : Fin m, a y j • x' j) =
∑ i : Fin m, (a w i - A w m * a y i) • x' i + A w m • b y
```

The rest is a simple manipulation with sums.

### 3.1.4 A few corollaries

In this subsection, we will work with the following context:

```
variable {I J F : Type} [Fintype I] [Fintype J] [LinearOrderedField F]
```

For the corollaries, we start with the matrix version, which we have already reviewed:

```
theorem equalityFarkas (A : Matrix I J F) (b : I → F) :
    (∃ x : J → F, 0 ≤ x ∧ A *ᵥ x = b) ≠
    (∃ y : I → F, 0 ≤ Aᵀ *ᵥ y ∧ b ⬝ᵥ y < 0)
```

Fredholm [93] established that a system of linear equalities has a solution iff we cannot obtain a contradiction by taking a linear combination of the equalities. We state his theorem as follows (the first version is more aligned with our surrounding theorems; the second version is more aligned with textbooks on linear algebra):

```
theorem basicLinearAlgebra_lt (A : Matrix I J F) (b : I → F) :
    (∃ x : J → F, A *ᵥ x = b) ≠
    (∃ y : I → F, Aᵀ *ᵥ y = 0 ∧ b ⬝ᵥ y < 0)
```

```
theorem basicLinearAlgebra (A : Matrix I J F) (b : I → F) :
    (∃ x : J → F, A *ᵥ x = b) ≠
    (∃ y : I → F, Aᵀ *ᵥ y = 0 ∧ b ⬝ᵥ y ≠ 0)
```

Geometric interpretation of `basicLinearAlgebra` is straightforward. The column vectors of `A` generate a hyperplane in the |`I`|-dimensional Euclidean space that contains the origin. The point `b` either lies in this hyperplane (in this case, the entries of `x` give coefficients which, when applied to the column vectors of `A`, give a vector from the origin to the point `b`), or there exists a line through the origin that is orthogonal to all the column vectors of `A` (i.e., orthogonal to the entire hyperplane) such that `b` projected onto this line falls outside of the origin (in which case, `y` gives a direction of this line), i.e., to a different point from where all column vectors of `A` get projected.

This theorem can be given in much more general settings [93]. In our library, however, it is the only version we provide. Instead of diving deeper into linear algebra, we focus more on inequalities and prove the following corollary by Hermann Minkowski [94]. A system of linear inequalities has a nonnegative solution iff we cannot obtain a contradiction by taking a nonnegative linear combination of the inequalities:

```
theorem inequalityFarkas (A : Matrix I J F) (b : I → F) :
    (∃ x : J → F, 0 ≤ x ∧ A *ᵥ x ≤ b) ≠
    (∃ y : I → F, 0 ≤ y ∧ 0 ≤ Aᵀ *ᵥ y ∧ b ⬝ᵥ y < 0)
```

Geometric interpretation of `inequalityFarkas` is a bit more convoluted. The column vectors of `A` generate a cone in the |`I`|-dimensional Euclidean space from the origin to some infinity. The point `b` determines an orthogonal cone that starts in `b` and goes to negative infinity in the direction of all coordinate axes. Either these two cones intersect (in which case, the entries of `x` give nonnegative coefficients which, when applied to the column vectors of `A`, give a vector from the origin to a point in the intersection), or there exists a hyperplane that contains the origin and that strictly separates `b` from the cone generated by `A` but does not cut through the positive orthant, i.e., the origin is the only nonnegative point contained in the hyperplane (in which case, `y` gives a normal vector of this hyperplane).

```
theorem inequalityFarkas_neg (A : Matrix I J F) (b : I → F) :
    (∃ x : J → F, 0 ≤ x ∧ A *ᵥ x ≤ b) ≠
    (∃ y : I → F, 0 ≤ y ∧ -Aᵀ *ᵥ y ≤ 0 ∧ b ⬝ᵥ y < 0)
```

This theorem is nearly identical to `inequalityFarkas`, but `inequalityFarkas_neg` replaces the condition `0 ≤ Aᵀ *ᵥ y` by an equivalent condition `-Aᵀ *ᵥ y ≤ 0` for convenience in later extended versions (it will be in such algebra that these two conditions will not be equivalent).

### 3.1.5 Linear Programming

Before we define linear programming in generality, we start with a simple illustrative example:

$$\text{minimize} \quad 6 * x_0 + 6 * x_1$$
$$2 * x_0 + x_1 \geq 4$$
$$x_0 + 2 * x_1 \geq 5$$
$$x_0 \geq 0$$
$$x_1 \geq 0$$

Imagine that everything is a real number. Such a problem is called linear program because it is an optimization problem where the objective function (i.e., what should be minimized) is linear and all constraints (i.e., what must hold) are linear inequalities. One solution is:

$$x_0 = 10$$
$$x_1 = 10$$

The resulting objective value is `120`, which is too much. A different solution is:

$$x_0 = 5$$
$$x_1 = 0$$

The resulting objective value is `30`, which is way better. Another solution is:

$$x_0 = 0$$
$$x_1 = 4$$

The resulting objective value is `24`, which is even better. Yet another solution is:

$$x_0 = 1$$
$$x_1 = 2$$

The resulting objective value is `18`, which is the best solution so far. However, we would like to know better. We would like to know that it is the best solution — that no better solution exists. How can we know that any solution (assignment of numbers to variables that satisfy all constraints) yields an objective value that is greater or equal to something?

One way to obtain a lower bound is to multiply the first inequality by three (note that variables must be nonnegative, hence the first inequality below holds):

$$6 * x_0 + 6 * x_1 \geq 6 * x_0 + 3 * x_1 \geq 12$$

A stronger lower bound appears when we multiply the second inequality by three:

$$6 * x_0 + 6 * x_1 \geq 3 * x_0 + 6 * x_1 \geq 15$$

Given what we have observed so far, the optimum must be somewhere between `15` and `18`. Let's try to find another lower bound and be smarter about it this time. The following lower bound multiplies both inequalities by two and adds them up:

$$6 * x_0 + 6 * x_1 = (4 * x_0 + 2 * x_1) + (2 * x_0 + 4 * x_1) \geq 8 + 10 = 18$$

We have found a solution that yields the objective value `18` and a lower bound that any solution must yield an objective value at least `18`. Therefore, `18` is the real optimum.

Is there any systematic way to search for the best lower bound, or is it matter of trial and error every time? In the linear program above, we can consider multiplying the first inequality by a nonnegative coefficient $y_0$ and the second inequality by a nonnegative coefficient $y_1$ such that:

$$
\begin{aligned}
&2 * y_0 + y_1 \leq 6 \\
&y_0 + 2 * y_1 \leq 6
\end{aligned}
$$

It results in a lower bound:

$$4 * y_0 + 5 * y_1$$

To summarize, to search for the best (i.e., highest) lower bound is to solve the following optimization problem:

$$
\begin{aligned}
&\text{maximize} \quad 4 * y_0 + 5 * y_1 \\
&2 * y_0 + y_1 \leq 6 \\
&y_0 + 2 * y_1 \leq 6 \\
&y_0 \geq 0 \\
&y_1 \geq 0
\end{aligned}
$$

It is a linear program, again! One solution for the new linear program is:

$$
\begin{aligned}
&y_0 = 1 \\
&y_1 = 1
\end{aligned}
$$

The resulting objective value is 9, which we know isn't maximal. The optimal solution is:

$$
\begin{aligned}
&y_0 = 2 \\
&y_1 = 2
\end{aligned}
$$

The resulting objective value is 18, which is equal to what a solution for the previous linear program gave, and therefore, it must be optimal.

If we were to express the search for the best (i.e., lowest) upper bound of the new linear program as another linear program, we would obtain:

$$
\begin{aligned}
&\text{minimize} \quad 6 * z_0 + 6 * z_1 \\
&2 * z_0 + z_1 \geq 4 \\
&z_0 + 2 * z_1 \geq 5 \\
&z_0 \geq 0 \\
&z_1 \geq 0
\end{aligned}
$$

For comparison, the initial problem we started with was:

$$
\begin{aligned}
&\text{minimize} \quad 6 * x_0 + 6 * x_1 \\
&2 * x_0 + x_1 \geq 4 \\
&x_0 + 2 * x_1 \geq 5 \\
&x_0 \geq 0 \\
&x_1 \geq 0
\end{aligned}
$$

We see that it is the same linear program.

Note that the same linear program can be expressed in terms of vectors and a matrix:

$$
\begin{gathered}
\text{minimize} \quad \begin{pmatrix} 6 & 6 \end{pmatrix} \cdot_v \begin{pmatrix} x_0 \\ x_1 \end{pmatrix} \\
\begin{pmatrix} 2 & 1 \\ 1 & 2 \end{pmatrix} *_v \begin{pmatrix} x_0 \\ x_1 \end{pmatrix} \geq \begin{pmatrix} 4 \\ 5 \end{pmatrix} \\
\begin{pmatrix} x_0 \\ x_1 \end{pmatrix} \geq \begin{pmatrix} 0 \\ 0 \end{pmatrix}
\end{gathered}
$$

Equivalently, it can be written as follows:

$$\textsf{minimize}\quad \begin{pmatrix} 6 & 6 \end{pmatrix} \cdot_v \begin{pmatrix} x_0 \\ x_1 \end{pmatrix}$$

$$\begin{pmatrix} -2 & -1 \\ -1 & -2 \end{pmatrix} *_v \begin{pmatrix} x_0 \\ x_1 \end{pmatrix} \leq \begin{pmatrix} -4 \\ -5 \end{pmatrix}$$

$$\begin{pmatrix} x_0 \\ x_1 \end{pmatrix} \geq \begin{pmatrix} 0 \\ 0 \end{pmatrix}$$

This form is less convenient for the sake of the illustratory example but more convenient for the purpose of future development.

Now, recall the other linear program:

```
maximize   4 * y₀ + 5 * y₁
2 * y₀ + y₁ ≤ 6
y₀ + 2 * y₁ ≤ 6
y₀ ≥ 0
y₁ ≥ 0
```

This linear program can be expressed as follows:

$$\textsf{minimize}\quad \begin{pmatrix} -4 & -5 \end{pmatrix} \cdot_v \begin{pmatrix} y_0 \\ y_1 \end{pmatrix}$$

$$\begin{pmatrix} 2 & 1 \\ 1 & 2 \end{pmatrix} *_v \begin{pmatrix} y_0 \\ y_1 \end{pmatrix} \leq \begin{pmatrix} 6 \\ 6 \end{pmatrix}$$

$$\begin{pmatrix} y_0 \\ y_1 \end{pmatrix} \geq \begin{pmatrix} 0 \\ 0 \end{pmatrix}$$

Now, both linear programs are written in the exactly same form and only their data differ. And it is the form that will be captured by our "standard" LPs.

The formal definition of LP comes now.

A *linear program* is defined by a matrix `A` and vectors `b` and `c` of compatible dimensions:

```
structure StandardLP (I J R : Type) where
  A : Matrix I J R
  b : I → R
  c : J → R
```

Variables are of type `J`. Constraints are indexed by type `I`. The implicit objective function is intended to be minimized (unlike in the classical literature, where both the minimization and the maximization are defined, as we hinted in the introductory example).

In the rest of this subsection, we will assume:

```
variable {I J R : Type} [Fintype J]
```

A vector `x` made of nonnegative values is a *solution* to a linear program iff its multiplication by the matrix `A` from the left yields a vector whose all entries are less or equal to the corresponding entries of the vector `b`:

```
def StandardLP.IsSolution [OrderedSemiring R]
    (P : StandardLP I J R) (x : J → R≥0) :
    Prop :=
  P.A *ᵥ x ≤ P.b
```

A linear program *reaches* an objective value `r` iff it has a solution `x` such that, when its entries are elementwise multiplied by the the coefficients `c` and summed up, the result is the value `r`:

```
def StandardLP.Reaches [OrderedSemiring R]
    (P : StandardLP I J R) (r : R) :
    Prop :=
  ∃ x : J → R≥0, P.IsSolution x ∧ P.c ·ᵥ x = r
```

A linear program is *feasible* iff[16] it reaches a value:

```
def StandardLP.IsFeasible [OrderedSemiring R]
    (P : StandardLP I J R) :
    Prop :=
  ∃ r : R, P.Reaches r
```

A linear program is *bounded by* a value `r` (from below — we always minimize) iff it reaches only values greater or equal to `r`:

```
def StandardLP.IsBoundedBy [OrderedSemiring R]
    (P : StandardLP I J R) (r : R) :
    Prop :=
  ∀ p : R, P.Reaches p → r ≤ p
```

A linear program is *unbounded* iff it has no lower bound:

```
def StandardLP.IsUnbounded [OrderedSemiring R]
    (P : StandardLP I J R) :
    Prop :=
  ¬∃ r : R, P.IsBoundedBy r
```

To *dualize* a linear program, we transpose the matrix and flip all its signs, and we swap the right-hand-side vector with the vector of objective function coefficients:

```
def StandardLP.dualize [Ring R] (P : StandardLP I J R) :
    StandardLP J I R :=
  ⟨-P.Aᵀ, P.c, P.b⟩
```

Note that this definition requires `R` to be a ring, because the unary minus is needed. Keep in mind that the implicit intention still is to minimize the objective function.

One result we prove is the weak duality of linear programming:

```
theorem StandardLP.weakDuality [Fintype I] [OrderedCommRing R]
    {P : StandardLP I J R}
    {p : R} (_ : P.Reaches p) {q : R} (_ : P.dualize.Reaches q) :
    0 ≤ p + q
```

Note that the nonstandard conclusion `0 ≤ p + q` corresponds to the fact that both LPs are minimized (whereas literature usually minimizes one LP and maximizes the other LP).

[16] Here, we could equivalently say that a linear program is feasible iff it has a solution. However, it will not be the same for extended linear programs, which we will define later.

In the rest of this subsection, we will assume:

```
variable [LinearOrderedField R]
```

While the weak duality theorem talks about any pair of `p` and `q`, including the optimal ones, before we state the strong duality theorem, we need to define the notion of *optimum* of an LP. The following definition depends on the notion of an extended linearly ordered field, which will be explained in the next subsection, but for our purposes, it should be understandable already hopefully:

```
noncomputable def StandardLP.optimum (P : StandardLP I J R) : Option R∞ :=
  if ¬P.IsFeasible then
    some ⊤
  else
    if P.IsUnbounded then
      some ⊥
    else
      if hr : ∃ r : R, P.Reaches r ∧ P.IsBoundedBy r then
        some hr.choose
      else
        none
```

The definition says that, if the LP is infeasible, its optimum is `⊤` (i.e., the positive infinity, i.e., the worst value). Else, if the LP is unbounded, its optimum is `⊥` (i.e., the negative infinity, i.e., the best value). Else, if the LP has a lower bound that it reaches, it has this finite optimum. Otherwise, the LP doesn't have an optimum, according to the definition above. A detailed breakdown how exactly the definition of optimum works can be found in Section 3.1.6 (after the extended linearly ordered field is formally defined).

One theorem we prove is that the optimum always exists (that is, there cannot be an LP with a finite infimum that isn't attained):

```
theorem StandardLP.optimum_neq_none (P : StandardLP I J R) :
    P.optimum ≠ none
```

We define what it means when two optima are *opposites*:

```
def OppositesOpt : Option R∞ → Option R∞ → Prop
| (p : R∞), (q : R∞) => p = -q
| _       , _        => False
```

For example:

```
OppositesOpt 5 (-5) = True
OppositesOpt (-3) 3 = True
OppositesOpt 0 0 = True
OppositesOpt ⊤ ⊥ = True
OppositesOpt ⊥ ⊤ = True
OppositesOpt none none = False
OppositesOpt none 0 = False
OppositesOpt 6 (-4) = False
OppositesOpt (-3) 4 = False
```

```
OppositesOpt (-5) (-5) = False
OppositesOpt 2 2 = False
OppositesOpt ⊤ 7 = False
OppositesOpt 0 ⊥ = False
OppositesOpt ⊤ ⊤ = False
```

Finally, we can state the strong duality theorem (consider an LP and its dual; if at least one of them is feasible, the optima of these two LPs are opposites) :

```
theorem StandardLP.strongDuality (P : StandardLP I J R)
    (_ : P.IsFeasible ∨ P.dualize.IsFeasible) :
    OppositesOpt P.optimum P.dualize.optimum
```

The strong duality theorem was originally discussed in a different form in the context of zero-sum games by George Dantzig and John von Neumann [95]; later by Gale, Kuhn, Tucker [96]. Our proof of the strong duality theorem is obtained from the extended version, which will be discussed later.

### 3.1.6 Extended linearly ordered fields

Until now, we have talked about known results. What follows is a new extension of the theory.

Let `F` be a linearly ordered field. We define an *extended* linearly ordered field `F∞` as `F ∪ {⊥, ⊤}` with the following properties. Let `p` and `q` be numbers from `F`. We have `⊥ < p < ⊤`. We define addition, negation, and scalar action on `F∞` as follows:

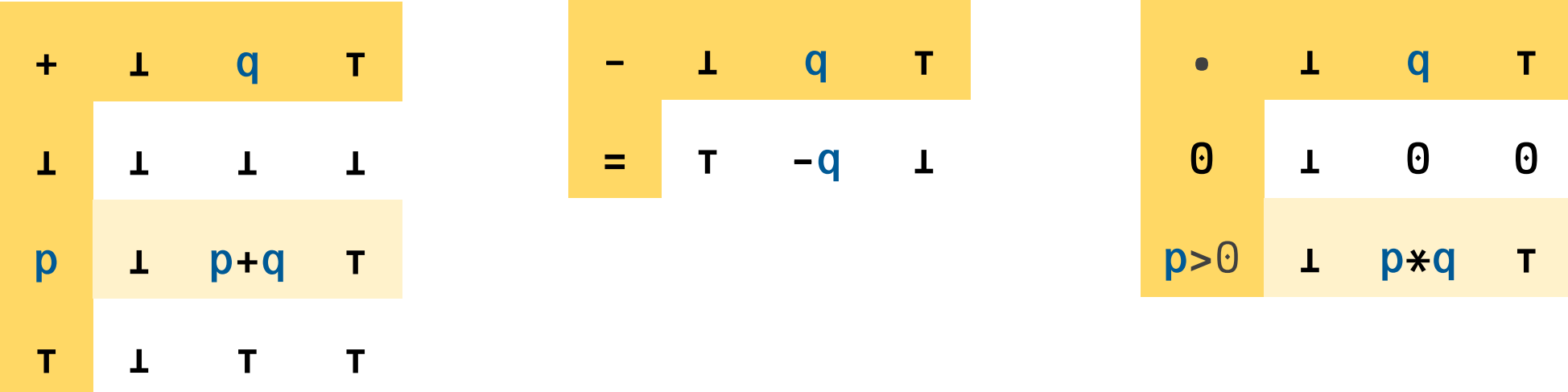

| + | ⊥ | q | ⊤ |
| --- | --- | --- | --- |
| ⊥ | ⊥ | ⊥ | ⊥ |
| p | ⊥ | p+q | ⊤ |
| ⊤ | ⊥ | ⊤ | ⊤ |

| - | ⊥ | q | ⊤ |
| --- | --- | --- | --- |
| = | ⊤ | -q | ⊥ |

| • | ⊥ | q | ⊤ |
| --- | --- | --- | --- |
| 0 | ⊥ | 0 | 0 |
| p>0 | ⊥ | p*q | ⊤ |

When we talk about elements of `F∞`, we say that values from `F` are *finite*.

Informally speaking, `⊤` represents the positive infinity, `⊥` represents the negative infinity, and we say `⊥` is "stronger" than `⊤` in all arithmetic operations. The surprising parts are `⊥ + ⊤ = ⊥` and `0 * ⊥ = ⊥`. Because of them, `F∞` is not a field. In fact, `F∞` is not even a group. However, `F∞` is still a densely linearly ordered abelian monoid with characteristic zero. Note that `⊥ + ⊤ = ⊥` is a standard convention in Mathlib [1] but `0 * ⊥ = ⊥` is ad hoc.

The implementation of the rules above is as follows:

```
def Extend (F : Type) := WithBot (WithTop F)

variable {F : Type} [LinearOrderedField F]

instance : LinearOrderedAddCommMonoid (Extend F) :=
  inferInstanceAs (LinearOrderedAddCommMonoid (WithBot (WithTop F)))

instance : AddCommMonoidWithOne (Extend F) :=
  inferInstanceAs (AddCommMonoidWithOne (WithBot (WithTop F)))

@[coe] def toE : F → (Extend F) := some ∘ some
instance : Coe F (Extend F) := ⟨toE⟩
```

```
def neg : Extend F → Extend F
| ⊥ => ⊤
| ⊤ => ⊥
| (x : F) => toE (-x)
```

```
syntax:max ident noWs "∞" : term
```

```
def EF.smulNN (c : F≥0) : F∞ → F∞
| ⊥ => ⊥
| ⊤ => if c = 0 then 0 else ⊤
| (f : F) => toE (c.val * f)
```

For working with vectors and matrices over extended linearly ordered fields, ordinary notions of multiplication, as they are defined in Mathlib, are not sufficient. We need to generalize the operations `dotProduct` and `Matrix.mulVec` as follows:

```
variable {α γ I : Type} [AddCommMonoid α] [SMul γ α] [Fintype I]
```

```
def dotWeig (v : I → α) (w : I → γ) : α :=
  ∑ i : I, w i • v i
```

```
def Matrix.mulWeig {J : Type} [Fintype J]
    (M : Matrix I J α) (w : J → γ) (i : I) : α :=
  M i ᵥ⬝ w
```

```
infixl:72 " ᵥ⬝ " => dotWeig
infixr:73 " ₘ* " => Matrix.mulWeig
```

We start by declaring that `α` and `γ` are types such that `α` forms an abelian monoid and `γ` has a scalar action on `α`. In this setting, we can instantiate `α` with `F∞` and `γ` with `F≥0` for any linearly ordered field `F`.

For explicit vectors `v : I → α` and `w : I → γ`, we define their product of type `α` as follows. Every element of `v` gets multiplied from left by an element of `w` on the same index. Then we sum them all together (in unspecified order). For a matrix `M` and a vector `w`, we define their product as a function that takes an index `i` and outputs the dot product (in the generalized sense we have just defined) between the `i`-th row of `M` and the vector `w`.

Beware that the arguments (both in the function definition and in the infix notation) come in the opposite order from how scalar action is written. We recommend a mnemonic "vector times weights" for `v ᵥ⬝ w` and "matrix times weights" for `M ₘ* w` where arguments come in alphabetical order.

In the infix notation, you can distinguish between the standard Mathlib [1] definitions and our definitions by observing that Mathlib operators put the letter `ᵥ` to the right of the symbol whereas our operators put a letter to the left of the symbol. In the case `α = γ` our definitions coincide with the Mathlib definitions.

Since we have new definitions, we have to rebuild all API (a lot of lemmas) for `dotWeig` and `Matrix.mulWeig` from scratch. This process was very tiresome. We decided not to develop a full reusable library, but prove only those lemmas we wanted to use in our project. For similar reasons, we did not generalize the Mathlib definition of "vector times matrix", as "matrix times vector" was all we needed. It was still a lot of lemmas.

### 3.1.7 Extended results

With extended linearly ordered fields, `inequalityFarkas_neg` is generalized as follows:

```
theorem extendedFarkas {I J F : Type} [LinearOrderedField F]
    (A : Matrix I J F∞) (b : I → F∞)
    (_ : ¬∃ i : I, (∃ j : J, A i j = ⊥) ∧ (∃ j : J, A i j = ⊤))
    (_ : ¬∃ j : J, (∃ i : I, A i j = ⊥) ∧ (∃ i : I, A i j = ⊤))
    (_ : ¬∃ i : I, (∃ j : J, A i j = ⊤) ∧ b i = ⊤)
    (_ : ¬∃ i : I, (∃ j : J, A i j = ⊥) ∧ b i = ⊥) :
    (∃ x : J → F≥0, A ₘ* x ≤ b) ≠ (∃ y : I → F≥0, -Aᵀ ₘ* y ≤ 0 ∧ b ᵥ⬝ y < 0)
```

The assumptions can be informally recapitulated as follows:

- `A` does not have `⊥` and `⊤` in the same row
- `A` does not have `⊥` and `⊤` in the same column
- `A` does not have `⊤` in any row where `b` has `⊤`
- `A` does not have `⊥` in any row where `b` has `⊥`

The idea of the proof is to do the following steps in the given order:

1. Delete all rows of both `A` and `b` where `A` has `⊥` or `b` has `⊤` (they are tautologies).
2. Delete all columns of `A` that contain `⊤` (they force respective variables to be zero).
3. If `b` contains `⊥`, then `A ₘ* x ≤ b` cannot be satisfied, but `y = 0` satisfies `(-Aᵀ) * y ≤ 0` and `b * y < 0`. Stop here.
4. Assume there is no `⊥` in `b`. Call `inequalityFarkas_neg`. In either case, extend `x` or `y` with zeros on all deleted positions.

*Extended linear programs* are defined similarly to standard linear programs. However, we need to review them separately, because they are an independent part of the trusted code. While results about LP are reused, all definitions were "written twice". So here we go…

An *extended linear program* is defined by a matrix `A` and vectors `b` and `c` (all of them over an extended linearly ordered field):

```
structure ExtendedLP (I J F : Type) [LinearOrderedField F] where
  A : Matrix I J F∞
  b : I → F∞
  c : J → F∞
```

A *valid* extended linear program is defined as follows:

```
structure ValidELP (I J F : Type) [LinearOrderedField F] extends
    ExtendedLP I J F where
  hAi : ¬∃ i : I, (∃ j : J, A i j = ⊥) ∧ (∃ j : J, A i j = ⊤)
  hAj : ¬∃ j : J, (∃ i : I, A i j = ⊥) ∧ (∃ i : I, A i j = ⊤)
  hbA : ¬∃ i : I, (∃ j : J, A i j = ⊥) ∧ b i = ⊥
  hcA : ¬∃ j : J, (∃ i : I, A i j = ⊤) ∧ c j = ⊥
  hAb : ¬∃ i : I, (∃ j : J, A i j = ⊤) ∧ b i = ⊤
  hAc : ¬∃ j : J, (∃ i : I, A i j = ⊥) ∧ c j = ⊤
```

Informally speaking, the validity conditions are as follows:

- `A` does not have `⊥` and `⊤` in the same row
- `A` does not have `⊥` and `⊤` in the same column
- `A` does not have `⊥` in any row where `b` has `⊥`
- `A` does not have `⊤` in any column where `c` has `⊥`
- `A` does not have `⊤` in any row where `b` has `⊤`
- `A` does not have `⊥` in any column where `c` has `⊤`

For the rest of this subsection, we will assume:

```
variable {I J F : Type} [LinearOrderedField F]
```

A vector `x` made of finite nonnegative values is a *solution* iff its multiplication by the matrix `A` from the left yields a vector whose all entries are less or equal to the corresponding entries of the vector `b`:

```
def ExtendedLP.IsSolution [Fintype J]
    (P : ExtendedLP I J F) (x : J → F≥0) :
    Prop :=
  P.A ₘ* x ≤ P.b
```

An extended linear program *reaches* a value `r` iff it has a solution `x` such that, when its entries are elementwise multiplied by the the coefficients `c` and summed up, the result is the value `r`:

```
def ExtendedLP.Reaches [Fintype J]
    (P : ExtendedLP I J F) (r : F∞) :
    Prop :=
  ∃ x : J → F≥0, P.IsSolution x ∧ P.c ᵥ⬝ x = r
```

An extended linear program is *feasible* iff it reaches a value different from `⊤`:

```
def ExtendedLP.IsFeasible [Fintype J]
    (P : ExtendedLP I J F) :
    Prop :=
  ∃ p : F∞, P.Reaches p ∧ p ≠ ⊤
```

An extended linear program is *bounded by* a value `r` (from below — we always minimize) iff it reaches only values greater or equal to `r`:

```
def ExtendedLP.IsBoundedBy [Fintype J]
    (P : ExtendedLP I J F) (r : F) :
    Prop :=
  ∀ p : F∞, P.Reaches p → r ≤ p
```

An extended linear program is *unbounded* iff it has no finite lower bound:

```
def ExtendedLP.IsUnbounded [Fintype J]
    (P : ExtendedLP I J F) :
    Prop :=
  ¬∃ r : F, P.IsBoundedBy r
```

To *dualize* an extended linear program, we transpose the matrix and flip all its signs, and we swap the right-hand-side vector with the vector of objective function coefficients:

```
abbrev ExtendedLP.dualize (P : ExtendedLP I J F) : ExtendedLP J I F :=
  ⟨-P.Aᵀ, P.c, P.b⟩
```

This dualization is inherited to define how valid extended linear programs are dualized:

```
def ValidELP.dualize (P : ValidELP I J F) : ValidELP J I F where
  toExtendedLP := P.toExtendedLP.dualize
  hAi := by aeply P.hAj
  hAj := by aeply P.hAi
  hbA := by aeply P.hcA
  hcA := by aeply P.hbA
  hAb := by aeply P.hAc
  hAc := by aeply P.hAb
```

The last six lines generate proofs that our six conditions stay satisfied after dualization, where `aeply` is a custom tactic based on `apply` and `aesop` [97].

All of the most important results will furthermore require that we have a finite amount of variables and a finite amount of conditions. Weak duality for valid extended LPs is stated as follows:

```
theorem ValidELP.weakDuality [Fintype I] [Fintype J] (P : ValidELP I J F)
    {p : F∞} (_ : P.Reaches p) {q : F∞} (_ : P.dualize.Reaches q) :
    0 ≤ p + q
```

Without validity, there is no weak duality, as we will soon demonstrate.

Again, before we proceed to strong duality, we need to define what the optimum is (the definition does not require validity, but the results do):

```
noncomputable def ExtendedLP.optimum [Fintype J]
    (P : ExtendedLP I J F) :
    Option F∞ :=
  if ¬P.IsFeasible then
    some ⊤
  else
    if P.IsUnbounded then
      some ⊥
    else
      if hr : ∃ r : F, P.Reaches (toE r) ∧ P.IsBoundedBy r then
        some (toE hr.choose)
      else
        none
```

The type `Option F∞`, which is implemented as `Option (Option (Option F))` after unfolding definitions, allows the following values:

- `none`
- `some ⊥` implemented as `some none`
- `some ⊤` implemented as `some (some none)`
- `some (toE r)` implemented as `some (some (some r))` for any `r : F`

We assign the following semantics to `Option F∞` values:

- `none` … invalid finite value (infimum is not attained)
- `some ⊥` … feasible unbounded
- `some ⊤` … infeasible
- `some (toE r)` … the minimum is a finite value `r`

The definition `ExtendedLP.optimum` first asks if `P` is feasible; if not, it returns `some ⊤` (i.e., the worst value). When `P` is feasible, it asks whether `P` is unbounded; if yes, it returns `some ⊥` (i.e., the best value). When `P` is feasible and bounded, it asks if there is a finite value `r` such that `P` reaches `r` and, at the same time, `P` is bounded by `r`; if so, it returns `some (toE r)`. Otherwise, it returns `none`. Note that we use the verbs “ask” and “return” metaphorically; `ExtendedLP.optimum` is not a computable function; it is just a mathematical definition (see the keyword `noncomputable` above) we can prove things about.

Again, we prove that the optimum always exists (that is, there cannot be a valid extended LP with a finite infimum that isn’t attained):

```
theorem ValidELP.optimum_neq_none [Fintype I] [Fintype J]
    (P : ValidELP I J F) :
  P.optimum ≠ none
```

The existence of optimum could be proved more generally, but we didn’t do it.

Finally, we state the holy grail, extended strong duality:

```
theorem ValidELP.strongDuality [Fintype I] [Fintype J]
    (P : ValidELP I J F) (_ : P.IsFeasible ∨ P.dualize.IsFeasible) :
    OppositesOpt P.optimum P.dualize.optimum
```

### 3.1.8 Example — cheap lunch

According to Nutritionix[17] a kilogram of boiled white rice contains 27 g protein and 1300 kcal. According to Nutritionix[18] a kilogram of boiled lentils contains 90 g protein and 1150 kcal. Let's say that a kilogram of boiled white rice costs 0.92 euro and that a kilogram of boiled lentils costs 1.75 euro. We want to cook a lunch, as cheap as possible, that contains at least 30 g protein and at least 700 kcal. The choice of white rice and lentils isn't random — I ate this lunch at a mathematical camp — at the moment, I didn't like the lunch — it wasn't very tasty — but later I realized what an awesome nutritional value the lunch had for its low price. This is a simple example of the well-known diet problem [98].

```
minimize   0.92 * r + 1.75 * l
(-27) * r + (-90) * l ≤ -30
(-1300) * r + (-1150) * l ≤ -700
r ≥ 0
l ≥ 0
```

[17] https://www.nutritionix.com/food/white-rice/1000-g

[18] https://www.nutritionix.com/food/lentils

Using a numerical LP solver, we obtain the optimal solution `r` = 0.331588 and `l` = 0.233857 which satisfies our dietary requirements for 0.714311 euro.

The dual of this problem, in the sense of `ExtendedLP.strongDuality`, is the following LP:

```
minimize   (-30) * p + (-700) * k
27 * p + 1300 * k ≤ 0.92
90 * p + 1150 * k ≤ 1.75
p ≥ 0
k ≥ 0
```

Using a numerical LP solver, we obtain the optimal solution `p`=0.0141594 and `k`=0.000413613 leading to the objective value -0.714311 here. To summarize, the cheapest lunch that contains at least 30 g protein and at least 700 kcal consists of 332 g white rice and 234 g lentils; the shadow cost of a gram of protein is 0.014 euro and the shadow cost of a kcal is 0.00041 euro in our setting.

Now consider a setting when lentils are out of stock. We still want to obtain at least 30 g protein and at least 700 kcal. What is the price of our lunch now? On paper, it is most natural to entirely remove lentils out of the picture and recompute the LP. However, computer proof systems generally don't like it when sizes of matrices change, so let's do the modification of input in place. One could suggest that, in order to model the newly arisen situation, the price of lentils be increased to 999999 euro, so that the optimal solution will assign 0 to it. This is, however, not mathematically elegant, as it requires engineering insight into choosing an appropriately large constant. We claim that the proper mathematical approach is to increase the price of lentils to infinity! Let's see how our original LP will look now:

```
minimize   0.92 * r + ⊤ * l
(-27) * r + (-90) * l ≤ -30
(-1300) * r + (-1150) * l ≤ -700
r ≥ 0
l ≥ 0
```

The optimal solution is now `r` = 1.111111 and `l` = 0 and costs 1.022222 euro. The dual of this problem is the following LP:

```
minimize   (-30) * p + (-700) * k
27 * p + 1300 * k ≤ 0.92
90 * p + 1150 * k ≤ ⊤
p ≥ 0
k ≥ 0
```

At this point, the shadow price of a gram of protein is 0.034074 euro but the shadow price of kcal is 0 euro. The objective value is -1.022222 in agreement with our theoretical prediction.

### 3.1.9 Counterexamples in the extended settings

Recall that `extendedFarkas` has four preconditions on the matrix `A` and the vector `b`. The following examples show that omitting any of these preconditions makes the theorem false.

If `A` has ⊥ and ⊤ in the same row, it may happen that both `x` and `y` exist:

$$A = \begin{pmatrix} \bot & \top \\ 0 & -1 \end{pmatrix} \qquad b = \begin{pmatrix} 0 \\ -1 \end{pmatrix} \qquad x = \begin{pmatrix} 1 \\ 1 \end{pmatrix} \qquad y = \begin{pmatrix} 0 \\ 1 \end{pmatrix}$$

If `A` has ⊥ and ⊤ in the same column, it may happen that both `x` and `y` exist:

$$A = \begin{pmatrix} \bot \\ \top \end{pmatrix} \qquad b = \begin{pmatrix} -1 \\ 0 \end{pmatrix} \qquad x = \begin{pmatrix} 0 \end{pmatrix} \qquad y = \begin{pmatrix} 1 \\ 1 \end{pmatrix}$$

If `A` has ⊤ in a row where `b` has ⊤, it may happen that both `x` and `y` exist:

$$A = \begin{pmatrix} \top \\ -1 \end{pmatrix} \qquad b = \begin{pmatrix} \top \\ -1 \end{pmatrix} \qquad x = \begin{pmatrix} 1 \end{pmatrix} \qquad y = \begin{pmatrix} 0 \\ 1 \end{pmatrix}$$

If `A` has ⊥ in a row where `b` has ⊥, it may happen that both `x` and `y` exist:

$$A = \begin{pmatrix} \bot \end{pmatrix} \qquad b = \begin{pmatrix} \bot \end{pmatrix} \qquad x = \begin{pmatrix} 1 \end{pmatrix} \qquad y = \begin{pmatrix} 0 \end{pmatrix}$$

Recall that `ValidELP.strongDuality` is formulated for valid LPs, and the definition of a valid LP has six conditions. We show that omitting any of these six conditions makes the theorem false.

Omitting `hAj` or `hAi` allows the following LPs, respectively:

$$P = \left\langle \begin{pmatrix} \bot \\ \top \end{pmatrix}, \begin{pmatrix} -1 \\ 0 \end{pmatrix}, \begin{pmatrix} 0 \end{pmatrix} \right\rangle \qquad Q = \left\langle \begin{pmatrix} \top & \bot \end{pmatrix}, \begin{pmatrix} 0 \end{pmatrix}, \begin{pmatrix} -1 \\ 0 \end{pmatrix} \right\rangle$$

Omitting `hbA` or `hcA` allows the following LPs, respectively:

$$P = \left\langle \begin{pmatrix} \bot \end{pmatrix}, \begin{pmatrix} \bot \end{pmatrix}, \begin{pmatrix} 0 \end{pmatrix} \right\rangle \qquad Q = \left\langle \begin{pmatrix} \top \end{pmatrix}, \begin{pmatrix} 0 \end{pmatrix}, \begin{pmatrix} \bot \end{pmatrix} \right\rangle$$

Omitting `hAb` or `hAc` allows the following LPs, respectively:

$$P = \left\langle \begin{pmatrix} \top \\ -1 \end{pmatrix}, \begin{pmatrix} \top \\ -1 \end{pmatrix}, \begin{pmatrix} 0 \end{pmatrix} \right\rangle \qquad Q = \left\langle \begin{pmatrix} \bot & 1 \end{pmatrix}, \begin{pmatrix} 0 \end{pmatrix}, \begin{pmatrix} \top \\ -1 \end{pmatrix} \right\rangle$$

It can be checked that `Q` is the dual of `P` in all three pairs. The strong duality fails in all three cases, since the optimum of `P` is `0` and the optimum of `Q` is ⊥ in all three pairs.

Unfortunately, we didn't formally verify any the counterexamples.

### 3.1.10 Proving extended weak duality

Our context continues having:

```
variable {I J F : Type} [LinearOrderedField F]
```

We start with a lemma:

```
lemma ValidELP.weakDuality_of_no_bot [Fintype I] [Fintype J]
    (P : ValidELP I J F)
    (_ : ¬∃ i : I, P.b i = ⊥)
    (_ : ¬∃ j : J, P.c j = ⊥)
    {p : F∞} (_ : P.Reaches p) {q : F∞} (_ : P.dualize.Reaches q) :
    0 ≤ p + q
```

We will also need this easy lemma:

```
lemma ValidELP.no_bot_of_reaches [Fintype J]
    (P : ValidELP I J F) {p : F∞} (_ : P.Reaches p) (i : I) :
    P.b i ≠ ⊥
```

Combining the last two lemmas together gives us the weak duality theorem:

```
theorem ValidELP.weakDuality [Fintype I] [Fintype J] (P : ValidELP I J F)
    {p : F∞} (_ : P.Reaches p) {q : F∞} (_ : P.dualize.Reaches q) :
    0 ≤ p + q
```

### 3.1.11 Proving extended strong duality

Our context still contains:

```
variable {I J F : Type} [LinearOrderedField F]
```

We start with proving several properties of valid extended LP.

First, we establish that dualization is a dual operation:

```
lemma ValidELP.dualize_dualize (P : ValidELP I J F) :
    P = P.dualize.dualize
```

Next, we establish that a feasible LP has no ⊥ in its right-hand-side vector:

```
lemma ValidELP.no_bot_of_feasible [Fintype J]
    (P : ValidELP I J F) (hP : P.IsFeasible) (i : I) :
    P.b i ≠ ⊥ :=
  P.no_bot_of_reaches hP.choose_spec.left i
```

Now, we provide a direct description of an unbounded LP:

```
lemma ValidELP.isUnbounded_iff [Fintype J] (P : ValidELP I J F) :
    P.IsUnbounded ↔ ∀ r : F, ∃ p : F∞, P.Reaches p ∧ p < r
```

Another description of an unbounded LP is as follows:

```
lemma ValidELP.unbounded_of_reaches_le [Fintype J] (P : ValidELP I J F)
    (_ : ∀ r : F, ∃ p : F∞, P.Reaches p ∧ p ≤ r) :
    P.IsUnbounded
```

Now, we provide a sufficient condition for an LP to be unbounded:

```
lemma ValidELP.unbounded_of_feasible_of_neg [Fintype J]
    (P : ValidELP I J F) (_ : P.IsFeasible) {x' : J → F≥0}
    (_ : P.c ᵥ⬝ x' < 0) (_ : P.A ₘ* x' + (0 : F≥0) • (-P.b) ≤ 0) :
    P.IsUnbounded
```

Next, we establish that a feasible LP whose dual is infeasible must be unbounded:

```
lemma ValidELP.unbounded_of_feasible_of_infeasible [Fintype I] [Fintype J]
    (P : ValidELP I J F) (_ : P.IsFeasible) (_ : ¬P.dualize.IsFeasible) :
    P.IsUnbounded
```

With the help of weak duality, we establish that an unbounded LP has an infeasible dual:

```
lemma ValidELP.infeasible_of_unbounded [Fintype I] [Fintype J]
    (P : ValidELP I J F) (_ : P.IsUnbounded) :
    ¬P.dualize.IsFeasible
```

Now, we confront the most difficult part of the proof of the strong duality:

```
lemma ValidELP.strongDuality_aux [Fintype I] [Fintype J]
    (P : ValidELP I J F) (_ : P.IsFeasible) (_ : P.dualize.IsFeasible) :
    ∃ p q : F, P.Reaches p ∧ P.dualize.Reaches q ∧ p + q ≤ 0
```

Next, we combine the almost-strong duality above with the weak duality:

```
lemma ValidELP.strongDuality_of_both_feasible [Fintype I] [Fintype J]
    (P : ValidELP I J F) (_ : P.IsFeasible) (_ : P.dualize.IsFeasible) :
    ∃ r : F, P.Reaches (toE (-r)) ∧ P.dualize.Reaches (toE r)
```

Before celebrating our success, we recall that the desired strong duality theorem talks about optima of both LPs; it isn't a mere existential statement about a value reached. Before we can prove the true strong duality, we need a lemma about uniqueness of the to-be optimum:

```
lemma ExtendedLP.optimum_unique [Fintype J] {P : ExtendedLP I J F}
    {r s : F}
    (_ : P.Reaches (toE r) ∧ P.IsBoundedBy r)
    (_ : P.Reaches (toE s) ∧ P.IsBoundedBy s) :
    r = s
```

Using the axiom of choice together with the uniqueness we have just proved, we obtain a very natural lemma, thanks to which we can finally start assessing that the optimum of an LP is equal to some value:

```
lemma ExtendedLP.optimum_eq_of_reaches_bounded [Fintype J]
    {P : ExtendedLP I J F} {r : F}
    (_ : P.Reaches (toE r)) (_ : P.IsBoundedBy r) :
    P.optimum = some r
```

It is convenient that `OppositesOpt` commutes:

```
lemma oppositesOpt_comm (p q : Option F∞) :
    OppositesOpt p q ↔ OppositesOpt q p
```

Combining all puzzle pieces together, we achieve the strong duality in the case where our LP is feasible:

```
lemma ValidELP.strongDuality_of_prim_feasible [Fintype I] [Fintype J]
    (P : ValidELP I J F) (_ : P.IsFeasible) :
    OppositesOpt P.optimum P.dualize.optimum
```

As a consequence, we have a guarantee that every valid extended LP with finite amount of variables and finite amount of conditions has an optimum:

```
theorem ValidELP.optimum_neq_none [Fintype I] [Fintype J]
    (P : ValidELP I J F) :
    P.optimum ≠ none
```

Combining `ValidELP.strongDuality_of_prim_feasible` with `oppositesOpt_comm` and `ValidELP.dualize_dualize`, we get the strong duality in the case where the dual of our LP is feasible:

```
lemma ValidELP.strongDuality_of_dual_feasible [Fintype I] [Fintype J]
    (P : ValidELP I J F) (_ : P.dualize.IsFeasible) :
    OppositesOpt P.optimum P.dualize.optimum
```

Finally, we complete the actual strong duality theorem in our extended setting:

```
theorem ValidELP.strongDuality [Fintype I] [Fintype J]
    (P : ValidELP I J F) (hP : P.IsFeasible ∨ P.dualize.IsFeasible) :
    OppositesOpt P.optimum P.dualize.optimum
```

### 3.1.12 Dependencies between theorems

Theorems are in black. Selected lemmas are in gray. What we consider to be the main theorems are denoted by blue background. The main corollaries are denoted by yellow background.

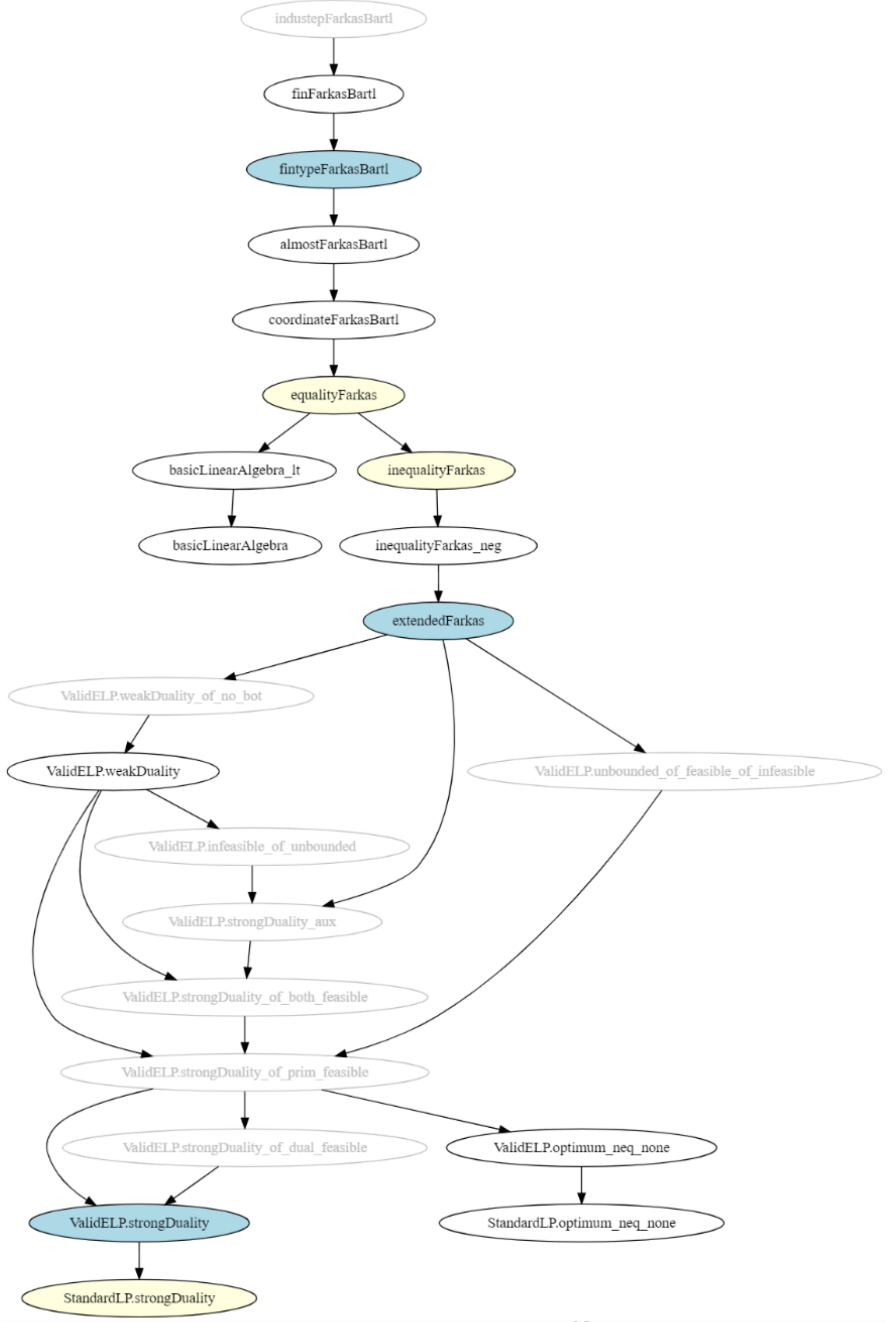

### 3.1.13 Related work

There is a substantial body of work on linear inequalities and linear programming formalized in Isabelle. While our work is focused only on proving mathematical theorems, the work in Isabelle is motivated by the development of SMT solvers.

- Bottesch, Haslbeck, Thiemann [99] proved a variant of `equalityFarkas` for δ-rationals by analyzing a specific implementation of the Simplex algorithm by Marić, Spasić, Thiemann [100].
- Bottesch, Raynaud, Thiemann [101] proved the Fundamental theorem of linear inequalities as well as both `equalityFarkas` and `inequalityFarkas` for all linearly ordered fields, alongside the Carathéodory's theorem and the Farkas-Minkowski-Weyl theorem. They also investigated systems of linear mixed-integer inequalities.
- Thiemann himself [102] then proved the strong duality for linear programming in the version where one LP has inequalities and unconstrained variables and the other LP has equalities and nonnegative variables.

Sakaguchi[19] proved a version of `equalityFarkas` for linearly ordered fields in Rocq, using the Fourier-Motzkin elimination. Allamigeon and Katz [103] made a large contribution to the study of convex polyhedra in Rocq—among other results, they proved a version of `equalityFarkas` for linearly ordered fields as well as the strong duality for LPs in the version where one LP has inequalities and unconstrained variables and the other has equalities and nonnegative variables.

### 3.1.14 Conclusion

We formally verified several Farkas-like theorems in Lean 4. We extended the existing theory to a new setting where some coefficient can carry infinite values.

We realized that the abstract work with modules over linearly ordered division rings and linear maps between them was fairly easy to carry on in Lean 4 thanks to the library Mathlib [1] that is perfectly suited for such tasks. In contrast, manipulation with matrices got tiresome whenever we needed a not-fully-standard operation. It turns out Lean 4 cannot automate case analyses unless they take place in the "outer layers" of formulas. Summation over subtypes and summation of conditional expression made us developed a lot of ad-hoc machinery which we would have preferred to be handled by existing tactics. Another area where Lean 4 is not yet helpful is the search for counterexamples. Despite these difficulties, we find Lean 4 to be an excellent tool for elegant expressions and organization of mathematical theorems and for proving them formally.

## *3.2 Valued Constraint Satisfaction Problems*

General-Valued Constraint Satisfaction Problems is a very broad class of problems in discrete optimization. General-Valued Constraint Satisfaction Problems subsumes Min-Cost-Hom (including 3-SAT for example) and Finite-Valued Constraint Satisfaction Problems.

[19] https://github.com/pi8027/vass

Constraint Satisfaction Problems have practical applications in artificial intelligence. They are useful because they are applicable in settings where we have partial information stemming from a local view of a problem [104]. Their declarative nature allows to separate problem modelling from algorithms that solve them [104]. Constraint Satisfaction Problems are used in planning, scheduling, and AI for games, among other things. Their usefulness extends to the optimization flavour of Constraint Satisfaction Problems, too, whether Weighted Constraint Satisfaction Problems or Finite-Valued Constraint Satisfaction Problems. Many traditional AI problems have a natural formulation in the VCSP framework. For example, in computer vision, tasks like image segmentation can be modeled as minimizing a cost function that balances data fidelity with smoothness constraints between neighboring pixels.

We focus on so-called fixed-template VCSP, that is, we are given a set of cost functions, and all terms in all VCSP instances can only use these cost functions. In the world of Crisp Constraint Satisfaction Problems (i.e., where answers can only be “yes” and “no”), it corresponds to the setting where we search for a homomorphism from an arbitrary relational structure to a fixed relational structure; for example, if the latter is a triangle, this fixed-template Crisp Constraint Satisfaction Problem corresponds to the decision problem of 3-coloring a graph on the input. A central question in the study of fixed-template VCSP is what classes of discrete functions admit an efficient minimization algorithm. Since there is no well-developed Lean library for asserting time complexity of algorithms, we couldn’t study complexity of VCSP per se. However, we studied the basic LP relaxation, which is motivated by the fact that linear programs can be efficiently solved; unfortunately, this motivation stay only on the informal level, as we didn’t implement any LP solver in Lean.

The theoretical literature usually culminates in so-called dichotomy theorems, which provide a sufficient and necessary condition for a class of problems to be solvable in polynomial time, with all the remaining problems being NP-hard. In particular:

- Schaefer [105] established the dichotomy for Boolean Constraint Satisfaction Problems.
- Hell and Nešetřil [106] established the dichotomy for the Graph Homomorphism Problems.
- Thapper and Živný [107] established the dichotomy for Finite-Valued Constraint Satisfaction Problems.
- Zhuk [108] and Bulatov [109] established the dichotomy for Crisp Constraint Satisfaction Problems.
- Kolmogorov, Rolínek, and Krokhin [110] established the dichotomy for General-Valued Constraint Satisfaction Problems.

Our project formalizes the following notions in Lean 4:

- VCSP template
- VCSP term
- VCSP instance
- Optimum solution of a VCSP instance
- Expressive power of a VCSP template
- Max-Cut property
- Symmetric fractional polymorphism
- Canonical LP
- Basic LP relaxation

Our project proves the following results in Lean 4:

- If a VCSP template over a linearly ordered cancellative abelian monoid can express Max-Cut, it cannot have any commutative fractional polymorphism.
- The basic LP relaxation for a VCSP template over any ordered ring of characteristic zero is valid.
- If a VCSP template over ℚ has symmetric fractional polymorphisms of all arities, then its basic LP relaxation is tight.

Our project thereby provides a small step towards a formally verified proof of the dichotomy for Finite-Valued Constraint Satisfaction Problems.

### 3.2.1 Templates and instances

A VCSP *template* is a set of cost functions that are allowed to be used in VCSP instances:

```
abbrev ValuedCSP (D C : Type) [OrderedAddCommMonoid C] :=
  Set (Σ (n : ℕ), (Fin n → D) → C)
```

Elements of `D` are usually called *labels*. The *domain* `D` is usually finite, but it can also be infinite. As a stupid artificial example, if our only objective function is the absolute value (e.g. we want to minimize $|x_0| + |x_1|$ where $x_0$ and $x_1$ are rational numbers), we first wrap the absolute value on rational numbers `(|·|) : ℚ → ℚ` into a cost function `absRat : (Fin 1 → ℚ) → ℚ` and then we create a singleton VCSP template out of it (by first wrapping it into a `Sigma` type and then into a set):

```
private def exampleAbs : Σ (n : ℕ), (Fin n → ℚ) → ℚ := ⟨1, absRat⟩
```

```
private def exampleFiniteValuedCSP : ValuedCSP ℚ ℚ := { exampleAbs }
```

From now on, we will assume:

```
variable {D C : Type} [OrderedAddCommMonoid C]
```

A VCSP *term* over a VCSP template `Γ` is:

```
structure ValuedCSP.Term (Γ : ValuedCSP D C) (ι : Type) where
  n : ℕ
  f : (Fin n → D) → C
  inΓ : ⟨n, f⟩ ∈ Γ
  app : Fin n → ι
```

This way we bypass the need for an indexing type for cost functions.

A single term is *evaluated*, where `x` is the entire *solution* being evaluated, as follows:

```
def ValuedCSP.Term.evalSolution {Γ : ValuedCSP D C} {ι : Type}
    (t : Γ.Term ι) (x : ι → D) : C :=
  t.f (x ∘ t.app)
```

A VCSP *instance* is then defined as a multiset of VCSP terms:

```
abbrev ValuedCSP.Instance (Γ : ValuedCSP D C) (ι : Type) :=
  Multiset (Γ.Term ι)
```

Continuing with our example, in which we want to minimize $|x_0| + |x_1|$ where $x_0$ and $x_1$ are rational numbers, we express this trivial problem as the following VCSP instance:

```
private lemma abs_in : ⟨1, absRat⟩ ∈ exampleFiniteValuedCSP := rfl
```

```
private def exampleFiniteValuedInstance :
    exampleFiniteValuedCSP.Instance (Fin 2) :=
  {⟨1, absRat, abs_in, ![0]⟩, ⟨1, absRat, abs_in, ![1]⟩}
```

The *value* of the VCSP instance for given solution (a.k.a. labelling) is defined as the sum of evaluations of all terms on the same labelling:

```
def ValuedCSP.Instance.evalSolution {Γ : ValuedCSP D C} {ι : Type}
    (I : Γ.Instance ι) (x : ι → D) : C :=
  (I.map (·.evalSolution x)).sum
```

For example, the following equality holds:

```
exampleFiniteValuedInstance.evalSolution ![0.9, -0.5] = 1.4
```

An *optimum* solution is defined as a solution that is below all solutions, i.e., a “minimum” of a given VCSP instance:

```
def ValuedCSP.Instance.IsOptimumSolution {Γ : ValuedCSP D C} {ι : Type}
    (I : Γ.Instance ι) (x : ι → D) : Prop :=
  ∀ y : ι → D, I.evalSolution x ≤ I.evalSolution y
```

For illustration, the following example can be proved:

```
example : exampleFiniteValuedInstance.IsOptimumSolution ![(0 : ℚ), (0 : ℚ)]
```

The concept of optimum solutions applies both to optimization problems (Finite-Valued Constraint Satisfaction Problems) and decision problems (Crisp Constraint Satisfaction Problems), as well as their combinations (General-Valued Constraint Satisfaction Problems). In case of decision (crisp) problems, `C` is `Bool` and we swap the meaning of `True` with `False` —in our interpretation, `True` means that a condition was broken whereas `False` means satisfied instance! It is defined this way because Lean defines `False < True` unfortunately, yet we want to minimize the objective function in all other types of problems, so we decided to stick with minimization. We acknowledge that it is confusing and we apologize for that.

The definition `exampleCrispCSP` in Mathlib [1] shows how the 3-coloring (decision) problem is modelled inside this framework. Afterwards, the 3-coloring of the graph $K_4^-$ is shown within this framework.

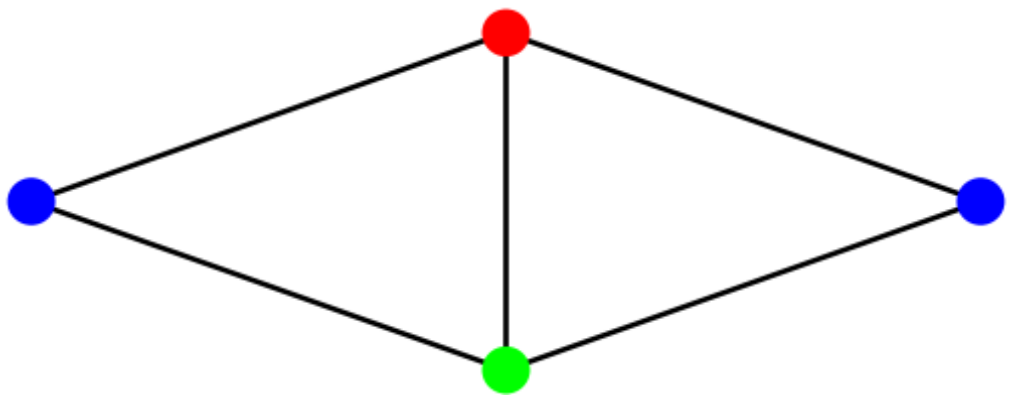

### 3.2.2 Properties

Our context continues having:

```
variable {D C : Type} [OrderedAddCommMonoid C]
```

We often characterize VCSP templates in terms of a so-called *Max-Cut* property. First, an auxiliary definition, we say that a function `f` has Max-Cut property at labels `a` and `b` when the argmin of `f` is { `![a, b]`, `![b, a]` } exactly:

```
def Function.HasMaxCutPropertyAt (f : (Fin 2 → D) → C) (a b : D) : Prop :=
  f ![a, b] = f ![b, a] ∧
  ∀ x y : D, f ![a, b] ≤ f ![x, y] ∧
    (f ![a, b] = f ![x, y] → a = x ∧ b = y ∨ a = y ∧ b = x)
```

Afterwards, we say that a function `f` has Max-Cut property iff it has the Max-Cut property on some two non-identical labels:

```
def Function.HasMaxCutProperty (f : (Fin 2 → D) → C) : Prop :=
  ∃ a b : D, a ≠ b ∧ f.HasMaxCutPropertyAt a b
```

An "orthogonal" concept to cost functions are fractional operations. A *fractional operation* is a finite unordered collection of `m`-ary operations on `D` (possibly with duplicates):

```
abbrev FractionalOperation (D : Type) (m : ℕ) :=
  Multiset ((Fin m → D) → D)
```

From now on, we add `{m : ℕ}` to the context. The *size* of a fractional operation is defined as the cardinality of the multiset:

```
def FractionalOperation.size (ω : FractionalOperation D m) : ℕ := ω.card
```

We say that a fractional operation is *valid* iff it is nonempty:

```
def FractionalOperation.IsValid (ω : FractionalOperation D m) : Prop :=
  ω ≠ ∅
```

We *apply* the fractional operation as follows:

```
def FractionalOperation.tt {ι : Type} (ω : FractionalOperation D m)
    (x : Fin m → ι → D) :
    Multiset (ι → D) :=
  ω.map (fun g : (Fin m → D) → D => fun i : ι => g (x.swap i))
```

For example, if we have `m = 3` and $\omega = \{g_0, g_1\}$ where we display $g_0$ by a green arrow and $g_1$ by a purple arrow, `ω.tt` is applied to input `x` giving output `y` as follows:

```
x 0  =  ( x 0 0   x 0 1   x 0 2   x 0 3  … )
x 1  =  ( x 1 0   x 1 1   x 1 2   x 1 3  … )
x 2  =  ( x 2 0   x 2 1   x 2 2   x 2 3  … )

y 0  =  ( y 0 0   y 0 1   y 0 2   y 0 3  … )
y 1  =  ( y 1 0   y 1 1   y 1 2   y 1 3  … )
```

Note that `x` could have been defined as a matrix, but we define it as a family of explicit vectors because it usually isn't useful to think about the collection of solutions `x 0` to `x (m - 1)` as of a matrix (and we don't apply any operation specific to matrices to it).

We say that a cost function `f` *admits* a fractional operation `ω` iff `ω` improves `f` in the ≤ sense:

```
def Function.AdmitsFractional {n : ℕ} (f : (Fin n → D) → C)
    (ω : FractionalOperation D m) :
    Prop :=
  ∀ x : (Fin m → (Fin n → D)),
    m • ((ω.tt x).map f).sum ≤ ω.size • Finset.univ.sum (f ∘ x)
```

The last line may look confusing. This hard-to-read definition is a price for its generality. If `C` happens to be an ordered ring, we can write the inequality as follows:

```
m * ((ω.tt x).map f).sum ≤ ω.size * Finset.univ.sum (f ∘ x)
```

Moreover, if `C` is a linearly ordered field (in this version of Mathlib, ordered field isn't defined but linearly ordered field is), `0 < m`, and `ω` is valid, then the same inequality can be equivalently stated as:

```
((ω.tt x).map f).sum / ω.size ≤ Finset.univ.sum (f ∘ x) / m
```

Now it is hopefully recognizable that the RHS is the arithmetic average of the `f` images of all vectors in `x` and that the LHS is the arithmetic average of the `f` images of all outputs of `ω` applied to `x` element-wise. In a picture:

```
x 0  =  ( x 0 0  x 0 1  x 0 2  x 0 3  … )  → f ┐
x 1  =  ( x 1 0  x 1 1  x 1 2  x 1 3  … )  → f ├ average
x 2  =  ( x 2 0  x 2 1  x 2 2  x 2 3  … )  → f ┘
                                                     ≥
y 0  =  ( y 0 0  y 0 1  y 0 2  y 0 3  … )  → f ┐
y 1  =  ( y 1 0  y 1 1  y 1 2  y 1 3  … )  → f ┘ average
```

The also exists a probabilistic interpretation; namely, if we apply a uniformly-chosen random operation from `ω` to `x` element-wise and then apply `f` to the output vector, we cannot do worse than choosing a vector from `x` uniformly at random and apply `f` to it. This formulation also foreshadows why it might be useful to have a fractional operation admitted by a lot of (ideally all) cost functions.

Note that in literature, fractional operation comes with a weight function—every operation has a nonnegative rational weight, hence the name "fractional operation". However, since any family of rational numbers can be expressed as ratios of natural numbers, we decided to replace the standard definition by multiset of operations—every operation is in the fractional operation as many times as its weight ought to be. In this way, the theory is more amendable to formalization, at the cost of being less believable. At the same time, it allows us to be more general, since `C` isn't required to have multiplication defined (only addition is necessary—and, for the purpose of determining whether the fractional operation is admitted by a cost function, we already know that the scalar action • can be interpreted as an iterated addition in any abelian monoid). Another advantage of our approach is that we don't have to distinguish between the fractional operation and its support—the multiset is all there is.

In the probabilistic interpretation above, instead of choosing an operation uniformly at random, we would be choosing the operation with probability proportional to its weight. Again, the resulting behaviour is the same as in our version where operations don't have weights but multiplicity instead.

Fractional operation `ω` is a *fractional polymorphism* for a VCSP template `Γ` iff every cost function in `Γ` admits `ω` (the `.snd` part only discards the arity; we work with each cost function itself):

```
def FractionalOperation.IsFractionalPolymorphismFor
    (ω : FractionalOperation D m) (Γ : ValuedCSP D C) :
    Prop :=
  ∀ f ∈ Γ, f.snd.AdmitsFractional ω
```

The fractional polymorphism generalizes the notion of multimorphism from [111].

We say that a fractional operation ω is *symmetric* iff all operations in ω depend only on the multiset of its inputs, disregarding in which order they come:

```
def FractionalOperation.IsSymmetric (ω : FractionalOperation D m) : Prop :=
  ∀ x y : (Fin m → D),
    List.ofFn x ~ List.ofFn y → ∀ g ∈ ω, g x = g y
```

We say that a fractional operation ω is a *symmetric fractional polymorphism* for a VCSP template Γ iff ω is a fractional polymorphism for Γ and ω is symmetric:

```
def FractionalOperation.IsSymmetricFractionalPolymorphismFor
    (ω : FractionalOperation D m) (Γ : ValuedCSP D C) : Prop :=
  ω.IsFractionalPolymorphismFor Γ ∧ ω.IsSymmetric
```

For example, if the VCSP template consists of convex functions only, the arithmetic average is a singleton symmetric fractional polymorphism. Another example is the pair {min, max} as a symmetric fractional polymorphism for the rank function in a matroid [112] (not to be confused with the rank of a matroid). A canonical example of a fractional polymorphism that isn't symmetric is the set of all projections (for any fixed arity), which works for any VCSP.

### 3.2.3 Expressive power

> Everything you say should be true, but not everything true should be said.
>
> — Voltaire (allegedly)

In this subsection, we will start with a fresh context containing only:

```
variable {D C : Type} [Nonempty D] [Fintype D]
  [LinearOrderedAddCommMonoid C]
```

We define *expressive power* of a VCSP template inductively, resulting in a definition that looks very different from how the expressive power is defined in literature:

```
inductive ValuedCSP.expresses (Γ : ValuedCSP D C) : ValuedCSP D C
| single {n : ℕ} {f : (Fin n → D) → C}
    (_ : ⟨n, f⟩ ∈ Γ) :
    Γ.expresses ⟨n, f⟩
| double {n : ℕ} {f g : (Fin n → D) → C}
    (_ : Γ.expresses ⟨n, f⟩) (_ : Γ.expresses ⟨n, g⟩) :
    Γ.expresses ⟨n, f+g⟩
| minimize {n : ℕ} {f : (Fin n.succ → D) → C}
    (_ : Γ.expresses ⟨n.succ, f⟩) :
    Γ.expresses ⟨n, fun x : Fin n → D =>
        Finset.univ.inf' sorry (fun z : D => f (z :: x))⟩
| remap {n m : ℕ} {f : (Fin n → D) → C}
    (_ : Γ.expresses ⟨n, f⟩) (τ : Fin n → Fin m) :
    Γ.expresses ⟨m, fun x : Fin m → D => f (x ∘ τ)⟩
```

These four cases can be summarized as follows:

- If `f ∈ Γ` then `Γ` can express `f`.
- If `Γ` can express `f` and `g` then `Γ` can express `f+g` as well.
- If `Γ` can express (`n+1`)-ary function `f` then `Γ` can express the `n`-ary function that shifts the arguments by one position to the right, plugs them into `f` and searches for the label that minimizes the cost when plugged in as the first argument.
- If `Γ` can express `n`-ary function `f` and `τ` is a function from `Fin n` to `Fin m` then `Γ` can express the `m`-ary function that applies `f` to arguments remapped by `τ`.

For comparison, a more familiar definition of expressive power could be written as follows:

```
def ValuedCSP.Instance.evalPartial {Γ : ValuedCSP D C} {ι μ : Type}
    (I : Γ.Instance (ι ⊕ μ)) (x : ι → D) :
    (μ → D) → C :=
  I.evalSolution ∘ Sum.elim x

def ValuedCSP.Instance.evalMinimize
    {Γ : ValuedCSP D C} {ι μ : Type} [Fintype μ]
    (I : Γ.Instance (ι ⊕ μ)) (x : ι → D) : C :=
  Finset.univ.inf' sorry (I.evalPartial x)

def ValuedCSP.expressivePower (Γ : ValuedCSP D C) : ValuedCSP D C :=
  { ⟨n, I.evalMinimize⟩ | (n : ℕ) (μ : Type) (_ : Fintype μ)
                           (I : Γ.Instance (Fin n ⊕ μ)) }
```

This definition (which is usually presented in literature as a single definition) is split into three layers for modularity:

- Partial evaluation of a `Γ` instance `I` for given partial solution `x`, waiting for rest.
- Evaluation of a `Γ` instance `I` for given partial solution `x`, minimizing over the rest.
- A new VCSP template made of all functions expressible by `Γ`.

In other words, `ValuedCSP.expressivePower` works by converting an entire VCSP instance into a new cost function, exposing some of the variables as arguments on the new cost function and minimizing over the remaining arguments.

Unfortunately, we didn't prove equivalence of these two definitions, so you will have to trust our word that they are equivalent. We deeply apologize for it.

We decided to ditch `ValuedCSP.expressivePower` and only keep `ValuedCSP.expresses` in the end. Inductive definitions are usually easier to handle in Lean, and especially so in this case. At the same time, this decision is a bit unsatisfactory. I believe it would be better to have both definitions, show their equivalence, use `ValuedCSP.expresses` in all later proofs, but have the option to hide it and make only `ValuedCSP.expressivePower` (with its prerequisites) a part of the trusted code.

We proved two lemmas that together establish that `ValuedCSP.expresses` acts as a closure on VCSP templates:

```
lemma ValuedCSP.subset_expresses (Γ : ValuedCSP D C) :
    Γ ⊆ Γ.expresses
```

```
lemma ValuedCSP.expresses_expresses (Γ : ValuedCSP D C) :
    Γ.expresses = Γ.expresses.expresses
```

Proving them was trivial with the inductive definition. Trying to prove the last lemma with the classical definition was a nightmare. Here, unfortunately, my laziness won over my desire to produce high-quality results.

Finally, we define when a VCSP template *can express Max-Cut*:

```
def ValuedCSP.CanExpressMaxCut (Γ : ValuedCSP D C) : Prop :=
  ∃ f : (Fin 2 → D) → C, ⟨2, f⟩ ∈ Γ.expresses ∧ f.HasMaxCutProperty
```

### 3.2.4 Basic LP relaxation

We introduce the basic LP relaxation of a VCSP instance. Let's start with a small example for illustration.

The cost function `f` is defined by a table on the domain `ae` as follows:

| f | a | e |
|---|---|---|
| a | 2 | 0 |
| e | 0 | 1 |

The task is to solve the following VCSP instance:

$$\text{minimize}\ f\ x_1\ x_2 + f\ x_2\ x_0 + f\ x_0\ x_1$$

Basic LP relaxation (all variables are nonnegative):

$$\text{minimize}\ 2 * y_{0aa} + y_{0ee} + 2 * y_{1aa} + y_{1ee} + 2 * y_{2aa} + y_{2ee}$$

$$y_{0aa} + y_{0ae} + y_{0ea} + y_{0ee} = 1$$

$$y_{1aa} + y_{1ae} + y_{1ea} + y_{1ee} = 1$$

$$y_{2aa} + y_{2ae} + y_{2ea} + y_{2ee} = 1$$

$$z_{0a} + z_{0e} = 1$$

$$z_{1a} + z_{1e} = 1$$

$$z_{2a} + z_{2e} = 1$$

$$y_{0aa} + y_{0ae} = z_{1a}$$

$$y_{0ea} + y_{0ee} = z_{1e}$$

$$y_{0aa} + y_{0ea} = z_{2a}$$

$$y_{0ae} + y_{0ee} = z_{2e}$$

$$y_{1aa} + y_{1ae} = z_{2a}$$

$$y_{1ea} + y_{1ee} = z_{2e}$$

$$y_{1aa} + y_{1ea} = z_{0a}$$

$$y_{1ae} + y_{1ee} = z_{0e}$$

$$y_{2aa} + y_{2ae} = z_{0a}$$

$$y_{2ea} + y_{2ee} = z_{0e}$$

$$y_{2aa} + y_{2ea} = z_{1a}$$

$$y_{2ae} + y_{2ee} = z_{1e}$$

This LP can be summarized as follows:

- $y_0$ variables … joint distribution of $x_1$ and $x_2$ in the term $f\ x_1\ x_2$
- $y_1$ variables … joint distribution of $x_2$ and $x_0$ in the term $f\ x_2\ x_0$
- $y_2$ variables … joint distribution of $x_0$ and $x_1$ in the term $f\ x_0\ x_1$
- $z_0$ variables … marginal distribution of $x_0$ in the VCSP instance
- $z_1$ variables … marginal distribution of $x_1$ in the VCSP instance
- $z_2$ variables … marginal distribution of $x_2$ in the VCSP instance

The joint distributions are there because they determine the cost of the (relaxed) solution. The marginal distributions are there because they ensure consistency between VCSP terms that share the same VCSP variable.

Optimal solution of the LP is:

$$y_{0aa} = y_{0ee} = y_{1aa} = y_{1ee} = y_{2aa} = y_{2ee} = 0$$

$$y_{0ae} = y_{0ea} = y_{1ae} = y_{1ea} = y_{2ae} = y_{2ea} = 0.5 = z_{0a} = z_{0e} = z_{1a} = z_{1e} = z_{2a} = z_{2e}$$

The resulting objective value is `0` (and, obviously, we cannot do any better).

The zero cost is not attainable by the VCSP instance. Its optimal solution is:

$$x_0 = a$$
$$x_1 = e$$
$$x_2 = e$$

Its cost is `1` (by checking all solutions, we make sure it is really optimal). We conclude that the basic LP relaxation isn't tight here.

Are there VCSP templates for which every VCSP instance has a tight basic LP relaxation? We will soon answer this question. Let's first formally define the basic LP relaxation in generality.

We start by introducing a slightly different definition of LP than what we used in the previous section. A *canonical* linear program consists of equations. Variables are of type `J`. Constraints are indexed by type `I`. The objective function is again intended to be minimized.

```
structure CanonicalLP (I J R : Type) where
  A : Matrix I J R
  b : I → R
  c : J → R
```

In the next three definitions, we will assume:

```
variable {I J R : Type} [Fintype J] [OrderedSemiring R]
```

Vector `x` is a *solution* to a canonical linear program `P` iff multiplying `x` by the matrix `A` from the left yields the vector `b` and all entries of `x` are nonnegative:

```
def CanonicalLP.IsSolution (P : CanonicalLP I J R) (x : J → R) : Prop :=
  P.A *ᵥ x = P.b ∧ 0 ≤ x
```

Canonical linear program `P` *reaches* objective value `r` iff there is a solution `x` such that, when its entries are elementwise multiplied by the coefficients `c` and summed up, `r` is the result:

```
def CanonicalLP.Reaches (P : CanonicalLP I J R) (r : R) : Prop :=
  ∃ x : J → R, P.IsSolution x ∧ P.c ⬝ᵥ x = r
```

Canonical linear program `P` has *minimum* `v` iff `P` reaches `v` but no lower objective value:

```
def CanonicalLP.Minimum (P : CanonicalLP I J R) (v : R) : Prop :=
  P.Reaches v ∧ ∀ r : R, P.Reaches r → v ≤ r
```

Let's define the basic LP relaxation of a VCSP instance. We now forget about `I`, `J`, and `R`. Instead, we will work with the following variables in the rest of this subsection:

```
{D : Type} [Fintype D]
{ι : Type} [Fintype ι]
{C : Type} [OrderedRing C]
{Γ : ValuedCSP D C}
```

The *basic LP relaxation* is defined as follows:

```
def ValuedCSP.Instance.RelaxBLP (I : Γ.Instance ι) :
    CanonicalLP
      ((Σ t : I, (Fin t.fst.n × D)) ⊕ (ι ⊕ I))
      ((Σ t : I, (Fin t.fst.n → D)) ⊕ (ι × D))
      C :=
  CanonicalLP.mk
    (Matrix.fromBlocks
      (fun ⟨cₜ, cₙ, cₐ⟩ => fun ⟨t, v⟩ =>
        if ht : cₜ = t
        then
          if v (Fin.cast (congr_arg (ValuedCSP.Term.n ∘ Sigma.fst) ht) cₙ)
             = cₐ
          then 1
          else 0
        else 0)
      (fun ⟨⟨cₜ, _⟩, cₙ, cₐ⟩ => fun ⟨i, a⟩ =>
        if cₜ.app cₙ = i ∧ cₐ = a then -1 else 0)
      (Matrix.fromRows
        0
        (fun cₜ : I => fun ⟨t, _⟩ => if cₜ = t then 1 else 0))
      (Matrix.fromRows
        (fun cᵢ : ι => fun ⟨i, _⟩ => if cᵢ = i then 1 else 0)
        0))
    (Sum.elim
      (fun _ : (Σ t : I, (Fin t.fst.n × D)) => 0)
      (fun _ : ι ⊕ I => 1))
    (Sum.elim
      (fun ⟨⟨t, _⟩, v⟩ => t.f v)
      (fun _ => 0))
```

Note that Σ denotes a dependent pair. As the definition above says, the LP variables are indexed by:

```
(Σ t : I, (Fin t.fst.n → D)) ⊕ (ι × D)
```

The left half of LP variables encodes the joint distribution of every VCSP term. The right half of LP variables encodes the marginal distribution of every VCSP variable.

The same LP in a slightly simplified picture (the vector `b` is to the right side of the matrix `A`; the vector `c` is underneath the matrix `A`) can be studied here:

| | (Σ t : I, (Fin t.fst.n → D)) ⊕ (ι × D) | | |
|---|---|---|---|
| (Σ t : I, (Fin t.fst.n × D)) ⊕ | fun ⟨$c_t$, $c_n$, $c_a$⟩ => fun ⟨t, v⟩ =><br>if $c_t$ = t<br>then<br>if v $c_n$ = $c_a$ then 1 else 0<br>else 0 | fun ⟨⟨$c_t$, _⟩, $c_n$, $c_a$⟩ => fun ⟨i, a⟩ =><br>if $c_t$.app $c_n$ = i ∧ $c_a$ = a<br>then -1<br>else 0 | 0 |
| ι ⊕ | 0 | fun $c_i$ : ι => fun ⟨i, _⟩ =><br>if $c_i$ = i then 1 else 0 | 1 |
| I | fun $c_t$ : I => fun ⟨t, _⟩ =><br>if $c_t$ = t then 1 else 0 | 0 | |
| | fun ⟨⟨t, _⟩, v⟩ => t.f v | 0 | |

The upper half of the matrix encodes that the joint distributions are consistent with the marginal distributions. The lower half of the matrix has again two halves; the first half encodes the requirement that the marginal probabilities of every variable sum up to one; the second half encodes the requirement that the joint probabilities of every term in the instance sum up to one. The cost function has nonzero coefficients only for the LP variables that represent the joint distributions.

Note that the VCSP instance (the multiset) `I` may contain the same VCSP term multiple times. In the top left block, where ⟨$c_t$, $c_n$, $c_a$⟩ is matched, the condition $c_t$ = `t` tests both the equality of the VCSP terms and the equality of their identifiers — each copy of the VCSP term has its own joint distribution. In the top right block, however, only the contents of the VCSP term matter; therefore ⟨⟨$c_t$, _⟩, $c_n$, $c_a$⟩ ignores the attached identifier.

The construction could be simplified if cost functions of arity zero were forbidden (the only thing they do is that they increase the cost of the VCSP instance by a constant). In the current version, most of the time it is redundant to check both the totals of marginal distributions and the totals of joint distributions.

### 3.2.5 Results

We prove three theorems about VCSP, with proofs loosely following Kolmogorov et al. [113].

First, we prove that, if a VCSP template over a linearly ordered cancellative abelian monoid can express Max-Cut, it cannot have any commutative fractional polymorphism:

```
theorem
ValuedCSP.CanExpressMaxCut.forbids_commutativeFractionalPolymorphism
    {D C : Type} [Nonempty D] [Fintype D]
    [LinearOrderedCancelAddCommMonoid C]
    {Γ : ValuedCSP D C} (_ : Γ.CanExpressMaxCut)
    {ω : FractionalOperation D 2} (_ : ω.IsValid) :
    ¬ ω.IsSymmetricFractionalPolymorphismFor Γ
```

Next, we prove that the basic LP relaxation for a VCSP template over any ordered ring of characteristic zero is valid:

```
theorem ValuedCSP.Instance.RelaxBLP_reaches
    {D : Type} [Fintype D]
    {ι : Type} [Fintype ι]
    {C : Type} [OrderedRing C] [CharZero C]
    {Γ : ValuedCSP D C} (I : Γ.Instance ι) (x : ι → D) :
    I.RelaxBLP.Reaches (I.evalSolution x)
```

Finally, we prove that, if a VCSP template over ℚ has symmetric fractional polymorphisms of all arities, then its basic LP relaxation is tight:

```
theorem
ValuedCSP.Instance.RelaxBLP_improved_of_allSymmetricFractionalPolymorphisms
    {D : Type} [Fintype D]
    {ι : Type} [Fintype ι]
    {Γ : ValuedCSP D ℚ} (I : Γ.Instance ι)
    {o : ℚ} (_ : I.RelaxBLP.Reaches o)
    (_ : ∀ m : ℕ, ∃ ω : FractionalOperation D m,
      ω.IsValid ∧ ω.IsSymmetricFractionalPolymorphismFor Γ) :
    ∃ x : ι → D, I.evalSolution x ≤ o
```

### 3.2.6 Corollary

Let's additionally define the optimum of a VCSP instance as the cost of any optimum solution:

```
def ValuedCSP.Instance.Optimum {D C ι : Type} [OrderedAddCommMonoid C]
    {Γ : ValuedCSP D C} (I : Γ.Instance ι) (o : C) :
    Prop :=
  ∃ x : ι → D, I.IsOptimumSolution x ∧ I.evalSolution x = o
```

As a corollary of the last two theorems, we prove that, if a VCSP template over ℚ has symmetric fractional polymorphisms of all arities, then (assuming their existence) the optimum of every instance is equal to the minimum of its basic LP relaxation:

```
theorem
ValuedCSP.Instance.optimum_iff_relaxBLP_minimum_of_allSymmFracPolymorphisms
    {D ι : Type} [Fintype D] [Fintype ι] {Γ : ValuedCSP D ℚ}
    (I : Γ.Instance ι)
    (_ : ∀ m : ℕ, ∃ ω : FractionalOperation D m,
      ω.IsValid ∧ ω.IsSymmetricFractionalPolymorphismFor Γ)
    (v : ℚ) :
    I.Optimum v ↔ I.RelaxBLP.Minimum v
```

What we didn't prove, though, is the existence of the optimum and minimum. I think it is a gap in our library that should be addressed in the future with high priority.

### 3.2.7 Related work

There are works on formally verified constraint-satisfaction solvers [114] [115]. To the best of our knowledge, nobody has yet formally verified theoretical results about VCSP.

### 3.2.8 Conclusion

We developed the very basics of the VCSP theory in Lean 4. Our library encompasses VCSP templates, instances of fixed-template VCSP problems, a certain interpretation of fractional polymorphisms and what Max-Cut is, a certain interpretation of the expressive power of a VCSP template, and the basic LP relaxation together with a few results about it. However, our work would be more valuable did we stick to the standard theory of VCSP. It would be better if our definitions were a more literal formalization of the standard notions from the literature, but it would be more challenging to finish the proofs.

As a possible future continuation, apart from the optima existence, a complementary result to `ValuedCSP.Instance.RelaxBLP_improved_of_allSymmetricFractionalPolymorphisms` could be proved; if the basic LP relaxation is tight, then the VCSP template has symmetric fractional polymorphisms of all arities [113]. Our results could also be extended to the setting of General-Valued Constraint Satisfaction Problems, i.e., the codomain would be the extended rationals [113] (albeit the negative infinity wouldn't be used). Another possible improvement, which wouldn't be interesting theoretically but would be useful if the basic LP relaxation were to be used practically, would be decreasing the number of constraints. Currently, all terms are checked for the variables expressing their joint distribution summing up to one. Ideally, only the sums of marginal distributions would be explicitly checked, and terms of zero arity would be dealt with differently. Such a change would not only make the LP slightly smaller, but it would also align it better with existing literature.

# 4 Seymour project

Seymour's regular matroid decomposition theorem is a hallmark structural result in matroid theory [116] [117] [118] [119]. It states that, on the one hand, any 1-sum of two regular matroids is regular, any 2-sum of two regular matroids is regular, any 3-sum of two regular matroids is regular, and on the other hand, any regular matroid can be decomposed into matroids that are graphic, cographic, or isomorphic to R10 by repeating 1-sum, 2-sum, and 3-sum decompositions.

The interest in matroids comes from the fact that they capture and generalize many mathematical concepts, such as linear independence (captured by vector matroids), graphs (graphic matroids), and extensions of fields (algebraic matroids). Another advantage of matroids is that they have a relatively short definition, making them amenable to formalization.

As for Seymour's theorem, it not only presents a structural characterization of the class of regular matroids, but also leads to important applications, such as polynomial-time algorithms for testing whether a matrix is totally unimodular [120]. Additionally, Seymour's theorem can offer a structural approach for solving certain combinatorial optimization problems [121], for example, it leads to the characterization and efficient algorithms for the cycle polytope.

The goal of our work was to develop a general and reusable library proving a result that is at least as strong as the forward (composition) direction of classical Seymour's theorem (i.e., stated for finite matroids). Moreover, our aim was to make our library modular and extensible by ensuring compatibility with matroids in Mathlib [1].

To achieve our goals, we made the following compromises. First, we focused on the proof of the composition direction, while only stating the decomposition direction. Second, we assumed finiteness where it would simplify proofs, while making sure that the final results held for finite matroids (in the end, they hold for matroids that may potentially have infinite ground sets but they must have finite rank). Third, we tailored our implementation specifically to Seymour's theorem, avoiding introducing additional matroid notions when possible.

Our project makes the following contributions in Lean 4:

- We formalize the definition and selected properties of totally unimodular matrices, some of which were added to Mathlib.
- We formalize definitions and some basic properties of vector matroids, their standard representations, regular matroids, graphic matroids, and 1-sums, 2-sums, and 3-sums of vector matroids given by their standard representations.
- We prove the composition direction of Seymour's theorem — that any 1-sum of regular matroids is regular, that any 2-sum of regular matroids is regular, that any 3-sum of regular matroids is regular — all in the case where the matroids may have infinite ground sets but must have finite rank.
- We state the decomposition direction of Seymour's theorem — that any regular matroid of finite rank can be decomposed into graphic matroids, cographic matroids, and matroids isomorphic to R10 by repeated 1-sum, 2-sum, and 3-sum decompositions.

Our formalization is conceptually split into two parts — "implementation" and "presentation". Implementation is contained in the `Seymour` folder and encompasses all definitions and lemmas used to obtain our results. Presentation is contained in the `Seymour.lean` file, which repeats selected definitions and theorems comprising the key final results of our contribution. Every definition in the "presentation" file is checked to be definitionally equal to its counterpart from the "implementation" using the `recall` or the `example` command. Similarly, we `recall` every theorem presented here and then use the `#guard_msgs in #print axioms` command to check that the implementation of its proof (including the entire dependency tree) depends only on the three axioms `[propext, Classical.choice, Quot.sound]` which are standard for Lean projects that use classical logic. In other words, we identified what is the trusted code, and we repeated all nontrivial trusted code in the `Seymour.lean` file, so that our results can be believed.

While working on our project, we leveraged the LeanBlueprint[20] tool to help guide our formalization efforts. In particular, we used it to create an informal blueprint and a dependency graph, which allowed us to get a clearer overview of the results we were formalizing, as well as their dependencies. In our workflow, we first created a write-up encompassing the classical results from Truemper [117]. Based on this write-up, we developed a self-contained blueprint for our formalization by filling in gaps, fleshing out technical details, and sometimes reworking certain proofs. We followed this blueprint during the development of our library, keeping it up to date and turning it into documentation of our code. For a reader who prefers reading informal MathematiCS, we recommend reading our blueprint[21]. A reader who prefers reading formal MathematiCS should continue reading this chapter.

## *4.1 Matroids*

A *matroid* is usually described as a set of subsets of a finite ground set **`E`**, called *independent sets*, with the following properties [117]:

- The empty set is independent.
- Every subset of an independent set is independent.
- Whenever an independent set is smaller than another independent set, the smaller one can be augmented with an element of the bigger one.

Matroids defined using a finite ground set and a predicate "is an independent set" are captured by the constructor `IndepMatroid.ofFinite` in Mathlib, which isn't a part of our trusted code.

Equivalently, a matroid can be described as a set of subsets of a finite ground set **`E`**, called *bases*, with the following properties [117]:

- There is a base.
- For any two bases, we can exchange elements between them (one for one) and it is still a base.

Matroids defined using a finite ground set and a predicate "is a base" are captured by the constructor `Matroid.ofIsBaseOfFinite` in Mathlib, which isn't a part of our trusted code.

Informally speaking, the equivalence of these two definitions stems from:

- We can take the former structure, call all maximum independent sets "bases", and we obtain the latter structure.
- We can take the latter structure, call all subsets of all bases "independent", and we obtain the former structure.

However, these classical definitions are restricted to **`E`** being finite. From now on, we will refer to such structures as a *finite matroid*. If we don't want to restrict **`E`** to be a finite set, we must be more careful [122].

---

[20] https://github.com/PatrickMassot/leanblueprint

[21] https://ivan-sergeyev.github.io/seymour/blueprint.pdf

In Mathlib [1], the structure `Matroid` captures the definition of a matroid (not necessarily finite) [122] via the *base conditions* as follows:

```
def ExchangeProperty {α : Type} (P : Set α → Prop) : Prop :=
  ∀ X Y : Set α, P X → P Y → ∀ a ∈ X \ Y, ∃ b ∈ Y \ X, P (b ᓑ (X \ {a}))

def Maximal {α : Type} (P : α → Prop) (x : α) : Prop :=
  P x ∧ ∀ y : α, P y → x ≤ y → y ≤ x

def ExistsMaximalSubsetProperty {α : Type} (P : Set α → Prop) (X : Set α) :
    Prop :=
  ∀ I : Set α, P I → I ⊆ X →
    ∃ J : Set α, I ⊆ J ∧ Maximal (fun K : Set α => P K ∧ K ⊆ X) J

structure Matroid (α : Type) where
  (E : Set α)
  (IsBase : Set α → Prop)
  (Indep : Set α → Prop)
  (indep_iff' : ∀ I : Set α, Indep I ↔ ∃ B : Set α, IsBase B ∧ I ⊆ B)
  (exists_isBase : ∃ B : Set α, IsBase B)
  (isBase_exchange : ExchangeProperty IsBase)
  (maximality : ∀ X : Set α, X ⊆ E → ExistsMaximalSubsetProperty Indep X)
  (subset_ground : ∀ B : Set α, IsBase B → B ⊆ E)
```

Additionally, Mathlib [1] allows the user to construct matroids (not necessarily finite) in terms of the *independence conditions* using:

```
structure IndepMatroid (α : Type) where
  (E : Set α)
  (Indep : Set α → Prop)
  (indep_empty : Indep ∅)
  (indep_subset : ∀ I J : Set α, Indep J → I ⊆ J → Indep I)
  (indep_aug : ∀ I B : Set α, Indep I →
    ¬ Maximal Indep I → Maximal Indep B → ∃ x ∈ B \ I, Indep (x ᓑ I))
  (indep_maximal : ∀ X : Set α, X ⊆ E → ExistsMaximalSubsetProperty Indep X)
  (subset_ground : ∀ I : Set α, Indep I → I ⊆ E)
```

We can then obtain `Matroid α` from `IndepMatroid α` via:

```
def IndepMatroid.matroid {α : Type} (M : IndepMatroid α) : Matroid α where
  E := M.E
  IsBase := Maximal M.Indep
  Indep := M.Indep
  indep_iff' := sorry
  exists_isBase := sorry
  isBase_exchange := sorry
  maximality := sorry
  subset_ground := sorry
```

Though we generally work with matroids that may be infinite, our final results require that the matroids have finite rank. A *finite-rank* matroid is one that has a finite base, implemented in Mathlib [1] as:

```
class Matroid.RankFinite {α : Type} (M : Matroid α) : Prop where
  exists_finite_isBase : ∃ B : Set α, M.IsBase B ∧ B.Finite
```

We could also require that all bases be finite, but it isn't part of the definition.

On a related note, a *finitary* matroid is a matroid where independence of any set depends only on its finite subsets:

```
class Matroid.Finitary {α : Type} (M : Matroid α) : Prop where
  indep_of_forall_finite :
    ∀ I : Set α, (∀ J : Set α, J ⊆ I → J.Finite → M.Indep J) → M.Indep I
```

Note that the opposite implication holds in every matroid:

```
example {α : Type} (M : Matroid α) :
    ∀ I : Set α, M.Indep I → (∀ J : Set α, J ⊆ I → J.Finite → M.Indep J)
```

Obviously, every finite-rank matroid is finitary.

## 4.2 *Sets, subsets, and types*

Already in Section 2.4.2, we have reviewed how sets are handled in Lean. In this section, we will dive into several specific definitions that we need in the Seymour project.

First of them is *disjoint* sets. Since the abstract definition `Disjoint` is difficult to understand, we instead found our understanding of disjointness on the following characterization:

```
theorem Set.disjoint_iff_inter_eq_empty {α : Type} {s t : Set α} :
    Disjoint s t ↔ s ∩ t = ∅
```

Next, we need the *range* of a function:

```
def Set.range {α ι : Type} (f : ι → α) : Set α :=
  { x : α | ∃ y : ι, f y = x }
```

Both of them are equipped with a more convenient notation in our project, but we don't need the notation in the trusted code.

The last concept is a bit challenging to explain. Sometimes we need to consider the intersection of two sets not as a set of the ambient type but as a set of elements of the first set. It is the one situation where I don't exactly love type theory. The infix operator is `↓∩` and its most important property is

```
lemma Set.preimage_val_eq_univ_of_subset {α : Type} {A B : Set α}
    (_ : A ⊆ B) :
    A ↓∩ B = Set.univ
```

where the RHS is the set of the entire `α`, i.e., `(Set.univ : Set α)` is the function that always returns `True`. For very similar reasons, we declare

```
def HasSubset.Subset.elem {α : Type} {X Y : Set α} (hXY : X ⊆ Y)
    (x : X.Elem) : Y.Elem :=
  ⟨x.val, hXY x.property⟩
```

which allows us to write `hXY.elem` as the function that converts elements of `X` to elements of `Y` without changing their `Set.Elem.val` output. Therefore, the range of `hXY.elem` is equal to `(Y ↓∩ X : Set Y)`.

## 4.3 Linear independence

`LinearIndependent` is a predicate taking a semiring and a vector family as explicit arguments, and it is defined so that the following theorem holds:

```
theorem linearIndependent_iff' {ι R M : Type} {v : ι → M}
    [Ring R] [AddCommGroup M] [Module R M] :
    LinearIndependent R v ↔
    ∀ s : Finset ι, ∀ g : ι → R, ∑ i ∈ s, g i • v i = 0 → ∀ i ∈ s, g i = 0
```

In words, a vector family is linearly independent iff the only way how to obtain the zero vector as a finite linear combination of given vectors is to take zero coefficients everywhere. However, because this theorem covers modules over a ring only, we should also check what the definition says about modules over a semiring:

```
theorem linearIndependent_iff'ₛ {ι R M : Type} {v : ι → M}
    [Semiring R] [AddCommMonoid M] [Module R M] :
    LinearIndependent R v ↔
    ∀ s : Finset ι, ∀ f g : ι → R,
      ∑ i ∈ s, f i • v i = ∑ i ∈ s, g i • v i → ∀ i ∈ s, f i = g i
```

In this (more general) case, two finite linear combinations result in the same vector only if they have the same coefficients. Linear independence over semirings will be used only tangentially, as almost anything interesting will require a division ring at least.

Furthermore, a vector family is linearly independent on a set iff the vector family with a domain restricted to given set is linearly independent:

```
def LinearIndepOn {ι : Type} (R : Type) {M : Type} (v : ι → M)
    [Semiring R] [AddCommMonoid M] [Module R M] (s : Set ι) :
    Prop :=
  LinearIndependent R (fun x : s.Elem => v x.val)
```

## 4.4 Total unimodularity

We say that a matrix `A` over a commutative ring `R` is *totally unimodular* iff every finite square submatrix of `A` (not necessarily contiguous, but no row or column taken twice) has determinant in `{-1, 0, 1}`. Mathlib [1] implements this definition as follows:

```
def Matrix.IsTotallyUnimodular {m n R : Type} [CommRing R]
    (A : Matrix m n R) :
    Prop :=
  ∀ k : ℕ, ∀ f : Fin k → m, ∀ g : Fin k → n,
    f.Injective → g.Injective →
      (A.submatrix f g).det ∈ Set.range SignType.cast
```

Here, `SignType` is an inductive type with three values. They are `zero`, `neg`, and `pos`, which `SignType.cast` maps to `(0 : R)`, `(-1 : R)`, and `(1 : R)`, respectively. Although this abstract definition is convenient to work with in Lean, we find it unfortunate that the implementation via `SignType` makes it harder to understand what the definition means and to believe that it is formalized correctly.

Note that the indexing functions `f` and `g` are required to be injective in the definition, but this condition can be lifted. Indeed, the lemma `Matrix.isTotallyUnimodular_iff` shows that we can equivalently check the determinants of all finite square submatrices, not just the finite square submatrices without repeated rows and columns.

Keep in mind that the determinant is computed over `R`, so for certain commutative rings, all matrices are trivially totally unimodular, for example, when `R` is `Z3` (where `1 + 1 = -1`).

## 4.5 *Retyping matrix dimensions*

When constructing matroids, we often need to convert a block matrix whose blocks are indexed by disjoint sets into a matrix indexed by unions of those index sets. Although the contents of the matrix stay the same, both its dimensions change their type from a sum of sets to a set union of those sets. To this end we declare:

```
def Subtype.toSum {α : Type} {X Y : Set α} (i : (X ∪ Y).Elem) :
    X.Elem ⊕ Y.Elem :=
  if hiX : i.val ∈ X then ◩⟨i, hiX⟩ else
  if hiY : i.val ∈ Y then ◪⟨i, hiY⟩ else
  (i.property.elim hiX hiY).elim
```

We therefore have:

```
lemma toSum_left {α : Type} {X Y : Set α}
    {x : (X ∪ Y).Elem} (hx : x.val ∈ X) :
    x.toSum = ◩⟨x.val, hx⟩
```

```
lemma toSum_right {α : Type} {X Y : Set α}
    {y : (X ∪ Y).Elem} (hyX : y.val ∉ X) (hyY : y.val ∈ Y) :
    y.toSum = ◪⟨y.val, hyY⟩
```

In practice, `Subtype.toSum` makes sense only when `X` and `Y` are disjoint sets.

Afterwards we declare:

```
def Matrix.toMatrixUnionUnion {α β R : Type} {T₁ T₂ : Set α} {S₁ S₂ : Set β}
    (A : Matrix (T₁.Elem ⊕ T₂.Elem) (S₁.Elem ⊕ S₂.Elem) R) :
    Matrix (T₁ ∪ T₂).Elem (S₁ ∪ S₂).Elem R :=
  ((A ∘ Subtype.toSum) · ∘ Subtype.toSum)
```

Therefore, in the appropriate context we have:

```
A.toMatrixUnionUnion i j = A i.toSum j.toSum
```

We also declare other conversion functions, but they are not part of the trusted code (the other conversions are used only in proofs). One conversion function we would like to highlight is:

```
def Matrix.toMatrixElemElem {α β R : Type}
    {T₁ T₂ T : Set α} {S₁ S₂ S : Set β}
    (A : Matrix (T₁ ⊕ T₂) (S₁ ⊕ S₂) R)
    (hT : T = T₁ ∪ T₂) (hS : S = S₁ ∪ S₂) :
    Matrix T S R :=
  hT ▸ hS ▸ A.toMatrixUnionUnion
```

It comes with the following characterization:

```
lemma Matrix.toMatrixElemElem_apply {α β R : Type}
    {T₁ T₂ T : Set α} {S₁ S₂ S : Set β}
    (A : Matrix (T₁ ⊕ T₂) (S₁ ⊕ S₂) R)
    (hT : T = T₁ ∪ T₂) (hS : S = S₁ ∪ S₂) (i : T) (j : S) :
    A.toMatrixElemElem hT hS i j = A (hT ▸ i).toSum (hS ▸ j).toSum
```

Its usefulness will be revealed later.

## *4.6 Vector matroids*

Vector matroids [116] [117] is the most fundamental matroid class formalized in our work, serving as the basis for binary and regular matroids in later sections. A *vector matroid* is constructed from a matrix `A` by taking the column index set as the ground set and declaring a set `I` to be independent iff the set of columns of `A` indexed by `I` is linearly independent. To capture this definition, we first implement the independence predicate:

```
def Matrix.IndepCols {α R : Type} {X Y : Set α} [Semiring R]
    (A : Matrix X Y R) (I : Set α) :
    Prop :=
  I ⊆ Y ∧ LinearIndepOn R Aᵀ (Y ↓∩ I)
```

Next, we construct an `IndepMatroid` as follows:

```
def Matrix.toIndepMatroid {α R : Type} {X Y : Set α} [DivisionRing R]
    (A : Matrix X Y R) :
    IndepMatroid α where
  E := Y
  Indep := A.IndepCols
  indep_empty := A.indepCols_empty
  indep_subset := A.indepCols_subset
  indep_aug := A.indepCols_aug
  indep_maximal S _ := A.indepCols_maximal S
  subset_ground _ := And.left
```

Finally, we convert `IndepMatroid` to `Matroid` by chaining the two conversions:

```
def Matrix.toMatroid {α R : Type} {X Y : Set α} [DivisionRing R]
    (A : Matrix X Y R) :
    Matroid α :=
  A.toIndepMatroid.matroid
```

Going forward, we use `Matrix.toMatroid` for constructing vector matroids from matrices.

As part of the construction above, we had to show that `Matrix.IndepCols` satisfies the so-called *augmentation property*:

```
∀ I B : Set α,
  Indep I → ¬ Maximal Indep I → Maximal Indep B →
    ∃ x ∈ B \ I, Indep (x ᕃ I))
```

It is worth noting that while we define `Matrix.IndepCols` over a semiring `R` for the sake of generality, the augmentation property requires `R` to be at least a division ring. Indeed, for example, let `R = ZMod 6`, which is in fact a ring, and consider

$$\begin{pmatrix} 0 & 1 & 2 & 3 \\ 1 & 0 & 3 & 2 \end{pmatrix}$$

with columns indexed by `{0, 1, 2, 3}`. Then `I = {0}` is a non-maximal independent set over `R` and `J = {2, 3}` is a maximal independent set over `R`, but they do not satisfy the augmentation property. For this reason, we require `R` to be a division ring in the augmentation property and all subsequent results.

Additionally, we show that vector matroids as defined above are finitary, i.e., an infinite set in a vector matroid is independent iff all its finite subsets are independent:

```
lemma Matrix.toMatroid_isFinitary {α R : Type} {X Y : Set α}
    [DivisionRing R] (A : Matrix X Y R) :
    A.toMatroid.Finitary
```

## 4.7 *Standard representations*

The *standard representation* [116] [117] of a vector matroid is the following structure:

```
structure StandardRepr (α R : Type) where
  X : Set α
  Y : Set α
  hXY : Disjoint X Y
  B : Matrix X Y R
  decmemX : ∀ a, Decidable (a ∈ X)
  decmemY : ∀ a, Decidable (a ∈ Y)
```

In essence, `StandardRepr` is a wrapper for the standard representation matrix `B` indexed by disjoint sets `X` and `Y`, bundled together with the membership decidability for `X` and `Y`. The standard representation matrix `B` corresponds to the full representation matrix (`1`  `B`) with the conversion implemented as:

```
def StandardRepr.toFull {α R : Type} [Zero R] [One R]
    (S : StandardRepr α R) :
    Matrix S.X (S.X ∪ S.Y).Elem R :=
  ((Matrix.fromCols 1 S.B) · ∘ Subtype.toSum)
```

Thus, the vector matroid given by its standard representation is constructed as follows:

```
def StandardRepr.toMatroid {α R : Type} [DivisionRing R]
    (S : StandardRepr α R) :
    Matroid α :=
  S.toFull.toMatroid
```

In this matroid, the ground set is `X ∪ Y`, and a set `I ⊆ X ∪ Y` is independent iff the columns of $\begin{pmatrix} 1 & B \end{pmatrix}$ indexed by `I` are linearly independent over `R`.

Below are several results we prove about standard representations. We provide each of them either because we need them in the proof of regularity of the 1-sum, the 2-sum, or the 3-sum of regular matroids, or because they could be useful for downstream projects.

First, we show that if the row index set `X` of a standard representation is finite, then `X` is a base in the resulting matroid:

```
lemma StandardRepr.toMatroid_isBase_X {α R : Type} [Field R]
    (S : StandardRepr α R) [Fintype S.X] :
    S.toMatroid.IsBase S.X
```

This lemma restricts which sets can serve as row index sets of standard representations and motivates the assumptions in lemmas below.

Next, we prove that a full representation of a vector matroid can be transformed into a standard representation of the same matroid, with a given base as the row index set:

```
lemma Matrix.exists_standardRepr_isBase {α R : Type} [DivisionRing R]
    {X Y G : Set α} (A : Matrix X Y R) (_ : A.toMatroid.IsBase G) :
    ∃ S : StandardRepr α R, S.X = G ∧ S.toMatroid = A.toMatroid
```

We also prove an analog of the above lemma that additionally preserves total unimodularity of the representation matrix:

```
lemma Matrix.exists_standardRepr_isBase_isTotallyUnimodular {α R : Type}
    [Field R] {X Y G : Set α} [Fintype G] (A : Matrix X Y R)
    (_ : A.toMatroid.IsBase G) (_ : A.IsTotallyUnimodular) :
    ∃ S : StandardRepr α R,
      S.X = G ∧ S.toMatroid = A.toMatroid ∧ S.B.IsTotallyUnimodular
```

Note that this lemma takes stronger assumptions than the previous lemma, namely that `G` has to be finite and that the multiplication in `R` has to commute.

Another result we prove is that two standard representations of the same vector matroid over `Z2` with the same finite row index set must be identical:

```
lemma ext_standardRepr_of_same_matroid_same_X {α : Type}
    {S₁ S₂ : StandardRepr α Z2} [Fintype S₁.X]
    (_ : S₁.toMatroid = S₂.toMatroid) (_ : S₁.X = S₂.X) :
    S₁ = S₂
```

Although this particular lemma is never employed later in our project, it captures an important result that a binary matroid has an essentially unique standard representation [116] [117]. Nevertheless, we make use of a very similar result:

```
lemma support_eq_support_of_same_matroid_same_X {α F₁ F₂ : Type}
    [Field F₁] {S₁ : StandardRepr α F₁}
    [Field F₂] {S₂ : StandardRepr α F₂} [Fintype S₂.X]
    (_ : S₁.toMatroid = S₂.toMatroid) (hXX : S₁.X = S₂.X) :
    let hYY : S₁.Y = S₂.Y := sorry
    hXX ▸ hYY ▸ S₁.B.support = S₂.B.support
```

This lemma states that two standard representations of a vector matroid with identical (finite) row index sets have the same support, i.e., the zeros in them appear on identical positions. Crucially, this lemma holds for any two standard representations over any two fields. We will employ it for `ℚ` and `Z2`.

## 4.8 Regular matroids

Regular matroids [116] [117] are the core subject of Seymour's theorem. A matroid is *regular* iff [117] it can be constructed (as a vector matroid) from a rational totally unimodular matrix:

```
def Matroid.IsRegular {α : Type} (M : Matroid α) : Prop :=
  ∃ X Y : Set α,
    ∃ A : Matrix X Y ℚ, A.IsTotallyUnimodular ∧ A.toMatroid = M
```

One key result we prove is that every regular matroid is in fact *binary*, i.e., can be constructed from `Z2` matrix:

```
lemma Matroid.IsRegular.isBinary {α : Type} {M : Matroid α}
    (_ : M.IsRegular) :
  ∃ X Y : Set α, ∃ A : Matrix X Y Z2, A.toMatroid = M
```

Another important lemma we prove about regular matroids is their equivalent characterization in terms of totally unimodular signings. First, let's introduce the necessary definitions. We say that a matrix `A` is a *signing* of a matrix `U` iff their values are identical up to signs:

```
def Matrix.IsSigningOf {X Y R : Type} [LinearOrderedRing R] {n : ℕ}
    (A : Matrix X Y R) (U : Matrix X Y (ZMod n)) :
    Prop :=
  ∀ i : X, ∀ j : Y, |A i j| = (U i j).val
```

We then say that a binary matrix `U` *has a totally unimodular signing* iff it has a signing matrix `A` that is rational and totally unimodular:

```
def Matrix.IsTuSigningOf {X Y : Type}
    (A : Matrix X Y ℚ) (U : Matrix X Y Z2) :
    Prop :=
  A.IsTotallyUnimodular ∧ A.IsSigningOf U
```

```
def Matrix.HasTuSigning {X Y : Type} (U : Matrix X Y Z2) : Prop :=
  ∃ A : Matrix X Y ℚ, A.IsTuSigningOf U
```

Now, we can state the characterization. A standard representation over `Z2` with a finite set of rows represents a regular matroid iff its matrix has a totally unimodular signing:

```
lemma StandardRepr.toMatroid_isRegular_iff_hasTuSigning {α : Type}
    (S : StandardRepr α Z2) [Finite S.X] :
  S.toMatroid.IsRegular ↔ S.B.HasTuSigning
```

Out of all definitions in this section, only `Matroid.IsRegular` is a part of the trusted code.

## 4.9 *The 1-sum*

The 1-sum, the 2-sum, and the 3-sum of matroids are defined [117] as specific ways to compose their standard representation matrices. We hereby define the 1-sum, the 2-sum, and the 3-sum only for binary matroids. It should be mentioned, however, that the 1-sum and the 2-sum can also be defined more generally [116], which we will not do.

All matroid sums are defined on three levels:

- `Matrix` level
- `StandardRepr` level
- `Matroid` level

Let's review the distribution of responsibilities between the three levels.

`Matrix` 1-sum:

```
def matrixSum1 {R : Type} [Zero R] {Xₗ Yₗ Xᵣ Yᵣ : Type}
    (Aₗ : Matrix Xₗ Yₗ R) (Aᵣ : Matrix Xᵣ Yᵣ R) :
    Matrix (Xₗ ⊕ Xᵣ) (Yₗ ⊕ Yᵣ) R :=
  Matrix.fromBlocks Aₗ 0 0 Aᵣ
```

The same matrix in a picture:

$$\begin{pmatrix} A_l & 0 \\ 0 & A_r \end{pmatrix}$$

The `Matrix` level defines the standard representation matrix of the output matroid as a matrix indexed by `Sum` of indexing types. This definition is so straightforward that it would be natural to inline it into the subsequent definition. However, we retained it as a separate declaration for consistency with the 2-sum and the 3-sum, whose matrix constructions are more elaborate.

`StandardRepr` 1-sum:

```
noncomputable def standardReprSum1 {α : Type} {Sₗ Sᵣ : StandardRepr α Z2}
    (_ : Disjoint Sₗ.X Sᵣ.Y)
    (_ : Disjoint Sₗ.Y Sᵣ.X) :
    Option (StandardRepr α Z2) :=
  open scoped Classical in if
    Disjoint Sₗ.X Sᵣ.X ∧ Disjoint Sₗ.Y Sᵣ.Y
  then
    some ⟨
      Sₗ.X ∪ Sᵣ.X,
      Sₗ.Y ∪ Sᵣ.Y,
      sorry,
      (matrixSum1 Sₗ.B Sᵣ.B).toMatrixUnionUnion,
      inferInstance,
      inferInstance⟩
  else
    none
```

The `StandardRepr` level builds on top of the `Matrix` level. It converts the output matrix from being indexed by `Sum` to being index by set unions, it provides a proof that the resulting standard representation again has row indices and column indices disjoint, and it checks whether the operation is valid — if the preconditions are not met, it outputs `none` instead of `some` standard representation.

`Matroid` 1-sum:

```
def Matroid.IsSum1of {α : Type} (M : Matroid α) (Mₗ Mᵣ : Matroid α) :
    Prop :=
  ∃ S Sₗ Sᵣ : StandardRepr α Z2,
  ∃ hXY : Disjoint Sₗ.X Sᵣ.Y,
  ∃ hYX : Disjoint Sₗ.Y Sᵣ.X,
  standardReprSum1 hXY hYX = some S
  ∧ S.toMatroid = M
  ∧ Sₗ.toMatroid = Mₗ
  ∧ Sᵣ.toMatroid = Mᵣ
```

The `Matroid` level builds on top of the standard representation level but talks about matroids, the combinatorial objects. On the `Matroid` level, we don't define a function; instead, we define a predicate — when `M` is a 1-sum of `Mₗ` and `Mᵣ`.

In addition to our basic API about the 1-sum (for example, `Matroid.IsSum1of.E_eq`), we also provide a theorem `Matroid.IsSum1of.eq_disjointSum` that establishes the equality between the disjoint sum (defined in Mathlib) and the 1-sum (defined in our project) of binary matroids.

## *4.10 The 2-sum*

The definition of the 2-sum is also implemented on the three levels.

`Matrix` 2-sum:

```
def matrixSum2 {R : Type} [Semiring R] {Xₗ Yₗ Xᵣ Yᵣ : Type}
    (Aₗ : Matrix Xₗ Yₗ R) (r : Yₗ → R)
    (Aᵣ : Matrix Xᵣ Yᵣ R) (c : Xᵣ → R) :
    Matrix (Xₗ ⊕ Xᵣ) (Yₗ ⊕ Yᵣ) R :=
  Matrix.fromBlocks Aₗ 0 (c · * r ·) Aᵣ
```

The `Matrix` level is pretty similar to the one of the 1-sum. Again, the two given matrices are placed along the main diagonal of the resulting block matrix. However, the resulting two blocks aren't both zero, as this time the bottom left matrix contains the outer product of the two given vectors. The same matrix in a picture:

$$\begin{pmatrix} A_l & 0 \\ c \otimes r & A_r \end{pmatrix}$$

`StandardRepr` 2-sum:

```
noncomputable def standardReprSum2 {α : Type}
    {Sₗ Sᵣ : StandardRepr α Z2} {x y : α}
    (_ : Sₗ.X ∩ Sᵣ.X = {x})
    (_ : Sₗ.Y ∩ Sᵣ.Y = {y})
    (_ : Disjoint Sₗ.X Sᵣ.Y)
    (_ : Disjoint Sₗ.Y Sᵣ.X) :
    Option (StandardRepr α Z2) :=
```

```
  let Aₗ : Matrix (Sₗ.X \ {x}).Elem Sₗ.Y.Elem Z2 :=
    Sₗ.B.submatrix Set.diff_subset.elem id
  let Aᵣ : Matrix Sᵣ.X.Elem (Sᵣ.Y \ {y}).Elem Z2 :=
    Sᵣ.B.submatrix id Set.diff_subset.elem
  let r : Sₗ.Y.Elem → Z2 := Sₗ.B ⟨x, sorry⟩
  let c : Sᵣ.X.Elem → Z2 := (Sᵣ.B · ⟨y, sorry⟩)
  open scoped Classical in if
    r ≠ 0 ∧ c ≠ 0
  then
    some ⟨
      (Sₗ.X \ {x}) ∪ Sᵣ.X,
      Sₗ.Y ∪ (Sᵣ.Y \ {y}),
      sorry,
      (matrixSum2 Aₗ r Aᵣ c).toMatrixUnionUnion,
      inferInstance,
      inferInstance⟩
  else
    none
```

The `StandardRepr` level is more complicated. We first need to slice the last row of the matrix `Sₗ.B` and the first column of the matrix `Sᵣ.B` as the two separate vectors (`r` and `c`), naming the two remaining matrices `Aₗ` and `Aᵣ` respectively. To identify the special row and the special column (remember that matrices don't have rows and columns ordered, as much as we like to draw certain canonical ordering or rows and columns on paper for helpful visuals), we need to be given a specific element `x` in `Sₗ.X ∩ Sᵣ.X` and a specific element `y` in `Sₗ.Y ∩ Sᵣ.Y` and promised that there is no other element in any pairwise intersection among the four indexing sets. The following picture shows how `Sₗ.B` and `Sᵣ.B` are taken apart:

$$S_l.B = \begin{pmatrix} A_l \\ r \end{pmatrix} \qquad\qquad S_r.B = \begin{pmatrix} c & A_r \end{pmatrix}$$

Now we know the arguments to be given to the `Matrix` level. Again, we convert the output matrix from being indexed by `Sum` to being index by set unions, we provide a proof that the resulting standard representation has row indices and column indices disjoint, and we check whether the operation is valid — this time, the condition is that neither `r` nor `c` is a zero vector.

`Matroid` 2-sum:

```
def Matroid.IsSum2of {α : Type} (M : Matroid α) (Mₗ Mᵣ : Matroid α) :
    Prop :=
  ∃ S Sₗ Sᵣ : StandardRepr α Z2,
  ∃ x y : α,
  ∃ hXX : Sₗ.X ∩ Sᵣ.X = {x},
  ∃ hYY : Sₗ.Y ∩ Sᵣ.Y = {y},
  ∃ hXY : Disjoint Sₗ.X Sᵣ.Y,
  ∃ hYX : Disjoint Sₗ.Y Sᵣ.X,
  standardReprSum2 hXX hYY hXY hYX = some S
  ∧ S.toMatroid = M
  ∧ Sₗ.toMatroid = Mₗ
  ∧ Sᵣ.toMatroid = Mᵣ
```

The `Matroid` level is again a predicate — when $M$ is a 2-sum of $M_l$ and $M_r$.

## 4.11 *The 3-sum*

Before we dive into the formal definition of the 3-sum of matroids, let's introduce the definition informally.

Assume that the left matroid has its standard representation matrix (for some base and some ordering of rows and columns) exactly:

$B_l$ =

$A_l$ 0

1 1 0

$D_l$ $D_0$ 1

Assume that the right matroid has its standard representation matrix (for some base and some ordering of rows and columns) exactly:

$B_r$ =

1 1 0 0

$D_0$ 1

$A_r$

$D_r$

Further assume that the 2×2 matrix $D_0$ is the same submatrix in both standard representation matrices and that it is invertible. Moreover assume that the shared 3×3 submatrix is indexed by the same three base elements and the same three nonbase elements in both matroids.

We say that their 3-sum is a matroid that has the following standard representation matrix:

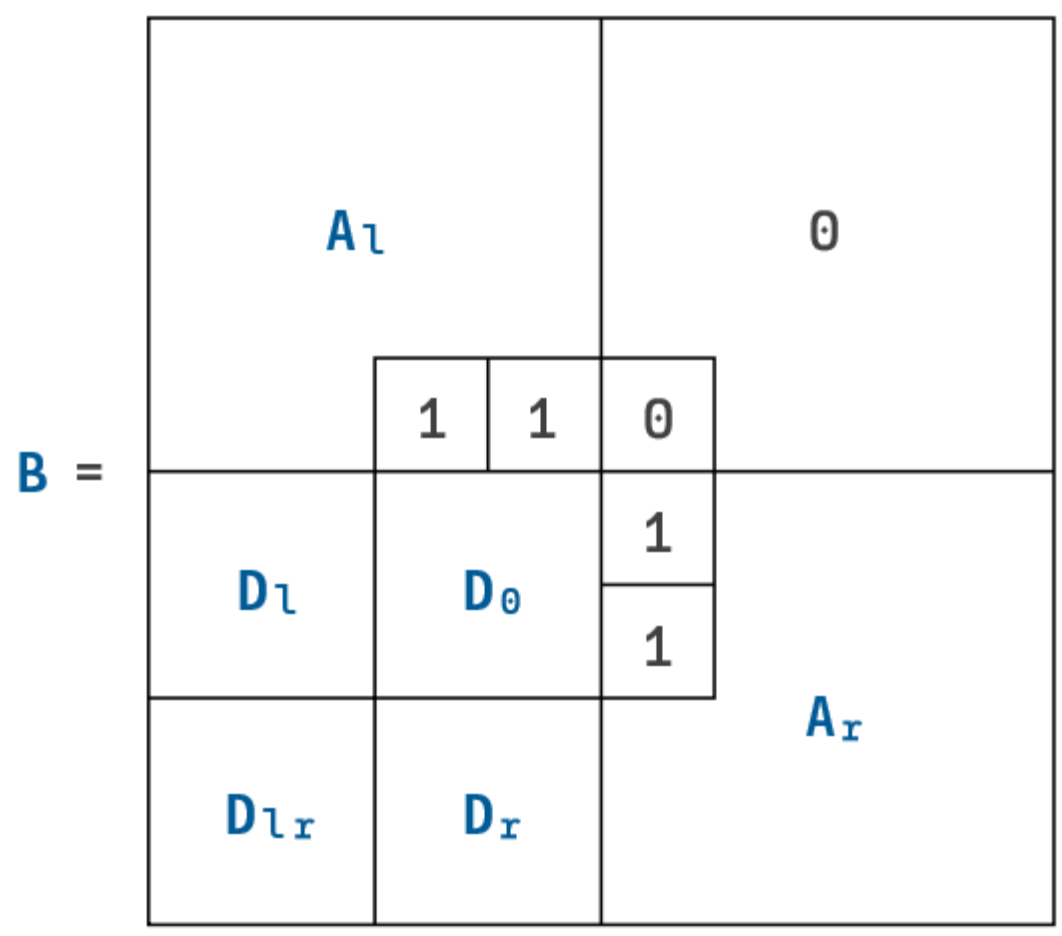

In the last picture, the not-yet-introduced part in the bottom left corner is:

$D_{lr} := D_r * D_0^{-1} * D_l$

The formal definition of the 3-sum is also implemented roughly on the three levels but much more complicated.

First, to define the 3-sum of matrices, we introduce a structure comprising the blocks of the summands:

```
structure MatrixSum3 (Xₗ Yₗ Xᵣ Yᵣ R : Type) where
  Aₗ  : Matrix (Xₗ ⊕ Unit) (Yₗ ⊕ Fin 2) R
  Dₗ  : Matrix (Fin 2) Yₗ R
  D₀ₗ : Matrix (Fin 2) (Fin 2) R
  D₀ᵣ : Matrix (Fin 2) (Fin 2) R
  Dᵣ  : Matrix Xᵣ (Fin 2) R
  Aᵣ  : Matrix (Fin 2 ⊕ Xᵣ) (Unit ⊕ Yᵣ) R
```

Our intention is that, in a valid 3-sum, the matrices $D_{0l}$ and $D_{0r}$ are equal, which is what we denote by $D_0$ in the informal definition.

We start building the 3-sum matrix from the bottom left part:

```
noncomputable abbrev MatrixSum3.D {Xₗ Yₗ Xᵣ Yᵣ R : Type} [CommRing R]
    (S : MatrixSum3 Xₗ Yₗ Xᵣ Yᵣ R) :
    Matrix (Fin 2 ⊕ Xᵣ) (Yₗ ⊕ Fin 2) R :=
  Matrix.fromBlocks S.Dₗ S.D₀ₗ (S.Dᵣ * S.D₀ₗ⁻¹ * S.Dₗ) S.Dᵣ
```

The resulting 3-sum matrix is then:

```
noncomputable def MatrixSum3.matrix {Xₗ Yₗ Xᵣ Yᵣ R : Type} [CommRing R]
    (S : MatrixSum3 Xₗ Yₗ Xᵣ Yᵣ R) :
    Matrix ((Xₗ ⊕ Unit) ⊕ (Fin 2 ⊕ Xᵣ)) ((Yₗ ⊕ Fin 2) ⊕ (Unit ⊕ Yᵣ)) R :=
  Matrix.fromBlocks S.Aₗ 0 S.D S.Aᵣ
```

Now the decomposition of the input matrices:

```
def blocksToMatrixSum3 {Xₗ Yₗ Xᵣ Yᵣ R : Type}
    (Bₗ : Matrix ((Xₗ ⊕ Unit) ⊕ Fin 2) ((Yₗ ⊕ Fin 2) ⊕ Unit) R)
    (Bᵣ : Matrix (Unit ⊕ (Fin 2 ⊕ Xᵣ)) (Fin 2 ⊕ (Unit ⊕ Yᵣ)) R) :
    MatrixSum3 Xₗ Yₗ Xᵣ Yᵣ R where
  Aₗ  := Bₗ.toBlocks₁₁
  Dₗ  := Bₗ.toBlocks₂₁.toCols₁
  D₀ₗ := Bₗ.toBlocks₂₁.toCols₂
  D₀ᵣ := Bᵣ.toBlocks₂₁.toRows₁
  Dᵣ  := Bᵣ.toBlocks₂₁.toRows₂
  Aᵣ  := Bᵣ.toBlocks₂₂
```

In our implementation, we frequently deal with sets with one, two, or three elements removed. To make our code more compact, we added abbreviations for removing one, two, and three elements from a set, as well as a definition for retyping an element of a set with three elements removed as an element of the original set:

```
abbrev Set.drop1 {α : Type} (Z : Set α) (z₀ : Z) : Set α :=
  Z \ {z₀.val}

abbrev Set.drop2 {α : Type} (Z : Set α) (z₀ z₁ : Z) : Set α :=
  Z \ {z₀.val, z₁.val}

abbrev Set.drop3 {α : Type} (Z : Set α) (z₀ z₁ z₂ : Z) : Set α :=
  Z \ {z₀.val, z₁.val, z₂.val}

def undrop3 {α : Type} {Z : Set α} {z₀ z₁ z₂ : Z} (i : Z.drop3 z₀ z₁ z₂) :
    Z :=
  ⟨i.val, i.property.left⟩
```

Furthermore, we declare a prefix operator downarrow to denote a function that ignores its (first) argument:

```
notation:max "↓"t:arg => (fun _ => t)
```

Writing `↓t` is slightly more general than writing `Function.const _ t` and, more importantly, our syntax `↓t` is much shorter and easier to read.

The input transformations from set indexing to sum indexing (to be applied to `Bₗ` and to `Bᵣ` respectively) is implemented as follows:

```
def Matrix.toBlockSummandₗ {α : Type} {Xₗ Yₗ : Set α} {R : Type}
    (Bₗ : Matrix Xₗ Yₗ R) (x₀ x₁ x₂ : Xₗ) (y₀ y₁ y₂ : Yₗ) :
    Matrix
      ((Xₗ.drop3 x₀ x₁ x₂ ⊕ Unit) ⊕ Fin 2)
      ((Yₗ.drop3 y₀ y₁ y₂ ⊕ Fin 2) ⊕ Unit)
      R :=
  Bₗ.submatrix
    (·.casesOn (·.casesOn undrop3 ↓x₂) ![x₀, x₁])
    (·.casesOn (·.casesOn undrop3 ![y₀, y₁]) ↓y₂)

def Matrix.toBlockSummandᵣ {α : Type} {Xᵣ Yᵣ : Set α} {R : Type}
    (Bᵣ : Matrix Xᵣ Yᵣ R) (x₀ x₁ x₂ : Xᵣ) (y₀ y₁ y₂ : Yᵣ) :
    Matrix
      (Unit ⊕ (Fin 2 ⊕ Xᵣ.drop3 x₀ x₁ x₂))
      (Fin 2 ⊕ (Unit ⊕ Yᵣ.drop3 y₀ y₁ y₂))
      R :=
  Bᵣ.submatrix
    (·.casesOn ↓x₂ (·.casesOn ![x₀, x₁] undrop3))
    (·.casesOn ![y₀, y₁] (·.casesOn ↓y₂ undrop3))
```

Now, to implement the 3-sums of standard representations, we also have to perform reindexing of the output matrix, to transform the dimensions of `MatrixSum3.matrix` to unions of sets:

```
def Matrix.toMatrixDropUnionDrop {α : Type}
    {Xₗ Yₗ Xᵣ Yᵣ : Set α} {R : Type}
    {x₀ₗ x₁ₗ x₂ₗ : Xₗ} {y₀ₗ y₁ₗ y₂ₗ : Yₗ}
    {x₀ᵣ x₁ᵣ x₂ᵣ : Xᵣ} {y₀ᵣ y₁ᵣ y₂ᵣ : Yᵣ}
    (A : Matrix
      ((Xₗ.drop3 x₀ₗ x₁ₗ x₂ₗ ⊕ Unit) ⊕ (Fin 2 ⊕ Xᵣ.drop3 x₀ᵣ x₁ᵣ x₂ᵣ))
      ((Yₗ.drop3 y₀ₗ y₁ₗ y₂ₗ ⊕ Fin 2) ⊕ (Unit ⊕ Yᵣ.drop3 y₀ᵣ y₁ᵣ y₂ᵣ))
      R) :
    Matrix
      (Xₗ.drop2 x₀ₗ x₁ₗ ∪ Xᵣ.drop1 x₂ᵣ).Elem
      (Yₗ.drop1 y₂ₗ ∪ Yᵣ.drop2 y₀ᵣ y₁ᵣ).Elem
      R :=
  A.submatrix
    (fun i : (Xₗ.drop2 x₀ₗ x₁ₗ ∪ Xᵣ.drop1 x₂ᵣ).Elem =>
      if hi₂ₗ : i.val = x₂ₗ then
        ◩◪0 else
      if hiXₗ : i.val ∈ Xₗ.drop3 x₀ₗ x₁ₗ x₂ₗ then
        ◩◩⟨i, hiXₗ⟩ else
      if hi₀ᵣ : i.val = x₀ᵣ then
        ◪◩0 else
      if hi₁ᵣ : i.val = x₁ᵣ then
        ◪◩1 else
      if hiXᵣ : i.val ∈ Xᵣ.drop3 x₀ᵣ x₁ᵣ x₂ᵣ then
        ◪◪⟨i, hiXᵣ⟩ else
      False.elim sorry)
    (fun j : (Yₗ.drop1 y₂ₗ ∪ Yᵣ.drop2 y₀ᵣ y₁ᵣ).Elem =>
      if hj₀ₗ : j.val = y₀ₗ then
        ◩◪0 else
      if hj₁ₗ : j.val = y₁ₗ then
        ◩◪1 else
      if hjYₗ : j.val ∈ Yₗ.drop3 y₀ₗ y₁ₗ y₂ₗ then
        ◩◩⟨j, hjYₗ⟩ else
      if hj₂ᵣ : j.val = y₂ᵣ then
        ◪◩0 else
      if hjYᵣ : j.val ∈ Yᵣ.drop3 y₀ᵣ y₁ᵣ y₂ᵣ then
        ◪◪⟨j, hjYᵣ⟩ else
      False.elim sorry)
```

Thanks to all the auxiliary definitions, we can define the 3-sum of standard representations as follows:

```lean
noncomputable def standardReprSum3 {α : Type}
    {Sₗ Sᵣ : StandardRepr α Z2} {x₀ x₁ x₂ y₀ y₁ y₂ : α}
    (_ : Sₗ.X ∩ Sᵣ.X = {x₀, x₁, x₂}) (_ : Sₗ.Y ∩ Sᵣ.Y = {y₀, y₁, y₂})
    (_ : Disjoint Sₗ.X Sᵣ.Y) (_ : Disjoint Sₗ.Y Sᵣ.X) :
    Option (StandardRepr α Z2) :=
  let x₀ₗ : Sₗ.X := ⟨x₀, sorry⟩
  let x₁ₗ : Sₗ.X := ⟨x₁, sorry⟩
  let x₂ₗ : Sₗ.X := ⟨x₂, sorry⟩
  let y₀ₗ : Sₗ.Y := ⟨y₀, sorry⟩
  let y₁ₗ : Sₗ.Y := ⟨y₁, sorry⟩
  let y₂ₗ : Sₗ.Y := ⟨y₂, sorry⟩
  let x₀ᵣ : Sᵣ.X := ⟨x₀, sorry⟩
  let x₁ᵣ : Sᵣ.X := ⟨x₁, sorry⟩
  let x₂ᵣ : Sᵣ.X := ⟨x₂, sorry⟩
  let y₀ᵣ : Sᵣ.Y := ⟨y₀, sorry⟩
  let y₁ᵣ : Sᵣ.Y := ⟨y₁, sorry⟩
  let y₂ᵣ : Sᵣ.Y := ⟨y₂, sorry⟩
  open scoped Classical in if
    ((x₀ ≠ x₁ ∧ x₀ ≠ x₂ ∧ x₁ ≠ x₂) ∧ (y₀ ≠ y₁ ∧ y₀ ≠ y₂ ∧ y₁ ≠ y₂))
    ∧ Sₗ.B.submatrix ![x₀ₗ, x₁ₗ] ![y₀ₗ, y₁ₗ] =
      Sᵣ.B.submatrix ![x₀ᵣ, x₁ᵣ] ![y₀ᵣ, y₁ᵣ]
    ∧ IsUnit (Sₗ.B.submatrix ![x₀ₗ, x₁ₗ] ![y₀ₗ, y₁ₗ])
    ∧ Sₗ.B x₀ₗ y₂ₗ = 1
    ∧ Sₗ.B x₁ₗ y₂ₗ = 1
    ∧ Sₗ.B x₂ₗ y₀ₗ = 1
    ∧ Sₗ.B x₂ₗ y₁ₗ = 1
    ∧ (∀ x : α, ∀ hx : x ∈ Sₗ.X, x ≠ x₀ ∧ x ≠ x₁ → Sₗ.B ⟨x, hx⟩ y₂ₗ = 0)
    ∧ Sᵣ.B x₀ᵣ y₂ᵣ = 1
    ∧ Sᵣ.B x₁ᵣ y₂ᵣ = 1
    ∧ Sᵣ.B x₂ᵣ y₀ᵣ = 1
    ∧ Sᵣ.B x₂ᵣ y₁ᵣ = 1
    ∧ (∀ y : α, ∀ hy : y ∈ Sᵣ.Y, y ≠ y₀ ∧ y ≠ y₁ → Sᵣ.B x₂ᵣ ⟨y, hy⟩ = 0)
  then
    some ⟨
      (Sₗ.X.drop2 x₀ₗ x₁ₗ) ∪ (Sᵣ.X.drop1 x₂ᵣ),
      (Sₗ.Y.drop1 y₂ₗ) ∪ (Sᵣ.Y.drop2 y₀ᵣ y₁ᵣ),
      sorry,
      (blocksToMatrixSum3
          (Sₗ.B.toBlockSummandₗ x₀ₗ x₁ₗ x₂ₗ y₀ₗ y₁ₗ y₂ₗ)
          (Sᵣ.B.toBlockSummandᵣ x₀ᵣ x₁ᵣ x₂ᵣ y₀ᵣ y₁ᵣ y₂ᵣ)
        ).matrix.toMatrixDropUnionDrop,
      inferInstance,
      inferInstance⟩
  else
    none
```

Finally, the `Matroid`-level predicate is defined in a familiar way:

```
def Matroid.IsSum3of {α : Type} (M : Matroid α) (Mₗ Mᵣ : Matroid α) :
    Prop :=
  ∃ S Sₗ Sᵣ : StandardRepr α Z2,
  ∃ x₀ x₁ x₂ y₀ y₁ y₂ : α,
  ∃ hXX : Sₗ.X ∩ Sᵣ.X = {x₀, x₁, x₂},
  ∃ hYY : Sₗ.Y ∩ Sᵣ.Y = {y₀, y₁, y₂},
  ∃ hXY : Disjoint Sₗ.X Sᵣ.Y,
  ∃ hYX : Disjoint Sₗ.Y Sᵣ.X,
  standardReprSum3 hXX hYY hXY hYX = some S
  ∧ S.toMatroid = M
  ∧ Sₗ.toMatroid = Mₗ
  ∧ Sᵣ.toMatroid = Mᵣ
```

The formal definition of the 3-sum is long because the 3-sum is a complicated concept.

## 4.12 *More on equiv*

Recall that an equiv is a bundled bijection (defined in Section 2.4.1). You might wonder why this section, which provides more tools for constructing equivs, comes so late in the text. It is because nothing in this section is a part of the trusted code. Definitions from this section will be used only in the proofs of regularity, mostly in the proof of regularity of the 3-sum of regular matroids. You can safely skip this section.

Identity is an equiv between every type and itself (and thus, also a permutation, even though `Equiv.Perm` isn't a part of the signature):

```
def Equiv.refl (α : Type) : α ≃ α :=
  ⟨id, id, sorry, sorry⟩
```

When we want `α` to be implicit, we can leverage our custom notation:

```
notation "=.≃" => Equiv.refl _
```

The sides of any equiv can be easily flipped:

```
def Equiv.symm {α β : Type} (e : α ≃ β) : β ≃ α :=
  ⟨e.invFun, e.toFun, e.right_inv, e.left_inv⟩
```

We can compose equivs the same way we compose functions:

```
def Equiv.trans {α β γ : Type} (e₁ : α ≃ β) (e₂ : β ≃ γ) : α ≃ γ :=
  ⟨e₂ ∘ e₁, e₁.symm ∘ e₂.symm, sorry, sorry⟩
```

If we have two equivs, we can create an equiv between respective sum types.

```
def Equiv.sumCongr {α₁ α₂ β₁ β₂ : Type} (a : α₁ ≃ α₂) (b : β₁ ≃ β₂) :
    α₁ ⊕ β₁ ≃ α₂ ⊕ β₂ :=
  ⟨Sum.map a b, Sum.map a.symm b.symm, sorry, sorry⟩
```

We also declare an abbreviation for the definition above when one of the equivs is the identity:

```
abbrev Equiv.leftCongr {α ι₁ ι₂ : Type} (e : ι₁ ≃ ι₂) : ι₁ ⊕ α ≃ ι₂ ⊕ α :=
  Equiv.sumCongr e (Equiv.refl α)
```

```
abbrev Equiv.rightCongr {α ι₁ ι₂ : Type} (e : ι₁ ≃ ι₂) : α ⊕ ι₁ ≃ α ⊕ ι₂ :=
  Equiv.sumCongr (Equiv.refl α) e
```

The type `α` is implicit here, to allow chaining the dot notation (similar to what programmers like to do in OOP).

Sometimes we have an equality on sets, and it would be nice to have an equiv between the corresponding types, even though they aren't definitionally equal (which would allow `=.≃` to be used there). Mathlib provides `Equiv.setCongr {α : Type} {s t : Set α} (_ : s = t) : s.Elem ≃ t.Elem` with the intended effect, and we equip it with a convenient notation:

```
postfix:max ".≃" => Equiv.setCongr
```

It results in the following behaviour:

```
lemma Equiv.setCongr_apply {α : Type} {s t : Set α} (hst : s = t) (a : s) :
  hst.≃ a = ⟨a.val, hst ▸ a.property⟩
```

The notation `.≃` will be used on the output of the following four lemmas:

```
variable {α : Type} {Z : Set α} {z₀ z₁ z₂ : Z}

private lemma drop3_union_pair (_ : z₀ ≠ z₂) (_ : z₁ ≠ z₂) :
    Z.drop3 z₀ z₁ z₂ ∪ {z₀.val, z₁.val} = Z.drop1 z₂

private lemma pair_union_drop3 (_ : z₀ ≠ z₂) (_ : z₁ ≠ z₂) :
    {z₀.val, z₁.val} ∪ Z.drop3 z₀ z₁ z₂ = Z.drop1 z₂

private lemma drop3_union_mem (_ : z₀ ≠ z₂) (_ : z₁ ≠ z₂) :
    Z.drop3 z₀ z₁ z₂ ∪ {z₂.val} = Z.drop2 z₀ z₁

private lemma mem_union_drop3 (_ : z₀ ≠ z₂) (_ : z₁ ≠ z₂) :
    {z₂.val} ∪ Z.drop3 z₀ z₁ z₂ = Z.drop2 z₀ z₁
```

## *4.13 Regularity of sums*

Our main results are the following three theorems…

If a finite-rank matroid is a 1-sum of regular matroids, it is a regular matroid:

```
theorem Matroid.IsSum1of.isRegular {α : Type} {M Mₗ Mᵣ : Matroid α} :
  M.IsSum1of Mₗ Mᵣ → M.RankFinite → Mₗ.IsRegular → Mᵣ.IsRegular → M.IsRegular
```

If a finite-rank matroid is a 2-sum of regular matroids, it is a regular matroid:

```
theorem Matroid.IsSum2of.isRegular {α : Type} {M Mₗ Mᵣ : Matroid α} :
  M.IsSum2of Mₗ Mᵣ → M.RankFinite → Mₗ.IsRegular → Mᵣ.IsRegular → M.IsRegular
```

If a finite-rank matroid is a 3-sum of regular matroids, it is a regular matroid:

```
theorem Matroid.IsSum3of.isRegular {α : Type} {M Mₗ Mᵣ : Matroid α} :
  M.IsSum3of Mₗ Mᵣ → M.RankFinite → Mₗ.IsRegular → Mᵣ.IsRegular → M.IsRegular
```

This design has several advantages. All assumptions meet on the level of matroids. Matroid is the central notion. It is therefore much more interesting than if we proved properties of their representations only. It has also practical advantages. The user of our library can provide matroids `Mₗ`, `Mᵣ`, and `M`, and they can use three representations for witnessing that `M` is a k-sum of `Mₗ` and `Mᵣ`, a different representation for witnessing that `M` has finite rank, and two different representations for witnessing that `Mₗ` and `Mᵣ` are regular.

We split the proof of each of these theorems into three stages corresponding to the three abstraction layers used for the definitions:

- `Matrix`
- `StandardRepr`
- `Matroid`

The three final `Matroid`-level theorems are reduced to the respective lemmas for standard representations by applying `StandardRepr.toMatroid_isRegular_iff_hasTuSigning` and `StandardRepr.finite_X_of_toMatroid_rankFinite` in all three proofs (for the 1-sum, the 2-sum, and the 3-sum). These three proofs look nearly identical.

I will now elaborate on how the `StandardRepr`-level lemmas are reduced to the `Matroid`-level lemmas because I worked on this part on my own.

Let's start with the easiest lemma:

```
lemma standardReprSum1_hasTuSigning {α : Type} {Sₗ Sᵣ S : StandardRepr α Z2}
    {hXY : Disjoint Sₗ.X Sᵣ.Y} {hYX : Disjoint Sₗ.Y Sᵣ.X}
    (hSₗ : Sₗ.B.HasTuSigning) (hSᵣ : Sᵣ.B.HasTuSigning)
    (hS : standardReprSum1 hXY hYX = some S) :
    S.B.HasTuSigning
```

First, we decompose `hSₗ` to:

```
Bₗ : Matrix Sₗ.X.Elem Sₗ.Y.Elem ℚ
hBₗ : Bₗ.IsTotallyUnimodular
hBBₗ : Bₗ.IsSigningOf Sₗ.B
```

Similarly, we decompose `hSᵣ` to:

```
Bᵣ : Matrix Sᵣ.X.Elem Sᵣ.Y.Elem ℚ
hBᵣ : Bᵣ.IsTotallyUnimodular
hBBᵣ : Bᵣ.IsSigningOf Sᵣ.B
```

Next, we establish the following facts:

```
hSX : S.X = Sₗ.X ∪ Sᵣ.X
hSY : S.Y = Sₗ.Y ∪ Sᵣ.Y
hSB : S.B = (matrixSum1 Sₗ.B Sᵣ.B).toMatrixElemElem hSX hSY
```

These three facts directly follow from `hS` using existing lemmas. The goal is still untouched:

```
S.B.HasTuSigning
```

Now, we provide the signing of the 1-sum matrix as the 1-sum of the signing matrices. Note that we cannot instantiate the existential quantifier with `matrixSum1 Bₗ Bᵣ` directly, so we use:

```
(matrixSum1 Bₗ Bᵣ).toMatrixElemElem hSX hSY
```

We have two obligations now. First, we need to prove that the matrix above is totally unimodular. Second, we need to prove that it is a signing of the matrix `S.B` from the goal. The heavy lifting for the first obligation is done by the `Matrix`-level lemma. The second obligation (after `hSB` substitution) accounts to proving:

```
((matrixSum1 Bₗ Bᵣ).toMatrixElemElem hSX hSY).IsSigningOf
((matrixSum1 Sₗ.B Sᵣ.B).toMatrixElemElem hSX hSY)
```

We prove it by case analysis. We introduce the row index `(i : S.X.Elem)` and the column index `(j : S.Y.Elem)`. We simplify the goal with `Matrix.toMatrixElemElem_apply` from Section 4.5, resulting in:

```
|matrixSum1 Bₗ Bᵣ (hSX ▸ i).toSum (hSY ▸ j).toSum| =
↑(ZMod.val (matrixSum1 Sₗ.B Sᵣ.B (hSX ▸ i).toSum (hSY ▸ j).toSum))
```

While the goal may look scary, we are calm because we have all the ingredients ready. Its proof is exactly:

```
(hSX ▸ i).toSum.casesOn
  (fun iₗ : Sₗ.X => (hSY ▸ j).toSum.casesOn (hBBₗ iₗ) ↓abs_zero)
  (fun iᵣ : Sᵣ.X => (hSY ▸ j).toSum.casesOn ↓abs_zero (hBBᵣ iᵣ))
```

The proof, as it is finished, is straightforward. My previous difficulties stemmed from working with `Matrix.toMatrixUnionUnion` directly, without the help of `Matrix.toMatrixElemElem` and related lemmas. It took me nearly a week to realize what the good approach was.

We proceed to the `StandardRepr`-level lemma for the 2-sum:

```
lemma standardReprSum2_hasTuSigning {α : Type} {Sₗ Sᵣ S : StandardRepr α Z2}
    {x y : α} {hx : Sₗ.X ∩ Sᵣ.X = {x}} {hy : Sₗ.Y ∩ Sᵣ.Y = {y}}
    {hXY : Disjoint Sₗ.X Sᵣ.Y} {hYX : Disjoint Sₗ.Y Sᵣ.X}
    (hSₗ : Sₗ.B.HasTuSigning) (hSᵣ : Sᵣ.B.HasTuSigning)
    (hS : standardReprSum2 hx hy hXY hYX = some S) :
    S.B.HasTuSigning
```

No surprises take place here. The proof is just a slightly more complicated version of what we did previously. The function `matrixSum2`, which is applied with `R = Z2` in the definition, is applied with `R = ℚ` in the proof to obtain a correct signing of the 2-sum matrix. Substitutions and case splits happen in the same places.

However, the `StandardRepr`-level lemma for the 3-sum is a very different story:

```
lemma standardReprSum3_hasTuSigning {α : Type} {Sₗ Sᵣ S : StandardRepr α Z2}
    {x₀ x₁ x₂ y₀ y₁ y₂ : α}
    {hXX : Sₗ.X ∩ Sᵣ.X = {x₀, x₁, x₂}} {hYY : Sₗ.Y ∩ Sᵣ.Y = {y₀, y₁, y₂}}
    {hXY : Disjoint Sₗ.X Sᵣ.Y} {hYX : Disjoint Sₗ.Y Sᵣ.X}
    (hSₗ : Sₗ.B.HasTuSigning) (hSᵣ : Sᵣ.B.HasTuSigning)
    (hS : standardReprSum3 hXX hYY hXY hYX = some S) :
    S.B.HasTuSigning
```

The proof of this lemma spans 1200 lines and takes nearly three million heartbeats to elaborate. The main difficulty is that `hS` doesn't tell us exactly what `S.B` looks like. We know that the central 2×2 submatrix is invertible; however, there are six possible invertible binary matrices. In order to reduce the `Matrix`-level proof from analyzing six cases to analyzing two cases, we employ the following lemma:

```
lemma Matrix.isUnit_2x2 (A : Matrix (Fin 2) (Fin 2) Z2) (_ : IsUnit A) :
  ∃ f g : Fin 2 ≃ Fin 2, A.submatrix f g = (1 0; 0 1) ∨ A.submatrix f g = (1 1; 0 1)
```

As a consequence, the `Matrix`-level proof for the 3-sum is bearable, but we pay the price of this simplification on the `StandardRepr` level, which I enthusiastically consented to work on.

In the proof (the following definition is not a part of the trusted code) we construct `MatrixSum3` terms from two standard representations using the following function:

```
private abbrev matrixSum3aux {α : Type} (Sₗ Sᵣ : StandardRepr α Z2)
    (x₀ₗ x₁ₗ x₂ₗ : Sₗ.X) (y₀ₗ y₁ₗ y₂ₗ : Sₗ.Y)
    (x₀ᵣ x₁ᵣ x₂ᵣ : Sᵣ.X) (y₀ᵣ y₁ᵣ y₂ᵣ : Sᵣ.Y) :
    MatrixSum3
      (Sₗ.X.drop3 x₀ₗ x₁ₗ x₂ₗ)
      (Sₗ.Y.drop3 y₀ₗ y₁ₗ y₂ₗ)
      (Sᵣ.X.drop3 x₀ᵣ x₁ᵣ x₂ᵣ)
      (Sᵣ.Y.drop3 y₀ᵣ y₁ᵣ y₂ᵣ)
      Z2 :=
  blocksToMatrixSum3
    (Sₗ.B.toBlockSummandₗ x₀ₗ x₁ₗ x₂ₗ y₀ₗ y₁ₗ y₂ₗ)
    (Sᵣ.B.toBlockSummandᵣ x₀ᵣ x₁ᵣ x₂ᵣ y₀ᵣ y₁ᵣ y₂ᵣ)
```

Before we can call this function inside our proof, we need to "double" each of the six special elements as two distinct terms. For example, the element `x₀` (which lies in both ground sets) becomes `x₀ₗ` when treated as a member of `Sₗ.X` and becomes `x₀ᵣ` when treated as a member of `Sᵣ.X` inside the proof. We thereby obtain twelve extra terms, and we establish all necessary inequalities between them.

From `hSₗ` and `hSᵣ` we obtain:

```
Bₗ : Matrix Sₗ.X.Elem Sₗ.Y.Elem ℚ
hBₗ : Bₗ.IsTotallyUnimodular
hSBₗ : Bₗ.IsSigningOf Sₗ.B
Bᵣ : Matrix Sᵣ.X.Elem Sᵣ.Y.Elem ℚ
hBᵣ : Bᵣ.IsTotallyUnimodular
hSBᵣ : Bᵣ.IsSigningOf Sᵣ.B
```

Decomposition of `hS` is more complicated. After a few lemma applications, we obtain `hSS` capturing the validity of the 3-sum, `hS'` characterizing what standard representation is `S` equal to, and explicit description of its indexing sets:

```
hXxxx : S.X = Sₗ.X \ {x₀, x₁} ∪ Sᵣ.X \ {x₂}
hYyyy : S.Y = Sₗ.Y \ {y₂} ∪ Sᵣ.Y \ {y₀, y₁}
```

With the help of `hSS` we call `Matrix.isUnit_2x2` to obtain the two equivs and the proof how `S.B` relates to them:

```
f g : Fin 2 ≃ Fin 2
hfg :
  !![Sₗ.B x₀ₗ y₀ₗ, Sₗ.B x₀ₗ y₁ₗ; Sₗ.B x₁ₗ y₀ₗ, Sₗ.B x₁ₗ y₁ₗ].submatrix f g
    = ⎛1  0⎞
      ⎝0  1⎠ ∨
  !![Sₗ.B x₀ₗ y₀ₗ, Sₗ.B x₀ₗ y₁ₗ; Sₗ.B x₁ₗ y₀ₗ, Sₗ.B x₁ₗ y₁ₗ].submatrix f g
    = ⎛1  1⎞
      ⎝0  1⎠
```

The case split on `f` and `g` immediately follows (each of them has two possibilities) and then `matrixSum3aux` is called accordingly (only the orderings of the arguments differ), with the four branches as follows:

```
M := matrixSum3aux Sₗ Sᵣ x₀ₗ x₁ₗ x₂ₗ y₀ₗ y₁ₗ y₂ₗ x₀ᵣ x₁ᵣ x₂ᵣ y₀ᵣ y₁ᵣ y₂ᵣ
M := matrixSum3aux Sₗ Sᵣ x₀ₗ x₁ₗ x₂ₗ y₁ₗ y₀ₗ y₂ₗ x₀ᵣ x₁ᵣ x₂ᵣ y₁ᵣ y₀ᵣ y₂ᵣ
M := matrixSum3aux Sₗ Sᵣ x₁ₗ x₀ₗ x₂ₗ y₀ₗ y₁ₗ y₂ₗ x₁ᵣ x₀ᵣ x₂ᵣ y₀ᵣ y₁ᵣ y₂ᵣ
M := matrixSum3aux Sₗ Sᵣ x₁ₗ x₀ₗ x₂ₗ y₁ₗ y₀ₗ y₂ₗ x₁ᵣ x₀ᵣ x₂ᵣ y₁ᵣ y₀ᵣ y₂ᵣ
```

In each case, we construct a so-called canonical signing (a technique invented by Ivan Sergeev to streamline Truemper's proof so that we don't have to deal with many intermediate signings), which requires many conditions to be checked on the `StandardRepr` level, with only minor alterations between the four branches.

The canonical signing is represented by a ℚ-valued matrix `B` (its precise dimensions depend on which of the four branches we view it from) accompanied with:

```
hB : B.IsTotallyUnimodular
hBM : B.IsSigningOf M.matrix
```

In order to make the rest of the proof feasible, we need to introduce another auxiliary function. `Matrix.toMatrixDropUnionDropInternal` is in fact a `Matrix.toMatrixDropUnionDrop` reimplementation. The definition starts by building two auxiliary bijections.

Equiv for row indices:

```
private def equiv₃X {α : Type} {Xₗ Xᵣ : Set α}
    {x₀ₗ x₁ₗ x₂ₗ : Xₗ} {x₀ᵣ x₁ᵣ x₂ᵣ : Xᵣ}
    (hx₀ₗ : x₁ₗ ≠ x₂ₗ) (hx₁ₗ : x₀ₗ ≠ x₂ₗ)
    (hx₀ᵣ : x₁ᵣ ≠ x₂ᵣ) (hx₁ᵣ : x₀ᵣ ≠ x₂ᵣ)
    (hx₂ᵣ : x₀ᵣ ≠ x₁ᵣ) :
    (Xₗ.drop3 x₀ₗ x₁ₗ x₂ₗ ⊕ Unit) ⊕ (Fin 2 ⊕ Xᵣ.drop3 x₀ᵣ x₁ᵣ x₂ᵣ) ≃
    (Xₗ.drop2 x₀ₗ x₁ₗ).Elem ⊕ (Xᵣ.drop1 x₂ᵣ).Elem :=
  Equiv.sumCongr
    (((equivFin1 x₂ₗ).rightCongr.trans
      (Xₗ.drop3_disjoint₂ x₀ₗ x₁ₗ x₂ₗ).equivSumUnion).trans
      (drop3_union_mem hx₁ₗ hx₀ₗ).≃)
    (((equivFin2 hx₂ᵣ).leftCongr.trans
      (Xᵣ.drop3_disjoint₀₁ x₀ᵣ x₁ᵣ x₂ᵣ).symm.equivSumUnion).trans
      (pair_union_drop3 hx₁ᵣ hx₀ᵣ).≃)
```

Equiv for column indices:

```
private def equiv₃Y {α : Type} {Yₗ Yᵣ : Set α}
    {y₀ₗ y₁ₗ y₂ₗ : Yₗ} {y₀ᵣ y₁ᵣ y₂ᵣ : Yᵣ}
    (hy₀ₗ : y₁ₗ ≠ y₂ₗ) (hy₁ₗ : y₀ₗ ≠ y₂ₗ)
    (hy₂ₗ : y₀ₗ ≠ y₁ₗ) (hy₀ᵣ : y₁ᵣ ≠ y₂ᵣ)
    (hy₁ᵣ : y₀ᵣ ≠ y₂ᵣ) :
    (Yₗ.drop3 y₀ₗ y₁ₗ y₂ₗ ⊕ Fin 2) ⊕ (Unit ⊕ Yᵣ.drop3 y₀ᵣ y₁ᵣ y₂ᵣ) ≃
    (Yₗ.drop1 y₂ₗ).Elem ⊕ (Yᵣ.drop2 y₀ᵣ y₁ᵣ).Elem :=
```

```
  Equiv.sumCongr
    (((equivFin2 hy₂ₗ).rightCongr.trans
      (Yₗ.drop3_disjoint₀₁ y₀ₗ y₁ₗ y₂ₗ).equivSumUnion).trans
      (drop3_union_pair hy₁ₗ hy₀ₗ).≃)
    (((equivFin1 y₂ᵣ).leftCongr.trans
      (Yᵣ.drop3_disjoint₂ y₀ᵣ y₁ᵣ y₂ᵣ).symm.equivSumUnion).trans
      (mem_union_drop3 hy₁ᵣ hy₀ᵣ).≃)
```

Afterwards, a new function `Matrix.toIntermediate` is defined as reindexing a given matrix by `equiv₃X` for the row index and by `equiv₃Y` for the column index; therefore, its output is a hybrid between set-based and type-based indexing (which would be bad to expose to the user, but this is all private code). Finally, we can show the second definition of the aforementioned conversion:

```
private def Matrix.toMatrixDropUnionDropInternal {α : Type}
    {Xₗ Yₗ Xᵣ Yᵣ : Set α} {R : Type}
    {x₀ₗ x₁ₗ x₂ₗ : Xₗ} {y₀ₗ y₁ₗ y₂ₗ : Yₗ}
    {x₀ᵣ x₁ᵣ x₂ᵣ : Xᵣ} {y₀ᵣ y₁ᵣ y₂ᵣ : Yᵣ}
    (A : Matrix
      ((Xₗ.drop3 x₀ₗ x₁ₗ x₂ₗ ⊕ Unit) ⊕ (Fin 2 ⊕ Xᵣ.drop3 x₀ᵣ x₁ᵣ x₂ᵣ))
      ((Yₗ.drop3 y₀ₗ y₁ₗ y₂ₗ ⊕ Fin 2) ⊕ (Unit ⊕ Yᵣ.drop3 y₀ᵣ y₁ᵣ y₂ᵣ))
      R)
    (hx₀ₗ : x₁ₗ ≠ x₂ₗ) (hx₁ₗ : x₀ₗ ≠ x₂ₗ)
    (hx₀ᵣ : x₁ᵣ ≠ x₂ᵣ) (hx₁ᵣ : x₀ᵣ ≠ x₂ᵣ)
    (hx₂ᵣ : x₀ᵣ ≠ x₁ᵣ) (hy₀ₗ : y₁ₗ ≠ y₂ₗ)
    (hy₁ₗ : y₀ₗ ≠ y₂ₗ) (hy₂ₗ : y₀ₗ ≠ y₁ₗ)
    (hy₀ᵣ : y₁ᵣ ≠ y₂ᵣ) (hy₁ᵣ : y₀ᵣ ≠ y₂ᵣ) :
    Matrix
      (Xₗ.drop2 x₀ₗ x₁ₗ ∪ Xᵣ.drop1 x₂ᵣ).Elem
      (Yₗ.drop1 y₂ₗ ∪ Yᵣ.drop2 y₀ᵣ y₁ᵣ).Elem
      R :=
  (A.toIntermediate
    hx₀ₗ hx₁ₗ hx₀ᵣ hx₁ᵣ hx₂ᵣ
    hy₀ₗ hy₁ₗ hy₂ₗ hy₀ᵣ hy₁ᵣ
  ).toMatrixUnionUnion
```

Understanding what `Matrix.toMatrixDropUnionDrop` does is much easier than reading `Matrix.toMatrixDropUnionDropInternal` with its dependencies and checking that it does the thing it is supposed to do. The extra conditions on elements being distinct are not the main reason why `Matrix.toMatrixDropUnionDropInternal` was excluded from the trusted code — these conditions would need to be satisfied either way. The main reason against it is the length of the definition (counted with its dependencies) and the difficulty of understanding what those intricate constructions involving `Equiv.sumCongr` and `Equiv.trans` actually do. While `Matrix.toMatrixDropUnionDropInternal` was excluded from the trusted code, it has been preserved in the 3-sum file as a private definition because it is very useful for proving `standardReprSum3_hasTuSigning` here. The reason is compositionality. By calling `Matrix.toMatrixUnionUnion` on the outermost layer, we can use convenient lemmas such as `Matrix.IsTotallyUnimodular.toMatrixElemElem` in the subsequent parts of the proof. By implementing `equiv₃X` and `equiv₃Y` as a composition of equivs, we make it possible to inject the swapping between `x₀` and `x₁` and/or the swapping between `y₀` and `y₁` on the innermost layer.

The price of having both conversion functions `Matrix.toMatrixDropUnionDrop` and `Matrix.toMatrixDropUnionDropInternal` in the code is that we needed to prove their equality (circa 100 lines) so that, while `Matrix.toMatrixDropUnionDrop` is in the 3-sum definition in the trusted code, we can rewrite `Matrix.toMatrixDropUnionDrop` to `Matrix.toMatrixDropUnionDropInternal` and continue the proof in the convenient way. We believe that it is a reasonable price to pay for having the trusted code cleaner.

After expressing `hS'` in terms of `Matrix.toMatrixDropUnionDropInternal`, our proof continues depending on which of the four branches we are in. In the first branch, `hB` and `hBM` are almost exactly what suffices to close the goal with our `.toMatrixElemElem` lemmas. In the remaining three branches, adjusting `hB` is easy but adjusting `hBM` requires a lot of work with `Equiv` compositions. This endgame is long but pretty straightforward (unless we flinch when `HEq` appears).

I will not elaborate on the `Matrix`-level proofs of regularity — the proofs that the resulting signed matrix is totally unimodular are too complicated. Nevertheless, I would like to mention a proof technique that helped us. In some proofs, we worked with large case splits, with up to 896 cases. To handle such situations, we wrote `all_goals try` followed by one or more tactics, discharging multiple goals at once without selecting them by hand or repeating the proof. We repeatedly applied this method to discharge the remaining goals in waves until the proof was complete. On paper (or in a language that has less proof automation than Lean has) we would need to search for more symmetries and smarter arguments to prove the same lemmas.

## *4.14 Graphic matroids*

We say that a vector is an *incidence vector* iff it is either a zero vector, or it has exactly one `+1` entry, exactly one `-1` entry, and `0` on all remaining positions:

```
def IsIncidenceMatrixColumn {m : Type} (v : m → ℚ) : Prop :=
  v = 0 ∨
  ∃ i₁ i₂ : m,
    i₁ ≠ i₂ ∧ v i₁ = 1 ∧ v i₂ = -1 ∧
    ∀ i : m, i ≠ i₁ → i ≠ i₂ → v i = 0
```

A matrix is a *node-edge incidence matrix of a directed graph* iff all of its columns are incidence vectors:

```
def Matrix.IsGraphic {m n : Type} (A : Matrix m n ℚ) : Prop :=
  ∀ y : n, IsIncidenceMatrixColumn (A · y)
```

A matroid is *graphic* iff it can be represented by a node-edge incidence matrix of a directed graph:

```
def Matroid.IsGraphic {α : Type} (M : Matroid α) : Prop :=
  ∃ X Y : Set α, ∃ A : Matrix X Y ℚ, A.IsGraphic ∧ A.toMatroid = M
```

All graphic matroids are regular:

```
theorem Matroid.IsGraphic.isRegular {α : Type} {M : Matroid α}
    (_ : M.IsGraphic) :
    M.IsRegular
```

The proof of this theorem easily follows from properties of totally unimodular matrices.

## 4.15 Cographic matroids

Mathlib defines a *dual* matroid as a matroid whose independent sets are those subsets of the ground sets that are disjoint from some base:

```
def Matroid.dualIndepMatroid {α : Type} (M : Matroid α) :
    IndepMatroid α where
  E := M.E
  Indep I := I ⊆ M.E ∧ ∃ B : Set α, M.IsBase B ∧ Disjoint I B
  indep_empty := sorry
  indep_subset := sorry
  indep_aug := sorry
  indep_maximal := sorry
  subset_ground := sorry
```

```
def Matroid.dual {α : Type} (M : Matroid α) : Matroid α :=
  M.dualIndepMatroid.matroid
```

A matroid is *cographic* iff its dual is graphic:

```
def Matroid.IsCographic {α : Type} (M : Matroid α) : Prop :=
  M.dual.IsGraphic
```

In our near future, we would like to prove that all cographic matroids are regular. The proof will probably follow the outline[22] by Cameron Rampell.

## 4.16 Matroid R10

The matroid R10 is a specific matroid that plays a special role in Seymour's theorem [116] [117]. We define R10 via the following (binary) standard representation matrix:

$$\begin{pmatrix} 1 & 0 & 0 & 1 & 1 \\ 1 & 1 & 0 & 0 & 1 \\ 0 & 1 & 1 & 0 & 1 \\ 0 & 0 & 1 & 1 & 1 \\ 1 & 1 & 1 & 1 & 1 \end{pmatrix}$$

We prove by reflection (i.e., a formally verified brute-force procedure) that the matroid R10 is regular. We furthermore prove that `Matroid.mapEquiv` (i.e., matroid isomorphisms) preserve regularity (it holds generally, but we will need it to the matroid R10 only). The last two results are labelled as `simp` lemmas, so we will not have to call them explicitly.

[22] https://github.com/cappucher/Cographic-Matroids

## 4.17 *Statement of Seymour's theorem*

In order to state Seymour's theorem in the strongest possible sense, we first need to define the class of *good* matroids:

```
inductive Matroid.IsGood {α : Type} : Matroid α → Prop
| graphic {M : Matroid α} (_ : M.IsGraphic) : M.IsGood
| cographic {M : Matroid α} (_ : M.IsCographic) : M.IsGood
| isomorphicR10 {M : Matroid α} {e : α ≃ Fin 10}
    (_ : M.mapEquiv e = matroidR10.toMatroid) : M.IsGood
| is1sum {M Mₗ Mᵣ : Matroid α} (_ : M.IsSum1of Mₗ Mᵣ) (_ : M.RankFinite)
    (_ : Mₗ.IsGood) (_ : Mᵣ.IsGood) : M.IsGood
| is2sum {M Mₗ Mᵣ : Matroid α} (_ : M.IsSum2of Mₗ Mᵣ) (_ : M.RankFinite)
    (_ : Mₗ.IsGood) (_ : Mᵣ.IsGood) : M.IsGood
| is3sum {M Mₗ Mᵣ : Matroid α} (_ : M.IsSum3of Mₗ Mᵣ) (_ : M.RankFinite)
    (_ : Mₗ.IsGood) (_ : Mᵣ.IsGood) : M.IsGood
```

The first three constructors are "leaf constructors". They enumerate that graphic matroids, cographic matroids, and matroids isomorphic to R10 are all good matroids. The next three constructors are "fork constructors". They declare that the class of good matroids is closed under the 1-sum, the 2-sums, and the 3-sum. The ad-hoc terminology ("fork" and "leaf") is based on the image of the inductive type forming a (rooted) tree.

The final theorem is then stated as follows:

```
theorem Matroid.RankFinite.isRegular_iff_isGood {α : Type}
    {M : Matroid α} (_ : M.RankFinite) :
    M.IsRegular ↔ M.IsGood
```

WE DON'T HAVE A PROOF OF THIS THEOREM !!!!!!!

The only thing we provide in `HardDirection.lean` is the theorem statement above, which can be paraphrased as:

"A finite-rank matroid is regular iff it can be decomposed into graphic matroids & cographic matroids & matroids isomorphic to R10 using 1-sums & 2-sums & 3-sums."

Let me repeat: `Matroid.RankFinite.isRegular_iff_isGood` isn't proved in our project! I don't want to give a false impression of what has been completed in the Seymour project. The project is actually really far from proving Seymour's theorem.

## 4.18 *Related work*

In Lean 4, the largest library formalizing matroid theory is due to Peter Nelson[23]. It implements matroids that may be infinite, following Bruhn et al. [122], together with many key notions and results about them. The definition that is fully formalized and is the most related to our work is `Matroid.disjointSum` (a sum of two matroids). For binary matroids, this definition is equivalent to the 1-sum implemented in our repository. Moreover, `Matroid.disjointSum` can be used for any matroids with disjoint ground sets, while our implementation is restricted to vector matroids constructed from `Z2` matrices. Peter Nelson's repository also makes progress towards formalizing other related notions, such as representable matroids, though this work is still ongoing. It is also worth noting that the results in Mathlib[24] have been copied over from this repository and comprise a strict subset of it.

[23] https://github.com/apnelson1/lean-matroids

Building upon Peter Nelson's work, Alena Gusakov's thesis [123] formalized the proof of Tutte's excluded minor theorem and, to this end, implemented definitions and results about representable matroids. Gusakov formalized representations and standard representations of matroids, which we also do in our work, but it takes a different approach. In particular, instead of working with matrix representations, Gusakov implemented a representation of `Matroid α` as a mapping from the entire type `α` to a vector space, which maps non-elements of the matroid to the zero vector and independent sets to linearly independent vectors. The advantage of this approach is that certain proofs become easier to formalize, but it comes at a cost of making it harder to match the implementation with the theory and believe the correctness of the code.

There are also two Lean 3 repositories due to Artem Vasilyev[25] and Bryan Gin-ge Chen[26] dedicated to formalization of matroid theory. Both of them work with finite matroids following Oxley [116] and implement basic definitions and properties of matroids concerning circuits, bases, and rank functions. Their results are completely subsumed by the current implementation of matroids in Mathlib [1].

Jonas Keinholz [124] formalized the classical definition of finite matroids [116] [117] in Isabelle/HOL along with other basic ideas such as minors, bases, circuits, rank, and closure.

More recently, Wan et al. [125] used Keinholz's formalization to design a verification framework using a Locale that checks whether a given collection of subsets of a given set is a matroid. The authors then showcased the verification algorithm by checking that the 0-1 knapsack problem does not conform to the matroid structure, while the fractional knapsack problem does. In comparison, Lean 4's Mathlib implements a more general definition of matroids and formalizes more results about them than either Keinholz [124] or Wan et al. [125], but Lean lacks a procedure for formally verifying if a collection of sets has matroid structure. In the HOL Light GitHub repository[27], John Harrison formalized finitary matroids. The formalization closely follows the field theory notes of Pete L. Clark[28]. In particular, finitary matroids are defined in terms of a closure operator with similar properties as those proposed by Bruhn et al. [122]. This repository also includes a formal proof that this notion of (finitary) matroids is equivalent to the definition of a matroid using independent sets. Unlike Lean 4's Mathlib formalization (which includes formalizations of the closure operator and the notions of spanning sets), this notion of infinite matroids does not respect the notion of duality that is defined for matroids in Oxley [116] and Truemper [117] as noted by Bruhn et al. [122].

Grzegorz Bancerek and Yasunari Shidama [126] formalized matroids in Mizar. Their formalization includes basic notions like rank, base, and cycle as well as examples like the matroid of linearly independent subsets for a given vector space. Overall, the scope of the Mizar formalization is comparable to the Isabelle/HOL formalization, except that the Mizar formalization allows for infinite matroids. In this sense, it is comparable to the Lean definition in Mathlib, which also allows for infinite matroids. However, while Mizar uses independence conditions to define matroids, Lean uses base conditions for the main definition and provides an API for constructing matroids via independence conditions.

---

[24] https://github.com/leanprover-community/mathlib4/tree/master/Mathlib/Combinatorics/Matroid

[25] https://github.com/VArtem/lean-matroids

[26] https://github.com/bryangingechen/lean-matroids

[27] https://github.com/jrh13/hol-light/blob/master/Library/matroids.ml

[28] https://plclark.github.io/PeteLClark/Expositions/FieldTheory.pdf

### *4.19 Conclusion*

In this work, we formally stated Seymour's decomposition theorem for regular matroids and implemented a formally verified proof of the forward (composition) direction of this theorem in the setting where the matroids have a finite rank but may have infinite ground sets. To this end, we developed a modular and extensible library in Lean 4 formalizing definitions and lemmas about totally unimodular matrices, vector matroids, regular matroids, and the 1-sum, the 2-sum, and the 3-sums of matroids.

Our work demonstrates that people can effectively use Lean and Mathlib to formally verify advanced results from matroid theory and extend classical results to a more general setting.

The most natural continuation of our project is proving the decomposition direction of Seymour's theorem, stated as `Matroid.IsRegular.isGood_of_rankFinite` in our library. Given that the decomposition direction is much more difficult than the composition direction, a potential future work could be split into three papers:

1. Formally verified splitter theorem and its corollaries
2. Formally verified Kuratowski's theorem
3. Formally verified Seymour's theorem (all the remaining parts)

Before continuing, however, it would be worth updating our repository to the current Lean version and the newest stable Mathlib. It would allow us to use powerful new tactics such as `canonical` and `grind`, and more lemmas about matroids would become available to us. Also, more AI-powered tools could be usable with the new Lean versions.

# 5 Theory of grammars

> The oldest, shortest words — "yes" and "no" — are those which require the most thought.
>
> — Pythagoras (allegedly)

The notion of *formal languages* (also known as *decision problems*) lies at the heart of computer science. There are several formalisms that recognize formal languages, including Turing machines and grammars [127]. In particular, both Turing machines and general grammars (also called type-0 grammars or unrestricted grammars) are known to characterize the same class of languages, namely, the recursively enumerable or type-0 languages.

A grammar is a structured way to describe how strings in a language can be formed. It does so by introducing two kinds of symbols; terminal symbols, which appear in the final strings of the language, and nonterminal symbols, which represent intermediate stages or larger components of structure. A grammar provides rewriting rules that declare how each nonterminal (or a string containing at least one nonterminal symbol) can be expanded into a sequence of terminals and nonterminals. Beginning with a designated start symbol, one repeatedly applies these rules to generate strings. In this way, a grammar gives a clear and systematic description of the syntax of a language—it specifies exactly which strings are allowed and how they can be constructed.

To illustrate what grammars do, consider the following grammar with an initial nonterminal symbol `S`, a terminal alphabet consisting of `()[]{}`, and the following rewriting rules:

```
S → SS
S →
S → (S)
S → [S]
S → {S}
```

This grammar is designed to recognize the set of all correct bracketings. Example derivation:

```
S → SS → (S)S → ()S →
(){S} →
(){SS} →
(){S[S]} →
(){S[]} →
(){[S][]} →
(){[][]}
```

We observe that the word `(){[][]}` is correctly bracketed, and so is any other word (list of terminals) generated by this grammar (this sentence appeals to the reader's intuition for what is and what isn't a correctly bracketed string; we don't have any definition that our grammar should adhere to). This type of grammar is called a context-free grammar because the LHS of every rule contains exactly one nonterminal (named "context-free" because one can match a nonterminal symbol without regard for its context, i.e., surroundings), and the set of all words generated by a context-free grammar is called a context-free language.

In the example above, `S` was the only nonterminal symbol (there always has to be at least one nonterminal symbol because the initial symbol must be nonterminal), but there can be several nonterminal symbols. For example, consider the following grammar, where the initial nonterminal `S` is only one of many nonterminal symbols, over the terminal alphabet `abc` intended to represent the multiplication of natural numbers (#`a` times #`b` equals #`c` in this order):

S → LR
L → aLX
R → BR
L → M
R → E
XB → BCX
CB → BC
XC → CX
XE → E
MB → bM
M → K
KC → cK
KE →

For example, two times three equals six:

S → LR → aLXR → aaLXXR → aaLXXBR → aaLXXBBR → aaLXXBBBR → aaMXXBBBR →
aaMXXBBBE →
aaMXBCXBBE →
aaMBCXCXBBE →
aaMBCXCBCXBE →
aaMBCXCBCBCXE →
aaMBCXBCCBCXE →
aaMBCXBCBCCXE →
aaMBCXBBCCCXE →
aaMBCBCXBCCCXE →
aaMBCBCBCXCCCXE →
aaMBBCCBCXCCCXE →
aaMBBCBCCXCCCXE →
aaMBBBCCCXCCCXE →
aaMBBBCCCCXCCXE →
aaMBBBCCCCCXCXE →
aaMBBBCCCCCCXXE →
aaMBBBCCCCCCXE →
aaMBBBCCCCCCE →
aabMBBCCCCCCE →
aabbMBCCCCCCE →
aabbbMCCCCCCE →
aabbbKCCCCCCE →
aabbbcKCCCCCE →
aabbbccKCCCCE →
aabbbcccKCCCE →
aabbbccccKCCE →
aabbbcccccKCE →
aabbbccccccKE →
aabbbcccccc

Rewriting ends precisely when only terminals are left. Casually speaking, the initial `S` always becomes `LR` in the first step, then we have any number of `a` on the left with the same number of `X` after `L`, then we have any number of `B`, after which `R` replaced by `E` ends the string, afterwards each `X` bubbles to the right while generating extra `C` every time `X` passes `B`, while all `B` are pushed to the left of all `C`, then `X` also bubbles to the right through all `C`, then the final `E` annihilates every `X`, then `M` transforms every `B` to the terminal `b`, then `K` transforms every `C` to the terminal `c`, finally `KE` gets removed at the end, leaving some amount of `a`, followed by some amount of `b`, followed by the amount of `c` that is equal to the product of the previous two amounts.

The grammar above represents an extremely inefficient algorithm for multiplying natural numbers because it has cubic complexity in the value of the larger factor.

Grammars are inherently nondeterministic in the matter of what rule is applied when and on which position; a word belongs to the language generated by a given grammar iff there is a sequence of choices that results in rewriting the initial symbol to a string containing only the desired terminal symbols.

Grammars are similar to L-systems [128]; in particular, context-free grammars exhibit the greatest resemblance; however, L-systems rewrite all symbols simultaneously, whereas grammars always apply their rules locally, allowing one part of the string to grow faster than other parts. As a consequence, the word generated by an L-system is determined solely by the rewriting rules and the number of iterations, whereas grammars allow many nondeterministic choices to be made when generating a word. One could argue that, in both cases, the number of words generated by a given system is countable, so there shouldn't be such a big difference; however, general grammars are capable of emulating L-systems, but L-systems cannot emulate grammars.

Grammars are also similar to semi-Thue systems, also known as string rewriting systems [129]. The key difference is that grammars distinguish terminal symbols from nonterminal symbols, whereas semi-Thue systems have only one collection of symbols, and the moment when the generation ends isn't uniquely determined.

Our project formalizes the following things in Lean 4:

- We define general grammars and context-free grammars.
- We prove four closure properties of type-0 languages.
    - Union
    - Reversal
    - Concatenation
    - Kleene star
- We prove three closure properties of context-free languages.
    - Union
    - Reversal
    - Concatenation

## 5.1 Languages

A *word* is a list of characters (the only difference between a list and a word is that, in case of a "word", the type of elements is informally called an *alphabet*; there is no formal difference). A *language* is a set of words over the same alphabet:

```
def Language (α : Type) := Set (List α)
```

We can ask whether a word belongs to a language the same way we ask whether a term belongs to a set.

The *union* of two languages is denoted by the symbol + (for reasons that will be soon apparent):

```
theorem Language.add_def {α : Type} (l m : Language α) :
    l + m = (l ∪ m : Set (List α))
```

The *concatenation* of two languages (the language that consists of all words that can be split into two parts, the first of which belongs to the first language, the second of which belongs to the second language) is denoted by the symbol * (we will soon see why):

```
def Set.image2 {α β γ : Type} (f : α → β → γ) (s : Set α) (t : Set β) :=
  { c : γ | ∃ a ∈ s, ∃ b ∈ t, f a b = c }
```

```
theorem Language.mul_def {α : Type} (l m : Language α) :
    l * m = Set.image2 (· ++ ·) l m
```

The reason behind the notation is that languages form a semiring. In this semiring, addition is union, multiplication is concatenation, "zero" is the empty language, and "one" is the language that contains only the empty word. Mathlib proves that all conditions of a semiring are satisfied.

The *reverse* of a language is the set of its words backwards:

```
def Language.reverse {α : Type} (l : Language α) : Language α :=
  { w : List α | w.reverse ∈ l }
```

The last operation on languages we will examine is more complicated. The *Kleene star* of a language is the language that consists of all words that can be split into any number of parts, each of which is a word from the original language:

```
lemma Language.kstar_def {α : Type} (l : Language α) :
    KStar.kstar l =
    { x : List α | ∃ L : List (List α), x = L.flatten ∧ ∀ y ∈ L, y ∈ l }
```

Note that `L` can be empty; therefore, every language has the empty word in its Kleene star.

## 5.2 Chomsky hierarchy

When studying properties of natural languages, Noam Chomsky proposed a classification of grammars according to their expressive power, now known as the Chomsky hierarchy. The key idea is that some grammars are more powerful than others in the sense that they can describe more complex patterns.

Noam Chomsky arranged them into levels based on how restrictive their rewriting rules are. At the most restrictive level, grammars capture only the simplest patterns. At the most permissive level, grammars can express any enumerable language (i.e., any language where a computer could be programmed to check whether a given word belongs to the language — in the positive case, the computer has to say "yes" in a finite time, whereas in the negative case, the computer must never say "yes" — saying "no" explicitly isn't required though). Ordered from the most general to the most restrictive:

- General grammars (a.k.a. unrestricted grammars, a.k.a. type-0 grammars, a.k.a. recursively enumerable grammars, a.k.a. phrase-structure grammars, a.k.a. grammars) allow any rewriting rules. The class of languages they generate is known as type-0 or (recursively) enumerable languages.

- Context-sensitive grammars allow any length of LHS, but only a single symbol from the LHS can be changed in the RHS, and it must be a nonterminal symbol that changes (but it can be replaced by multiple symbols). The class of languages they generate is called context-sensitive languages. The same class of languages can be described by essentially-noncontracting grammars (a.k.a. monotonic grammars), which allow any rules whose RHS is at least as long as their LHS, with the only exception that, if the initial symbol doesn't appear on any RHS, there may be a rule that rewrites the initial symbol to the empty string (which is how languages that contain the empty string are included in this class).
- Context-free grammars allow only rules whose LHS is a single nonterminal. The class of languages they generate is called context-free languages.
- Regular grammars (a.k.a. right-regular grammars) allow only rules that rewrite (LHS) a single nonterminal to (RHS) either the empty string, a single terminal, or a single terminal followed by a single nonterminal. The class of languages they generate is called regular languages. Regular grammars are very rarely used to speak about regular languages. Much more often, finite automata or regular expressions are used to describe regular languages.

Assuming that all terminals are lowercase letters and all nonterminals are uppercase letters, which is a common convention in computer science, we outline the Chomsky hierarchy using example rules that are allowed in some classes of grammars.

Rules that are allowed in all classes, including regular grammars:

```
A → aA
A → cB
```

Rules that are allowed in context-free grammars (and higher) but not in regular grammars:

```
A → aBc
A → abcA
A → Aabc
A → abcAabc
A → BB
A → BCDE
A → aBCDe
A → BeeeeeeeC
A → aAeAa
```

Rules that are allowed in context-sensitive (and general) grammars but not in context-free (and regular) grammars:

```
ABC → AEC
ABC → AeC
ABC → AeeC
ABC → AabcdeFGHabcdeC
ABC → aaBC
ABC → ABccFe
AB → AD
AB → EB
AA → aaaA
```

```
AA → AeBCD
ABCDE → ABCCDE
aB → ac
abcdE → abcde
ABCDe → ABcDe
eAAAAAm → eAAaBBcAAm
```

Rules that are allowed in general grammars but not in context-sensitive grammars (and lower):

```
AB → C
ABC → ae
ABC → AC
aBc → B
AA → A
Ae → A
Ae → c
abcD → ee
abcD → aDE
```

Out of the aforementioned classes, we will define only the type-0 and context-free languages. In both definitions, we will refer to the concept of a *reflexive&transitive closure*:

```
inductive ReflTransGen {α : Type} (r : α → α → Prop) (a : α) : α → Prop
  | refl : ReflTransGen r a a
  | tail {b c : α} : ReflTransGen r a b → r b c → ReflTransGen r a c
```

### 5.2.1 General grammars

Symbols are essentially defined as a sum type of terminals `T` and nonterminals `N`. However, we want to refer to terminals and nonterminals by constructor name (using `Symbol.terminal` and `Symbol.nonterminal` instead of `Sum.inl` and `Sum.inr` respectively), so we define symbols as an inductive type:

```
inductive Symbol (T N : Type)
  | terminal    (t : T) : Symbol T N
  | nonterminal (n : N) : Symbol T N
deriving
  DecidableEq, Repr, Fintype
```

We don't require `T` and `N` to be finite. As a result, we don't need to copy the typeclass instances `[Fintype T]` and `[Fintype N]` alongside our type parameters (which would appear in almost every lemma statement). Instead, later we work in terms of a list of rewriting rules, which is finite by definition and from which we could infer that only a finite set of terminals and a finite set of nonterminals can occur.

The LHS of a general rewriting rule consists of three parts (an arbitrary string, a nonterminal, and another arbitrary string), but the RHS is represented by a single string:

```
structure Grule (T N : Type) where
  inputL : List (Symbol T N)
  inputN : N
  inputR : List (Symbol T N)
```

```
  output : List (Symbol T N)
```

For example, a rule that rewrites `AxyB` to `yBx` can be expressed in two equivalent ways:

- `⟨[A, x, y], ._B, [], [y, B, x]⟩`
- `⟨[], ._A, [x, y, B], [y, B, x]⟩`

Their equivalence stems from how rewriting rules are applied, which we will learn in a minute.

An advantage of the representation above is that we don't need to carry the proposition "LHS contains a nonterminal" around. A disadvantage is that we subsequently need to concatenate more terms. The non-uniqueness could also be considered to be a disadvantage, but we don't care about uniqueness (in fact, we don't even forbid two definitionally equal rewriting rules to be part of the same grammar).

A definition of a general grammar follows. Notice that only the type argument `T` is part of its type signature:

```
structure Grammar (T : Type) where
  nt : Type
  initial : nt
  rules : List (Grule T nt)
```

The next line adds an implicit type argument `T` to all declarations that come after:

```
variable {T : Type}
```

The following definition captures the application of a rewriting rule:

```
def Grammar.Transforms (g : Grammar T) (w₁ w₂ : List (Symbol T g.nt)) :
    Prop :=
  ∃ r : Grule T g.nt,
    r ∈ g.rules ∧
    ∃ u v : List (Symbol T g.nt),
      w₁ = u ++ r.inputL ++ [Symbol.nonterminal r.inputN] ++ r.inputR ++ v
    ∧ w₂ = u ++ r.output ++ v
```

We can view `Grammar.Transforms` as a function that takes a grammar `g` over the terminal type `T` and outputs a binary relation over strings of the type that `g` works internally with.

The derivation relation is defined from `Grammar.Transforms` using the reflexive & transitive closure:

```
def Grammar.Derives (g : Grammar T) :
    List (Symbol T g.nt) → List (Symbol T g.nt) → Prop :=
  Relation.ReflTransGen g.Transforms
```

Consequently, proofs about derivations will use structural induction.

Words generated by a grammar are exactly those lists of terminals that can be derived from the initial symbol:

```
def Grammar.language (g : Grammar T) : Language T :=
  { w : List T | g.Derives [Symbol.nonterminal g.initial]
                           (w.map Symbol.terminal) }
```

Finally, we define the class of type-0 languages ("grammar-generated languages") as follows:

```
def Language.IsGG (L : Language T) : Prop :=
  ∃ g : Grammar T, g.language = L
```

All top-level theorems about type-0 languages are expressed in terms of the `Language.IsGG` predicate.

We remarked that some rules can be expressed in two (or perhaps more) equivalent ways. If we wanted the encoding of rules to be unique, we could have accomplished that, for example, by requiring that `inputL` contains terminals only:

```
structure Grule' (T N : Type) where
  inputL : List T
  inputN : N
  inputR : List (Symbol T N)
  output : List (Symbol T N)

structure Grammar' (T : Type) where
  nt : Type
  initial : nt
  rules : List (Grule' T nt)

def Grammar'.Transforms (g : Grammar' T) (w₁ w₂ : List (Symbol T g.nt)) :
    Prop :=
  ∃ r : Grule' T g.nt,
    r ∈ g.rules ∧
    ∃ u v : List (Symbol T g.nt),
      w₁ = u ++ r.inputL.map Symbol.terminal ++
           [Symbol.nonterminal r.inputN] ++ r.inputR ++ v ∧
      w₂ = u ++ r.output ++ v
```

The definitions `Grule'`, `Grammar'`, and `Grammar'.Transforms` serve only for illustrative purposes and cannot be found in any repository. We also didn't prove that they would lead to the same class of languages in the end; you will have to believe our claim without evidence (or consider it sufficiently unimportant to not care about it).

### 5.2.2 Context-free grammars

Since the context-free rewriting rules are just a single nonterminal on the LHS and any string on the RHS, we don't declare a special type for them, and we directly define the context-free grammar:

```
structure CFG (T : Type) where
  nt : Type
  initial : nt
  rules : List (nt × List (Symbol T nt))
```

The finiteness conditions are not written. The reason is the same as with general grammars.

From now on we have:

```
variable {T : Type}
```

One step of context-free rewriting is defined as follows:

```
def CFG.Transforms (g : CFG T) (w₁ w₂ : List (Symbol T g.nt)) : Prop :=
  ∃ r : g.nt × List (Symbol T g.nt),
    r ∈ g.rules ∧
    ∃ u v : List (Symbol T g.nt),
      w₁ = u ++ [Symbol.nonterminal r.fst] ++ v ∧ w₂ = u ++ r.snd ++ v
```

Any number of steps of context-free rewriting is defined as follows:

```
def CFG.Derives (g : CFG T) :
    List (Symbol T g.nt) → List (Symbol T g.nt) → Prop :=
  Relation.ReflTransGen g.Transforms
```

The language generated by a context-free grammar is defined as follows:

```
def CFG.language (g : CFG T) : Language T :=
  { w : List T | g.Derives [Symbol.nonterminal g.initial]
                 (w.map Symbol.terminal) }
```

Finally, we define the class of context-free languages:

```
def Language.IsCF (L : Language T) : Prop :=
  ∃ g : CFG T, g.language = L
```

All of our theorems about context-free languages are expressed in terms of the `Language.IsCF` predicate.

One basic theorem we prove is that context-free languages are a subclass of type-0 languages:

```
theorem CF_subclass_GG (L : Language T) :
    L.IsCF → L.IsGG
```

## 5.3 *Closure properties of general grammars*

Our main result from the theory of grammars is that we formally verified four closure properties of general grammars. We found that closure under union and closure under reversal were straightforward to prove, whereas we had to invest considerable effort to prove closure under concatenation and closure under Kleene star.

### 5.3.1 Union

In this subsection, we prove the following theorem:

```
theorem GG_of_GG_u_GG {T : Type} (L₁ : Language T) (L₂ : Language T) :
    L₁.IsGG ∧ L₂.IsGG → (L₁ + L₂).IsGG
```

Its proof consists of three main ingredients:

(1) a construction of a new grammar `g` from any two given grammars $g_1$ and $g_2$
(2) a proof that any word generated by $g_1$ or $g_2$ can also be generated by `g`
(3) a proof that any word generated by `g` can be equally generated by $g_1$ or $g_2$

Proofs of the other closure properties are organized analogously. We describe the proof of closure under union in more detail; it allows us to outline the main ideas of proving closure properties formally in a simple setting. Since (3) usually turns out to be much more difficult than (1) and (2), we refer to (2) as the "easy direction" and to (3) as the "hard direction".

The proof of the closure of type-0 languages under union follows the standard construction, which usually states only (1) explicitly, and leaves (2) and (3) to the reader. Proving (2) is easy because we can choose when is each new rule applied and when the original rules are applied. This comfort is not available when proving (3) because `g` can apply its rules in any order. It is up to us to come up with an invariant that `g` preserves and sufficiently restricts what is generated once only terminal symbols are there. In case of the construction for union, it is fortunately straightforward.

Note that nothing from the code in the rest of this subsection is a part of the trusted code, even though the upcoming definitions and lemmas aren't marked as private (they are public because we employ them in several files, not because we want the reader to pay attention to them).

The new grammar `g := unionGrammar g₁ g₂` is constructed as follows:

```
def liftSymbol {N N₀ T : Type} (f : N₀ → N) : Symbol T N₀ → Symbol T N
  | Symbol.terminal t => Symbol.terminal t
  | Symbol.nonterminal n => Symbol.nonterminal (f n)

def liftString {N N₀ T : Type} (f : N₀ → N) :
    List (Symbol T N₀) → List (Symbol T N) :=
  List.map (liftSymbol f)

def liftRule {N N₀ T : Type} (f : N₀ → N) : Grule T N₀ → Grule T N :=
  fun r : Grule T N₀ => Grule.mk
    (liftString f r.inputL)
    (f r.inputN)
    (liftString f r.inputR)
    (liftString f r.output)

def unionGrammar {T : Type} (g₁ g₂ : Grammar T) : Grammar T :=
  Grammar.mk (Option (g₁.nt ⊕ g₂.nt)) none (
    ⟨[], none, [], [Symbol.nonterminal (some ◩g₁.initial)]⟩ :: (
    ⟨[], none, [], [Symbol.nonterminal (some ◪g₂.initial)]⟩ :: (
    g₁.rules.map (liftRule (some ∘ Sum.inl)) ++
    g₂.rules.map (liftRule (some ∘ Sum.inr)))))
```

To illustrate how the construction works, consider two grammars over the alphabet made of all lowercase Latin letters. The first grammar has only one rewriting rule:

```
S → hello
```

The second grammar has two rewriting rules:

```
S → aS
S →
```

The new grammar, after local substitutions `S₁ := some ◩S`, `S₂ := some ◪S`, `S := none`, has the following rewriting rules:

```
S → S₁
S → S₂
S₁ → hello
S₂ → aS₂
S₂ →
```

Three example derivations (still written with our substitutions) performed by the new grammar:

```
S → S₁ → hello
S → S₂ → aS₂ → a
S → S₂ → aS₂ → aaS₂ → aaaS₂ → aaa
```

We need to prove that if $g_1$ generates $L_1$ and $g_2$ generates $L_2$ then (unionGrammar $g_1$ $g_2$) generates ($L_1 + L_2$).

To reduce the amount of repeated code in the proof, we developed lemmas that allow us to "lift" a grammar with a certain type of nonterminals to a grammar with a larger type of nonterminals while preserving what the grammar derives. Under certain conditions, we can also "sink" the larger grammar to the original grammar and preserve its derivations. The word "lift" doesn't refer to lifting in logic. The word "sink" doesn't refer to the kitchen wash basin.

To this end, we will need, in addition to the lifting functions defined above, the following two functions:

```
def sinkSymbol {N N₀ T : Type} (f : N → Option N₀) : Symbol T N →
      Option (Symbol T N₀)
  | Symbol.terminal t => some (Symbol.terminal t)
  | Symbol.nonterminal n => Option.map Symbol.nonterminal (f n)

def sinkString {N N₀ T : Type} (f : N → Option N₀) :
    List (Symbol T N) → List (Symbol T N₀) :=
  List.filterMap (sinkSymbol f)
```

The whole enterprise of lifting and sinking is organized around the following structure:

```
structure LiftedGrammar (T : Type) where
  g₀: Grammar T
  g : Grammar T
  liftNt : g₀.nt → g.nt
  sinkNt : g.nt → Option g₀.nt
  lift_inj : liftNt.Injective
  sink_inj : ∀ x y : g.nt, sinkNt x = sinkNt y → x = y ∨ sinkNt x = none
  sinkNt_liftNt : ∀ n₀ : g₀.nt, sinkNt (liftNt n₀) = some n₀
  corresponding_rules :
    ∀ r : Grule T g₀.nt,
      r ∈ g₀.rules → liftRule liftNt r ∈ g.rules
  preimage_of_rules :
    ∀ r : Grule T g.nt,
      (r ∈ g.rules ∧ ∃ n₀ : g₀.nt, liftNt n₀ = r.inputN) →
        (∃ r₀ ∈ g₀.rules, liftRule liftNt r₀ = r)
```

Thanks to this structure, we can abstract from the specifics of how the larger grammar is constructed in concrete proofs and care only about the properties that are required to follow analogous derivations. In particular, we will alternate between instantiating $g_0$ with $g_1$ and instantiating $g_0$ with $g_2$ in the proof for union.

For the rest of this subsection, we will have:

```
variable {T : Type}
```

The first lemma about `LiftedGrammar` we need to prove is that one step of general grammar transformation can be lifted:

```
lemma lift_tran {G : LiftedGrammar T} {w₁ w₂ : List (Symbol T G.g₀.nt)}
    (_ : G.g₀.Transforms w₁ w₂) :
    G.g.Transforms (liftString G.liftNt w₁) (liftString G.liftNt w₂)
```

We start the proof by unpacking the assumption into the following parts:

```
r : Grule T G.g₀.nt
rin : r ∈ G.g₀.rules
u v : List (Symbol T G.g₀.nt)
bef : w₁ = u ++ r.inputL ++ [Symbol.nonterminal r.inputN] ++ r.inputR ++ v
aft : w₂ = u ++ r.output ++ v
```

We use the rule `liftRule G.liftNt r` and from `G.corresponding_rules r rin` we justify that it exists. The parts of the string that aren't matched in the rewriting are instantiated with `liftString G.liftNt u` and `liftString G.liftNt v` respectively. The remaining goals are discharged essentially by wrapping `bef` and `aft` in `liftString G.liftNt` both.

Subsequently, we prove that any number of steps of general grammar transformation can be lifted:

```
lemma lift_deri (G : LiftedGrammar T) {w₁ w₂ : List (Symbol T G.g₀.nt)}
    (_ : G.g₀.Derives w₁ w₂) :
    G.g.Derives (liftString G.liftNt w₁) (liftString G.liftNt w₂)
```

We prove it by inductive application of the previous lemma.

Before we prove lemmas about `LiftedGrammar` for the hard direction, we need two auxiliary definitions:

```
def GoodLetter {G : LiftedGrammar T} : Symbol T G.g.nt → Prop
  | Symbol.terminal _ => True
  | Symbol.nonterminal n => ∃ n₀ : G.g₀.nt, G.sinkNt n = n₀

def GoodString {G : LiftedGrammar T} (s : List (Symbol T G.g.nt)) : Prop :=
  ∀ a ∈ s, GoodLetter a
```

We are now ready for going from the bigger grammar to the smaller grammar:

```
lemma sink_tran {G : LiftedGrammar T} {w₁ w₂ : List (Symbol T G.g.nt)}
    (_ : G.g.Transforms w₁ w₂) (_ : GoodString w₁) :
    G.g₀.Transforms (sinkString G.sinkNt w₁) (sinkString G.sinkNt w₂) ∧
    GoodString w₂
```

The proof is much more complicated than other proofs regarding `LiftedGrammar`. As usual, the first step is to unpack the main assumption into:

```
r : Grule T G.g.nt
rin : r ∈ G.g.rules
u v : List (Symbol T G.g.nt)
bef : w₁ = u ++ r.inputL ++ [Symbol.nonterminal r.inputN] ++ r.inputR ++ v
aft : w₂ = u ++ r.output ++ v
```

With a bit of work, from `G.preimage_of_rules r` we obtain (note that `GoodString w₁` is needed here) the following terms:

```
r₀ : Grule T G.g₀.nt
pre_in : r₀ ∈ G.g₀.rules
preimage : liftRule G.liftNt r₀ = r
```

As for the goal, let's focus on the second conjunct `GoodString w₂` first.

The relationship between `w₁` and `w₂` is provided by `bef` and `aft` together. With a help from `preimage`, we reduce the local goal to:

```
∀ a ∈ (liftRule G.liftNt r₀).output, GoodLetter a
```

We essentially just need to perform a case analysis and apply `G.sinkNt_liftNt` in the end.

Now let's attack the main goal:

```
G.g₀.Transforms (sinkString G.sinkNt w₁) (sinkString G.sinkNt w₂)
```

We use the rule `r₀` and let `pre_in` justify its existence. Now it shouldn't come as a surprise that the nonmatched parts of the string are `sinkString G.sinkNt u` for the prefix and `sinkString G.sinkNt v` for the suffix. We observe:

```
correct_inverse : sinkSymbol G.sinkNt ∘ liftSymbol G.liftNt = Option.some
```

Now let's prove:

```
sinkString G.sinkNt w₁ =
sinkString G.sinkNt u ++ r₀.inputL ++ [Symbol.nonterminal r₀.inputN] ++
                         r₀.inputR ++ sinkString G.sinkNt v
```

The intuition to wrap `bef` in `sinkString G.sinkNt` screams loudly here. Unfortunately, after we substitute `preimage` and perform (mostly manual) simplifications, we are still left with the following mismatches:

```
r₀.inputL ?= (liftString G.liftNt r₀.inputL).filterMap (sinkSymbol G.sinkNt)
```

```
r₀.inputR ?= (liftString G.liftNt r₀.inputR).filterMap (sinkSymbol G.sinkNt)
```

```
[Symbol.nonterminal r₀.inputN] ?=
[Symbol.nonterminal r₀.inputN].map (liftSymbol G.liftNt) |>.filterMap
    (sinkSymbol G.sinkNt)
```

Fortunately, `correct_inverse` together with two standard lemmas `List.filterMap_map` and `List.filterMap_some` discharges all three goals.

Finally let's prove:

```
sinkString G.sinkNt w₂ =
sinkString G.sinkNt u ++ r₀.output ++ sinkString G.sinkNt v
```

A simplified version of the trick with `correct_inverse` we did above does the job here as well.

Having established what happens when sinking a single rewriting step, we can proceed to the final lemma about `LiftedGrammar`, which should be pretty much expected at this point:

```
lemma sink_deri (G : LiftedGrammar T) {w₁ w₂ : List (Symbol T G.g.nt)}
    (_ : G.g.Derives w₁ w₂) (_ : GoodString w₁) :
    G.g₀.Derives (sinkString G.sinkNt w₁) (sinkString G.sinkNt w₂)
```

To prove it, we inductively apply the previous lemma and forget `GoodString w₂` at the end.

### 5.3.2 Reversal

In this subsection, we prove the following theorem:

```
theorem GG_of_reverse_GG {T : Type} (L : Language T) :
    L.IsGG → L.reverse.IsGG
```

The proof is very easy. Simply speaking, everything gets reversed. We start with the rewriting rules:

```
variable {T : Type}

private def reversalGrule {N : Type} (r : Grule T N) : Grule T N :=
  Grule.mk r.inputR.reverse r.inputN r.inputL.reverse r.output.reverse
```

The new grammar is constructed as follows:

```
private def reversalGrammar (g : Grammar T) : Grammar T :=
  Grammar.mk g.nt g.initial (g.rules.map reversalGrule)
```

For example, the opening grammar, which was used for correct bracketing, would look after our reversing as follows:

```
S → SS
S →
S → )S(
S → ]S[
S → }S{
```

The rest is essentially a repeated application of the lemmas `List.reverse_append_append` and `List.reverse_reverse` until the proof is finished. It was really easy — the entire proof has less than 100 lines.

### 5.3.3 Concatenation

Another theorem we prove is the closure under concatenation:

```
theorem GG_of_GG_c_GG {T : Type} (L₁ : Language T) (L₂ : Language T) :
    L₁.IsGG ∧ L₂.IsGG → (L₁ * L₂).IsGG
```

We do not comment on how the proof works.

### 5.3.4 Kleene star

In this subsection, we prove the following theorem:

```
theorem GG_of_star_GG {T : Type} (L : Language T) :
    L.IsGG → (KStar.kstar L).IsGG
```

Again, nothing else in this subsection is a part of the trusted code, so you can stop reading here.

Closure of recursively enumerable languages under the Kleene star is demonstrated in folklore by a hand-waving argument about a two-tape nondeterministic Turing machine. The language to be iterated is given by a single-tape (nondeterministic) Turing machine. The new machine scans the input on the first tape while copying it onto the second tape as it progresses, and nondeterministically chooses where the first word ends. Next, the original machine is simulated on the second tape. If the simulated machine accepts the word on the second tape, the process is repeated with the current position of the first head instead of returning to the beginning of the input. Finally, when the first head reaches the end of the input, the second tape contains a suffix of the first tape. The original machine is simulated on the second tape for the last time. If it accepts, the new machine accepts.

I think that Gabriele Röger[29] is wrong when she claims that the proof that type-0 languages are closed under the Kleene star is similar to context-free closure properties.

Unfortunately, we didn't find any proof based on grammars. Therefore, we had to invent a new construction. Informal description of our proof of the closure of grammar-generated languages under the Kleene star can be found in our paper [3]. This text continues with an exposition of the formal proof:

```
private def nn (N : Type) : Type :=
  N ⊕ Fin 3
```

```
private abbrev ns (T N : Type) : Type :=
  Symbol T (nn N)
```

Here `nn` serves as a new nonterminal type and `ns` serves as a new symbol type (that will be used by the "bigger" grammar). We continue with special symbols (constants), where `S` is just a shortcut to refer to the original grammar's initial nonterminal:

```
variable {T : Type}
```

```
private def Z {N : Type} : ns T N :=
  Symbol.nonterminal ◪0
```

```
private def H {N : Type} : ns T N :=
  Symbol.nonterminal ◪1
```

```
private def R {N : Type} : ns T N :=
  Symbol.nonterminal ◪2
```

```
private def S {g : Grammar T} : ns T g.nt :=
  Symbol.nonterminal ◩g.initial
```

[29] https://ai.dmi.unibas.ch/_files/teaching/fs19/theo/slides/theory-c08.pdf (slide 19)

The new grammar expands the nonterminal type with three additional nonterminals:

- a new starting symbol `Z`
    - when referring to `Z` as a nonterminal rather than a symbol, it is `◪0`
- a delimiter `H`
    - when referring to `H` as a nonterminal rather than a symbol, it is `◪1`
- a marker `R` for final rewriting
    - when referring to `R` as a nonterminal rather than a symbol, it is `◪2`

`Sum.inl` prefixes nonterminals of the original grammar.

Given a grammar `g` that generates a language `L`, we construct the following grammar (about which we prove that it generates the language `KStar.kstar L`):

```
private def wrapSym {N : Type} : Symbol T N → ns T N :=
  liftSymbol Sum.inl

private def wrapGr {N : Type} : Grule T N → Grule T (nn N) :=
  liftRule Sum.inl

def asTerminal {N : Type} : Symbol T N → Option T
  | Symbol.terminal t => some t
  | Symbol.nonterminal _ => none

def allUsedTerminals (g : Grammar T) : List T :=
  (g.rules.map Grule.output).flatten.filterMap asTerminal

private def rulesThatScanTerminals (g : Grammar T) :
    List (Grule T (nn g.nt)) :=
  (allUsedTerminals g).map (fun t : T =>
      Grule.mk [] ◪2 [Symbol.terminal t] [Symbol.terminal t, R])

private def Grammar.star (g : Grammar T) : Grammar T :=
  Grammar.mk (nn g.nt) ◪0 (
    Grule.mk [] ◪0 [] [Z, S, H] :: (
    Grule.mk [] ◪0 [] [R, H] :: (
    Grule.mk [] ◪2 [H] [R] :: (
    Grule.mk [] ◪2 [H] [] :: (
    g.rules.map wrapGr ++
    rulesThatScanTerminals g)))))
```

Intuitively, `Z` can generate any amount of `S`, where `H` builds compartments that isolate the words from the language `L`, and then `R` acts as a cleaner that traverses the string from beginning to end and removes the compartment delimiters `H`, thereby ensuring that only terminals are present to the left of `R`.

We illustrate the construction on a simple example (which could be a context-free grammar, but we model it as a general grammar to illustrate what really happens in our construction). Consider a grammar with an initial (and the only) nonterminal `S`, terminal alphabet consisting of `ab`, and the following rewriting rules:

```
S → aSb
S →
```

The new grammar has the same terminal alphabet `ab`, but its nonterminal symbols are **`ZHRS`**, out of which **`Z`** is the new initial nonterminal, and its rewriting rules are:

```
Z → ZSH
Z → RH
RH → R
RH →
S → aSb
S →
Ra → aR
Rb → bR
```

The new grammar can derive a word for example as follows:

```
Z →
ZSH →
ZSHSH →
ZaSbHSH →
ZaaSbbHSH →
ZSHaaSbbHSH →
ZaSbHaaSbbHSH →
ZaSbHaaaSbbbHSH →
ZaSbHaaabbbHSH →
RHaSbHaaabbbHSH →
RaSbHaaabbbHSH →
aRSbHaaabbbHSH →
aRbHaaabbbHSH →
abRHaaabbbHSH →
abRaaabbbHSH →
abaRaabbbHSH →
abaaRabbbHSH →
abaaaRbbbHSH →
abaaabRbbHSH →
abaaabRbbHaSbH →
abaaabbRbHaSbH →
abaaabbRbHabH →
abaaabbbRHabH →
abaaabbbRabH →
abaaabbbaRbH →
abaaabbbabRH →
abaaabbbab
```

The easy direction is proved by splitting the generation into two phases, each of which has an easy proof by induction.

However, to prove the hard direction, we had to come up with a sophisticated invariant:

```
private lemma star_induction {g : Grammar T} {α : List (ns T g.nt)}
    (_ : g.star.Derives [Z] α) :
  (∃ x : List (List (Symbol T g.nt)),
    (∀ xᵢ ∈ x, g.Derives [Symbol.nonterminal g.initial] xᵢ) ∧
    α = [Z] ++ ((x.map (List.map wrapSym)).map (· ++ [H])).flatten) ∨
  (∃ x : List (List (Symbol T g.nt)),
    (∀ xᵢ ∈ x, g.Derives [Symbol.nonterminal g.initial] xᵢ) ∧
    α = [R, H] ++ ((x.map (List.map wrapSym)).map (· ++ [H])).flatten) ∨
  (∃ w : List (List T), ∃ β : List T, ∃ γ : List (Symbol T g.nt),
   ∃ x : List (List (Symbol T g.nt)),
    (∀ wᵢ ∈ w, wᵢ ∈ g.language) ∧
    g.Derives [Symbol.nonterminal g.initial] (β.map Symbol.terminal ++ γ) ∧
    (∀ xᵢ ∈ x, g.Derives [Symbol.nonterminal g.initial] xᵢ) ∧
    α = w.flatten.map Symbol.terminal ++ β.map Symbol.terminal
        ++ [R] ++ γ.map wrapSym ++ [H] ++
        ((x.map (List.map wrapSym)).map (· ++ [H])).flatten) ∨
  (∃ u : List T, u ∈ KStar.kstar g.language ∧ α = u.map Symbol.terminal) ∨
  (∃ σ : List (Symbol T g.nt), α = σ.map wrapSym ++ [R]) ∨
  (∃ ω : List (ns T g.nt), α = ω ++ [H]) ∧ Z ∉ α ∧ R ∉ α
```

We will not explain how `star_induction` is proved, but we will illustrate how the disjuncts appear (or not) in the concrete example where `abaaabbbab` is derived.

In the example above, the first case arises when `α = ZaaSbbHSH`. We can check that setting `x = [aaSbb, S]` makes it hold that:

```
(∀ xᵢ ∈ x, g.Derives [S] xᵢ) ∧
α = [Z] ++ ((x.map (List.map wrapSym)).map (· ++ [H])).flatten
```

In the example above, the second case arises when `α = RHaSbHaaabbbHSH`. We can check that setting `x = [aSb, aaabbb, S]` makes it hold that:

```
(∀ xᵢ ∈ x, g.Derives [S] xᵢ) ∧
α = [R, H] ++ ((x.map (List.map wrapSym)).map (· ++ [H])).flatten
```

In the example above, the third case arises when `α = abaaabRbbHaSbH`. We can check that setting `w = [ab]`, `β = aaab`, `γ = bb`, `x = [aSb]` makes it hold that:

```
(∀ wᵢ ∈ w, wᵢ ∈ g.language) ∧
g.Derives [S] (β.map Symbol.terminal ++ γ) ∧
(∀ xᵢ ∈ x, g.Derives [S] xᵢ) ∧
α = w.flatten.map Symbol.terminal ++ β.map Symbol.terminal
    ++ [R] ++ γ.map wrapSym ++ [H] ++
    ((x.map (List.map wrapSym)).map (· ++ [H])).flatten
```

The fourth case arises only at the end of a successful computation, which is `α = abaaabbbab` in the example above.

The remaining two cases do not arise in the example above because they describe an unsuccessful computation (like taking a one-way street ending in a blind alley). It is essential that the invariant includes these two cases.

The penultimate case arises if the rule `RH → R` is used in the final position (where `RH →` should be used instead). The nonterminal `R` in the final position prevents the derivation from terminating.

The last case arises if the rule `RH →` is used too early (that is, anywhere but the final `H` position). The nonterminal `H` in the final position during the absence of `R` and `Z` in `α` prevents the derivation from terminating.

Proving that the invariant is preserved spans 1827 lines of Lean 4 code (it is 3204 lines in the Lean 3 version, which is very spaciously formatted).

## 5.4 *Closure properties of context-free grammars*

We prove the following three theorems about context-free languages:

```
theorem CF_of_reverse_CF {T : Type} (L : Language T) :
    L.IsCF → L.reverse.IsCF
```

```
theorem CF_of_CF_u_CF {T : Type} (L₁ : Language T) (L₂ : Language T) :
    L₁.IsCF ∧ L₂.IsCF → (L₁ + L₂).IsCF
```

```
theorem CF_of_CF_c_CF {T : Type} (L₁ : Language T) (L₂ : Language T) :
    L₁.IsCF ∧ L₂.IsCF → (L₁ * L₂).IsCF
```

The closure under reversal is proved directly, using the same idea as we used in the closure of type-0 languages under reversal, which resulted in a very short proof.

The closure under union and the closure under concatenation are proved by observing that our constructions for general grammars never create a rule with more than one symbol on the LHS unless the input grammar has a rule with more than one symbol on the LHS. The translation of the general results to context-free results is easy.

Unfortunately, the proof of the closure of type-0 languages under the Kleene star adds rules with two symbols on the LHS regardless of the input. Therefore, our proof cannot be reused to prove the closure of context-free languages under the Kleene star. However, there exists an easier construction for context-free languages that could be formalized separately if desired.

## 5.5 *Related work*

To our knowledge, no one has formalized general grammars before. Context-free grammars were formalized by Carlson et al. [130] in Mizar, by Minamide [131] in Isabelle/HOL, by Barthwal and Norrish [132] in HOL4, by Firsov and Uustalu [133] in Agda, and by Ramos [134] in Rocq.

Finite automata have often been subjected to verification. In particular, Thompson and Dillies [1] formalized finite automata, which recognize regular languages, in Lean. Thomson [1] also formalized regular expressions, which recognize regular languages as well.

There is ample verification work also for other models of computation:

- Turing machines were formalized in Mizar [135], Matita [136], Isabelle/HOL [137], Lean [138], Rocq [139], and recently again Isabelle/HOL [140]. Out of them, the most impressive development is probably the last one, by Balbach. It uses multi-tape Turing machines and culminates with a proof of the Cook-Levin theorem, which states that SAT is `NP`-complete.

- The λ-calculus was formalized by Norrish [141] in HOL4 and later by Forster, Kunze, and their colleagues [142] [143] [144] [145] [146] [147] using Rocq. The latter group of authors proposed the untyped call-by-value λ-calculus as a convenient basis for computability and complexity theory because it naturally supports compositionality.
- The partial recursive functions were formalized by Norrish [141] in HOL4 and by Carneiro [138] in Lean.
- Random access machines were formalized by Coen [148] in Rocq.

## 5.6 Conclusion

We defined general grammars and context-free grammars in Lean and used them to establish selected closure properties of type-0 languages and of context-free languages. Despite the tedium of some of the proofs, we believe that grammars are probably a more convenient formalism than Turing machines for showing closure properties of language classes. On the other hand, since grammars cannot define any of the important complexity classes (such as P), formalization of Turing machines and other computational models is needed to further develop the formal theory of computer science.

As future work, results about context-sensitive and regular grammars could be incorporated into our library. A comprehensive Lean library encompassing the entire Chomsky hierarchy would be valuable. Mathlib [1] results about automata could be connected to our library. We have already added the definition of regular languages to Mathlib, but have not yet proved that all regular languages are context-free. As a more ambitious goal, we might attempt to prove the equivalence between general grammars and Turing machines. Unfortunately, even though Mathlib [1] long defines Turing machines, the community hasn't yet agreed on a definition of the language recognized by a Turing machine.

# 6 Conclusions

> "Hofstadter's Law: It always takes longer than you expect,
> even when you take into account Hofstadter's Law." [149]

We manifested truth and beauty in three areas of MathematiCS (optimization theory, matroid theory, and the theory of grammars). While doing so, we have learnt a lot about Lean and its libraries, as well as about said areas of interest. Those areas of interest have been the main focus of this text; Lean is a communication[30] medium rather than a focus of the thesis. The text emphasizes the trusted code of said projects and only occasionally mentions other parts.

The most valuable contribution of my Ph.D., however, does not lie in the libraries we produced or the papers we wrote, but the countless small contributions to Mathlib [1]. Often it is the smallest PRs that have the greatest impact on the user experience of future students and researchers who will build projects on top of Mathlib. Every time I was proving a lemma that made me wonder "why doesn't such a basic thing exist yet" I attempted to adapt the lemma for the style of Mathlib (which often meant generalizing it beyond the version I needed for my work) and open a PR. Adding new definitions to Mathlib is more complicated — and rightfully so — new definitions can easily happen to bring negative value for Mathlib. Only rarely did I add new definitions to Mathlib. I was generally trying to add as many lemmas as I could while trying to avoid PRs that would stir too much controversy.

When I talk about small and easy PRs with big impact, a good example is my recent PR[31] that added pigeonhole-like results for `Fin` to Mathlib. It might seem that adding the theorem

```
theorem Fin.le_of_injective {m n : ℕ}
    (f : Fin m → Fin n) (_ : f.Injective) :
    m ≤ n
```

has no value for Mathlib, as the older theorem

```
theorem Fintype.card_le_of_injective {α β : Type} [Fintype α] [Fintype β]
    (f : α → β) (_ : f.Injective) :
    Fintype.card α ≤ Fintype.card β
```

already subsumes it. However, consider what happens when a beginner imports Mathlib and writes the following code (which is pretty realistic because beginners seldom work in a fully general setting):

```
example (m n : ℕ) (f : Fin m → Fin n) (_ : f.Injective) :
    m ≤ n := by
  hint
```

Nothing happens! To be more precise, before merging my PR, `hint` would report a failure. Although `hint` is a very powerful tactic, which calls `exact?` among other tactics, no lemma is found. Lean doesn't know that `m` equals `Fintype.card (Fin m)`, which would be necessary to see through to match `Fintype.card_le_of_injective`'s output. When asked explicitly, Lean would of course provide a lemma for `m = Fintype.card (Fin m)`, but none of the tactics called by `hint` chains library searches in such a way. As a result, adding a "redundant" theorem `Fin.le_of_injective` saves beginners from a lot of frustration. At the same time, this blank spot is usually overlooked by experienced Lean users, who form the vast majority of Mathlib contributors. Adding such lemmas is a net positive for the Mathlib community.

[30] In [152], Freek Wiedijk wrote: "A formalization is completely useless for communicating the mathematics that is formalized in it." I hereby hope that Freek Wiedijk has been proved wrong, or rather, that his desire for formalized mathematics that is human-readable at the same time has been answered. Indeed, in 2007, Lean didn't exist yet, thus Freek Wiedijk couldn't know how ergonomic and how graceful the go-to system for formalized mathematics would be less than 20 years later.

[31] https://github.com/leanprover-community/mathlib4/pull/26400

During my Ph.D., I also collaborated on community projects downstream of Mathlib, but my contributions there were rather small.

- On 2024-06-21, Floris van Doorn launched[32] a project[33] to formalize the generalized Carleson's theorem (for a generalized Carleson operator on doubling metric measure spaces). In simple terms, the goal was to establish an as-large-as-possible class of functions such that, when you perform the Fourier transform and then the Fourier synthesis, you obtain the original function almost everywhere. I contributed[34] by finishing very easy parts of a few proofs and by refactoring existing code. I'd say the best way to describe my contributions is to say that I was helping with little things that didn't require understanding the big picture. This project is not a part of my thesis.
- On 2024-09-25, Terence Tao launched[35] a project[36] to explore the space of single-equation theories of general magmas, ordered by implications (with `x = y` being the bottom and `x = x` being the top). The project focuses on the 4694 single-equation theories that involve at most four magma operations, up to symmetry and renaming, which give rise to 22 million potential implications to be proven or disproven. More than 50 people contributed to the project. The success of the project results from the synergy between manual mathematical reasoning and automated theorem proving. In the end, all proofs and counterexamples (both human-made and computer-generated) were formally verified in Lean 4. I contributed[37] by developing API for magma homomorphisms, by providing one manual counterexample, by improving notation, by refactoring existing code, by reporting bugs in the frontend, and by fixing typos in the manuscript. This project is not a part of my thesis.

## *6.1 Is MathematiCS a religion?*

As I was working on the presented projects, I was repeatedly asking myself about whether MathematiCS should be considered to be a kind of religion. I found myself unable to dismiss the idea outright. The more I formalized, the more the analogy pressed itself upon me. My own experience of doing MathematiCS in Lean felt closer to obsessive devotion than to empirical investigation.

[32] Announcement: https://leanprover.zulipchat.com/#narrow/channel/113486-announce/topic/Carleson.20project

[33] https://florisvandoorn.com/carleson/

[34] https://github.com/fpvandoorn/carleson/commits?author=madvorak

[35] Announcement: https://terrytao.wordpress.com/2024/09/25/a-pilot-project-in-universal-algebra-to-explore-new-ways-to-collaborate-and-use-machine-assistance/

[36] https://teorth.github.io/equational_theories/

[37] https://github.com/teorth/equational_theories/commits?author=madvorak

In the sociological and institutional sense, MathematiCS is not a religion. MathematiCS has no gods or sacred revelation. No ecclesiastical authority has oversight over what Mathematicians may and may not do. MathematiCS doesn't make existential promises about the afterlife or cosmic justice. From the sociological and institutional point of view, MathematiCS can look more like science than religion — we write papers, we engage in peer review, and we travel to conferences.

However, in the philosophical and phenomenological sense, MathematiCS shares deep affinities with religion. We receive a foundational framework and explore its consequences, much like theology interprets scriptures. MathematiCS is not about the physical world but about a realm of ideal objects — numbers, spaces, structures, the concept of change, possibility and necessity — "eternal" truths that don't depend on nature. Many mathematicians describe MathematiCS as a form of transcendence — a contact with something vast, orderly, and outside space and time. And even though the mathematical community generally welcomes newcomers, there are long apprenticeships, initiations into technical methods, shared doctrines, and the unmistakable presence of an elite guild of "those who understand" as opposed to those who cannot see yet.

In religion, practitioners interpret sacred texts; in MathematiCS, we interpret the consequences of our own chosen premises. In both settings, the authority resides not in observation but in fidelity to a framework, and mastery comes from long initiations rather than casual insight. Both disciplines cultivate a devotion to something unseen but deeply felt — a structure of meaning that shapes perception, guides practice, and gives purpose beyond the immediate world. Just as the worshipper in ritual ecstasy begins to speak in tongues, we begin to speak in symbols — our utterances seem opaque to outsiders but, to the initiated ones, they serve as a medium for contact with something larger than ourselves.

## *6.2 Mathematical novelty*

Some people ask me whether, in addition to formalizing known mathematical results, my thesis brings any new mathematical results. By design, no; however, as a byproduct of our work, a few new results have been written:

- In Chapter 3, `extendedFarkas` and `ValidELP.strongDuality` are new mathematical results, though not very interesting.
- In Chapter 4, `standardReprSum3_hasTuSigning` is new in the sense of not requiring any finiteness. This generalization of known results is not surprising, but to the best of our knowledge, it has never been published before.
- In Chapter 5, `GG_of_star_GG` had been a folklore theorem when our work was being conducted.

## *6.3 Reflections on truth and beauty*

As our long journey draws to a close, what stands before us isn't just a list of theorems but a transformation of thought. What began as a search for truth and beauty has become an encounter with their unity — a realization that, in the precise language of Lean, both truth and beauty are found — not beyond rigor, but within it.

While Vladimir Nabokov [150] famously portrayed beauty as a seductive force that can lure us away from truth, I explored the opposite direction of inquiry in my thesis. In Lean, beauty becomes a guide towards truth, not a distraction from it.

In MathematiCS, beauty is inseparable from truth. There is a deep satisfaction in seeing an idea fully revealed, in watching the machinery of Lean confirm every detail of our reasoning. Perhaps there is an inherent elegance in confronting the truth directly, no matter what it may be, and in letting the formal system have the final word. Thanks to Lean, we get to experience the magnificent connection between what the mind conceives and what the world of logic affirms.

To formalize is, in a sense, to take the red pill [151], to awaken from the comforting illusion of understanding into the unyielding clarity of precision. Once the veil of informality is lifted, there is no return to the cheap satisfaction from "it is clear that…" or "one can easily see". Every assumption must stand in the open, and every claim be justified so that it can be meticulously checked. What was once intuitive becomes explicit, and the beauty that emerges is not the beauty of ease but of integrity. Formalization reminds us that truth is not declared but demonstrated, not intuited but earned.

Just as Neo [151] must relearn movement in the real world, the formalizer must relearn mathematical reasoning devoid of all narrative shortcuts. To work in Lean is to learn a new rhythm of thought — slower, more deliberate, but much clearer. Each claim demands its justification, each definition its boundary. In meeting these demands, our thoughts get significantly refined — stripped of pretense, disciplined by truth. The proof becomes not an argument but a mirror — revealing not just what we know, but how honestly we know it.

And yet, in the required rigor, a grace appears — the beauty of thought laid bare. It is no longer the external grace of presentation. Instead, it is the inner harmony of coherence. In Lean, precision becomes a kind of music — every definition, theorem, and proof a note in a grand composition where rigor and intention converge. The proof that holds, the type that fits, the abstraction that unites — all speak about the fulfilled beauty of necessity. Formalization becomes a kind of meditation — an act of attention so pure and creation so complete that truth and self momentarily coincide.

The journey through formalization is thus both intellectual and aesthetic, both rational and emotional, both logical and spiritual — an act of shaping both thoughts and self — a movement from wishing to expressing, from guessing to seeing, from disorientation to knowledge, from persuasion to proof, from reliance to independence, from improbity to integrity, from ornament to essence, from imagining to experiencing. In the end, the pursuit of truth and beauty are not two paths but one. Truth is what endures when all illusions are stripped away; beauty is the grace with which it endures. Without truth, there is not much value in beauty; without beauty, there is not enough pleasure in truth.

And, when the system at last accepts the proof, when Lean utters its silent *q.e.d.*, we witness thoughts crystallized into certainty, the rare moment when reasoning feels complete — because it truly is. What began as wonder before the stars finds its echo here, reflected in the luminous precision of formal reasoning, on the computer screen. The original wonder persists — only now, the answer compiles.